\documentclass[aps,pra,preprint,showpacs,preprintnumbers,amsmath,amssymb,footinbib]{revtex4-1}
\usepackage{graphicx}
\usepackage{bm}
\usepackage[colorlinks=true, pdfstartview=FitV, linkcolor=red, citecolor=blue, urlcolor=blue]{hyperref}
\newcommand{\bfq}{{\bf q}}
\newcommand{\beq}{\begin{equation}}
\newcommand{\eeq}{\end{equation}}
\newcommand{\bea}{\begin{eqnarray}}
\newcommand{\eea}{\end{eqnarray}}
\newcommand{\Gtwo}{{\rm G(2)}}
\newcommand{\SUseven}{{\rm SU(7)}}
\newcommand{\Vpt}{V_{\rm pt}}
\newcommand{\Vnpt}{V_{\rm npt}}
\newcommand{\bi}{\begin{itemize}}
\newcommand{\ei}{\end{itemize}}

\newcommand{\fig}{Fig.\ }

\newcommand{\la}{\label}

\newcommand{\V}{{\mathop{\cal V}}}
\newcommand{\W}{{\mathop{\cal W}}}
\newcommand{\Y}{{\bf Y}}
\def\q{\bf{q}}
\def\a{\alpha}      \def\b{\beta}   \def\g{\gamma}      
\def\d{\delta}      \def\D{\Delta}  
               
\def\m{\mu}

\newcommand\Tr{{\rm tr}}

\begin{document}
\preprint{BNL-96946-2012-JA, NIKHEF-2012-002, RBRC-943, RIKEN-MP-40, RIKEN-QHP-20}
\title{Effective matrix model for deconfinement in pure gauge theories}
\author{Adrian Dumitru}
\email{dumitru@quark.phy.bnl.gov}
\affiliation{
Department of Natural Sciences, Baruch College,
17 Lexington Avenue, New York, NY 10010, USA}
\affiliation{
RIKEN/BNL Research Center, Brookhaven National Laboratory, 
Upton, NY 11973, USA}
\author{Yun Guo}
\email{yun@fias.uni-frankfurt.de}
\affiliation{Department of Physics, Guangxi Normal University, 
Guilin 541004, China}
\author{Yoshimasa Hidaka}
\email{hidaka@riken.jp}
\affiliation{Quantum Hadron Physics Laboratory, RIKEN Nishina Center, Saitama 351-0198, Japan}
\author{Christiaan P. Korthals Altes}
\email{chrisaltes@gmail.com}
\affiliation{Centre Physique Th\'eorique au CNRS Case 907,
 Campus de Luminy F-13288 Marseille, France}
\affiliation
{NIKHEF Theory Group, Science Park 105, 1098 XG Amsterdam, The Netherlands}
\author{Robert D. Pisarski}
\email{pisarski@bnl.gov}
\affiliation{
Department of Physics, Brookhaven National Laboratory, 
Upton, NY 11973}
\affiliation{
RIKEN/BNL Research Center, Brookhaven National Laboratory, 
Upton, NY 11973, USA}
\begin{abstract}
We construct matrix models for the deconfining phase transition in
$SU(N)$ 
gauge theories, without dynamical quarks, at a nonzero temperature $T$.  We
generalize models with zero 
\cite{Meisinger:2001cq, *Meisinger:2001fi} and one \cite{Dumitru:2010mj} free parameter
to study a model with two free parameters: besides
perturbative terms $\sim T^4$, we introduce terms
$\sim T^2$ and $\sim T^0$.
The two $N$-dependent parameters are determined by fitting 
to data from numerical simulations on the lattice for the pressure,
including the latent heat.  Good agreement is found for
the pressure in the semi-quark gluon plasma (QGP), which is
the region from $T_c$, the critical
temperature, to about $\sim 4 \, T_c$.
Above $\sim 1.2 \, T_c$, the pressure 
is a sum of a perturbative term, $\sim + \, T^4$, and a 
simple non-perturbative term, essentially 
just a constant times $\sim - \, T_c^2 \, T^2$.
For the pressure, the 
details of the matrix model only enter within a very narrow window,
from $T_c$ to $\sim 1.2 \, T_c$, whose
width does not change significantly with $N$.
Without further adjustment, the model also agrees well with lattice
data for the 't Hooft loop.  This is notable, because
in contrast to the pressure,
the 't Hooft loop is sensitive to the details of the matrix 
model over the entire semi-QGP.  
For the (renormalized) Polyakov loop, though, our results disagree
sharply with those from the lattice.
Matrix models provide a natural and generic explanation for why the
deconfining phase transition in $SU(N)$ gauge theories is of first
order not just for three, but also for four or more colors.
Lastly, we consider gauge theories where there is no strict order
parameter for deconfinement, such as for a $G(2)$ gauge group.  To agree with
lattice measurements, in the $G(2)$ matrix
model it is essential to add
terms which generate complete eigenvalue repulsion in the confining phase.
\end{abstract}
\maketitle
\tableofcontents

\section{Introduction}
\label{sec:introduction}

Quantum ChromoDynamics (QCD), is a theory of great beauty. Only a handful of
input parameters uniquely define its behavior at all distance scales. For the
``pure'' glue theory, with no dynamical quarks, there is only a single
parameter, which sets the overall scale of length.

Conversely, with so few parameters, it is very difficult to compute from first
principles. A useful technique is to perform numerical simulations on the
lattice.  While at present simulations with dynamical quarks are extremely
challenging, in the pure glue theory results close to the continuum limit can be
obtained.

The behavior of gauge theories at a nonzero temperature, $T$, is of particular
interest.  There are lattice results for the thermodynamic behavior of pure
$SU(N)$ gauge theories 
\cite{DeTar:2009ef,Petreczky:2012rq} for two 
\cite{Engels:1988ph, *Engels:1989fz, *Engels:1992fs, *Engels:1994xj, *Engels:1998nv}, three
\cite{Boyd:1991fb,Beinlich:1996xg, Umeda:2008bd, Borsanyi2012}, and four or more colors
\cite{Lucini:2003zr, Lucini:2005vg, Teper:2008yi, Panero:2009tv, Datta:2010sq, Lucini:2012wq}. 
They show that in the pure
glue theory, the thermodynamics for small $N$ is like that for large $N$. The
lattice results find a pressure, $p(T)$, which is small in the confined phase,
below the critical temperature, $T_c$. Scaled by the 
pressure of an ideal gas of
gluons, the ratio $p(T)/p_{\rm ideal}(T)$ grows sharply in the range from $T_c$
to about $4.0 \, T_c$, and is then approximately constant above $T > 4 \, T_c$;
$p/p_{\rm ideal} \sim 0.85$ at $4 \, T_c$. We term the region over which the
pressure grows markedly, from $T_c$ to $\sim 4.0 \, T_c$, as the semi-quark
gluon plasma (QGP) \cite{Dumitru:2010mj, Hidaka:2008dr, *Hidaka:2009hs,
*Hidaka:2009xh, *Hidaka:2009ma}; see, also
\cite{Pisarski:2000eq, Dumitru:2003hp, Dumitru:2004gd, Oswald:2005vr}.

At large $N$, the sharp increase in the pressure at $T_c$ is elementary. In the
confined phase there are only colorless glueballs, so the pressure is small,
$\sim 1$. In the deconfined phase, the pressure is proportional to the number of
gluons, $= N^2 -1 \sim N^2$, and so large. Lattice simulations for three
\cite{Boyd:1991fb,Beinlich:1996xg} and even two colors
\cite{Engels:1988ph, *Engels:1989fz, *Engels:1992fs, *Engels:1994xj, *Engels:1998nv} 
also find that the pressure below $T_c$ is much smaller than that above.

The similarity between small and large $N$ can be made quantitative. To
parametrize the deviations from ideality, consider the conformal anomaly, which
is the energy density, $e(T)$, minus three times the pressure.  Dividing by the
number of gluons, lattice studies show that as a function of $T/T_c$, the
dimensionless ratio $(e-3p)/((N^2 -1) T^4)$ is similar for $N = 3$, $4$ and $6$
\cite{Panero:2009tv,Datta:2010sq}: above $T > 1.2 \, T_c$, this ratio falls with
increasing $T$.

\begin{figure}
\includegraphics{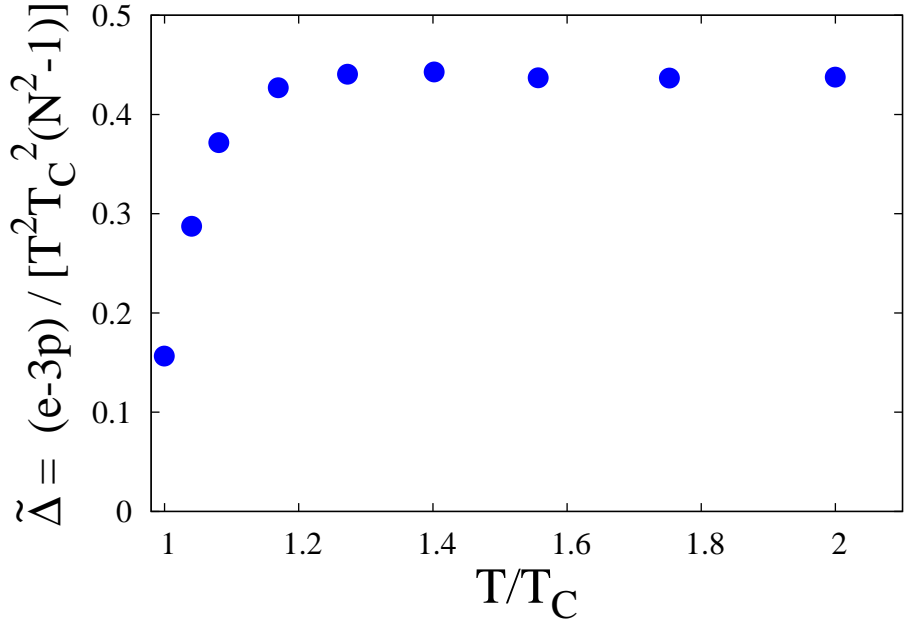} \caption{\label{figwhot}Plot of the
trace anomaly divided by $T^2$, $(e-3p)/(8 T^2 T_c^2)$, from the data of 
Umeda {\it et al.}, Ref. \cite{Umeda:2008bd}. } 
\end{figure}

Since the order of the transition changes with $N$, this similarity breaks down
close to the transition, below $1.2 \, T_c$. The deconfining transition is of
second order for two colors 
\cite{Engels:1988ph, *Engels:1989fz, *Engels:1992fs, *Engels:1994xj, *Engels:1998nv}, weakly first order for three
\cite{Boyd:1991fb,Beinlich:1996xg}, and first order for all $N \geq 4$
\cite{Lucini:2005vg,Panero:2009tv,Datta:2010sq}. While the ratio of the latent
heat to the number of gluons is a number of order one as $N \rightarrow \infty$,
this ratio increases significantly as $N$ does
\cite{Lucini:2005vg,Datta:2010sq}.

In this paper we use these detailed results from the lattice to develop an
effective theory for deconfinement in the pure glue theory. A common model for
deconfinement is to take an term like that of an ideal gas minus a MIT ``bag''
constant, $b$: $p(T) \sim c_1 T^4 - b$. If true, then above $\sim 1.2 \, T_c$,
the conformal anomaly$/T^4$ would fall off as $\sim b/T^4$. 

To understand the
fall off, consider the following quantity 
\cite{Meisinger:2001cq, Pisarski:2006hz,Andreev:2007zv,Panero:2009tv,Datta:2010sq,Dumitru:2010mj,Begun:2010eh,Borsanyi2012}:
\beq
\widetilde{\Delta}(T) = \frac{ e(T) - 3 p(T)}{(N^2 - 1) \; T_c^2 \; T^2} \; .
\label{rescaled_conf_anomaly}
\eeq
That is, we plot the conformal anomaly divided not by $T^4$, but by $T^2$ times
$T_c^2$, to form something dimensionless.  We also divide by
the number of perturbative gluons, to be able to compare different numbers
of colors.  If a bag constant dominated, this quantity would fall off
at large $T$ as $\widetilde{\Delta}(T) \sim b/T^2$.

For three colors, there is precise data from the
WHOT colloboration \cite{Umeda:2008bd}; see, also,
very recent results from Borsanyi {\it et. al.}  \cite{Borsanyi2012}.
We show $\widetilde{\Delta}(T)$,
extracted from the WHOT data, 
in Fig. (\ref{figwhot}). Between $1.2 \, T_c$ and $2.0 \, T_c$, this
ratio is constant to a remarkable degree, $\sim 1\%$. This implies that in this
range, the pressure can be approximated as 
\beq 
p(T) \approx c_1 \left( T^4 - c_2 \, T_c^2 \, T^2 \right) 
\;\; ; \;\; c_2 \approx 1.00 \pm 0.01 \;\; ; \;\;
T/T_c: 1.2 \rightarrow 2.0 \; . 
\label{approx_press} 
\eeq 
There is no data
from Ref. \cite{Umeda:2008bd} above $2.0 \, T_c$. 
The data of the Bielefeld collaboration \cite{Boyd:1991fb} 
can be used to show that $(e-3p)/T^2$ is approximately
constant in the entire semi-QGP. The same constancy is also seen for four and
six colors, albeit with larger error bars 
\cite{Panero:2009tv,Datta:2010sq}.  Notably, the
width of the window in which $\widetilde{\Delta}(T)$ is constant does 
{\it not} appear to change significantly 
as the number of colors increases, from
$N=3$ to $N = 4$ or $6$ \cite{Panero:2009tv,Datta:2010sq}.

The fact that non-ideal terms in the pressure are $\sim T^2$ was first noted in
Ref. \cite{Meisinger:2001cq} and then later in Ref. \cite{Pisarski:2006hz}; see
also \cite{Andreev:2007zv,Begun:2010eh,Borsanyi2012}. 
One implication, used previously by us in Ref.
\cite{Dumitru:2010mj} and also here, is that since the ideal gas term is $T^4$,
any non-perturbative terms which we introduce are assumed to be proportional
only to powers of $\sim T^2$, $\sim T^0$, {\it etc.}

Less obviously, with hindsight many features of our model 
can be understood from
Fig. (\ref{figwhot}). We use a $SU(N)$ matrix model, where the basic variables
are the eigenvalues of the thermal Wilson line.  The vacuum 
at a temperature $T$
is given by varying an effective Lagrangian with respect to these $N$ matrix
variables, the $q$'s; the pressure is (minus) the value of 
the potential at this
minimum.  Even without knowing what the $q$'s are, though, clearly the simplest
way of obtaining a constant term $\sim T^2$ in the pressure is simply to
introduce a similar constant in the potential for the $q$'s.

This implies that over most of the semi-QGP, above $1.2 \, T_c$ to 
$\sim 4.0 \, T_c$, 
the decrease of the pressure, relative to that of an ideal gluon gas, is
dominated by a ``trivial'' term, a pure number times $\sim - \, T_c^2 \, T^2$. 
In the deconfined phase, there is a non-trivial 
minimum of the effective theory,
in which the $q$'s are non-zero, only in a surprisingly narrow window, for 
$T \leq 1.2 \, T_c$.

This is not generic to matrix models, but is forced upon us by the lattice data
\cite{Boyd:1991fb,Beinlich:1996xg,Umeda:2008bd,Lucini:2005vg,Panero:2009tv,Datta:2010sq}. 
A matrix model for deconfinement was first introduced by
Meisinger, Miller, and Ogilvie \cite{Meisinger:2001cq}.  This model has no free
parameters, and, as we show later, a much broader result for $(e-3p)/T^4$ than
seen by the lattice. In Ref. \cite{Dumitru:2010mj} we introduced a model with
one free parameter, which allowed us to fit the narrow result for the conformal
anomaly observed on the lattice.  In the present work we show that in order to
agree with the lattice data near $T_c$, in particular for the latent heat,
requires a two parameter model.

Ignoring such details, the fundamental question remains:
{\it what is
the origin of this constant term in the pressure, 
$\sim - \, T_c^2 \, T^2$, which dominates the
corrections to non-ideality above $\sim 1.2 \, T_c$?}  

One natural
guess is a gluon ``mass'' $\sim T_c$.  Such a mass is typical, for example, in
solutions to the Schwinger-Dyson equations of QCD \cite{Cornwall2011}.  After
all, if we expand the pressure of massive gas about the massless limit, 
when $m \ll T$ the leading correction {\it is} $\sim - m^2 \, T^2$.

However, there is no simple form for the mass which will give 
such a flat result for 
the rescaled conformal anomaly of Eq. (\ref{rescaled_conf_anomaly}).
If we take $m \sim T$, then it is not difficult to see
that the ideal pressure of such a massive gas is a pure number times $T^4$. If
we take $m$ to have a constant mass, proportional to $T_c$, 
then numerically one
can check that the only way to obtain a correction $\sim T^2$ 
is if for small masses, $m \ll T_c$;
but the value of $c_2$ in Eq. (\ref{approx_press}) 
requires that $m$ is a number
of order one times $T_c$. The only way to fit the pressure is if the gluon mass
is an involved function of $T$.
This is what is done in quasiparticle models
\cite{Peshier:1995ty,Castorina:2011ja}; see, {\it e.g.}, Eq. (27) of
\cite{Castorina:2011ja}, where their $m(T)$ involves three parameters. In
contrast, in a matrix model we can fit the pressure, with a similar accuracy,
with one free parameter, the constant $\sim - T_c^2 \, T^2$.

It is also useful to note that for $SU(N)$ gauge theories
in $2+1$ dimensions \cite{Caselle:2011fy, Caselle:2011mn}, 
the pressure has a similar form to that in Eq.
(\ref{approx_press}). 
From Fig. (6) of Caselle {\it et al.} \cite{Caselle:2011mn}, 
above temperatures of $\sim 1.25 \, T_c$
the pressure is approximately $p(T) \approx T^3 - T_c T^2$.
That is, the non-ideal term is {\it again} $\sim T^2$;
this is not a mass term, since in $2+1$ dimensions this would be linear
in the temperature, $\sim m^2 T$.

Thus the term $\sim - \, T_c^2 \, T^2$ in the pressure 
does not appear to be just a
gluon mass.  Nor do we know why the window, from $T_c$ to $1.2 \, T_c$,
is so narrow.  One might guess that it is an effect $\sim 1/N^2$, but
to fit the lattice data, our model requires that the window has about the
same width for four and six colors as it does for three.
A term $\sim T^2$ is like the free
energy of massless fields in two dimensions, such as strings, but 
how can strings contribute to the free energy,  $\sim N^2$, 
in the {\it de}confined phase?

With our matrix model we also predict features which 
are not usually addressed by
other effective theories.  Taking the parameters from the fit to the pressure,
we compute the 't Hooft loop in the semi-QGP, and find good agreement with
lattice data. In contrast to the pressure, for the
't Hooft loop we find that the effects of ``non-ideal'' terms
matter not just below $1.2 \, T_c$, but over the {\it entire}
semi-QGP, from $T_c$ to $\sim 4.0 \, T_c$.  
For this reason, the computation of the 't Hooft loop is 
a sensitive and crucial test of the model.

There is one glaring discrepancy, though, between the lattice data and our
model. In our model the Polyakov loop only differs from one when the matrix
$q$'s are nonzero, below $1.2 \, T_c$. This is very different from the behavior
of the renormalized Polyakov loop from the lattice
\cite{Gupta:2007ax,Mykkanen:2012ri}, which varies over the entire semi-QGP.  We
do not understand the reason for this difference, and comment
further in the Conclusions, Sec. (\ref{sec:conclusion}).

Matrix models can help give further insight into deconfinement.
For $SU(N)$ theories without dynamical quarks, the existence
a global $Z(N)$ symmetry gives a rigorous
definition for the deconfining phase transition; it implies that
the confined phase is necessarily $Z(N)$ symmetric.  
In a $SU(N)$ matrix model, the confined 
phase is uniquely characterized by the complete repulsion of eigenvalues.
The structure of this point in the space of eigenvalues,
which is the Weyl group, is such that the $Z(N)$ symmetry 
of the confined phase is automatic.

Svetitsky and Yaffe showed that for three colors, that the
deconfining transition is generally of first order \cite{Svetitsky:1982gs}.
This is because for three colors, 
the $Z(3)$ symmetry allows one to form a cubic invariant of Polyakov loops.
As is typical of mean field theory, such a cubic
invariant ensures that the transition is of first order.
For four or more colors, though, $Z(N)$ invariant terms are of quartic
or higher order, and there is no prediction.  Note that this assumes
that the relevant variables are Polyakov loops, which are elements of
the Lie group.

As discussed above, though,
lattice simulations show that the deconfining transition is of
first order not just for three, but for four or more colors.  
This is explained naturally by matrix models.
In expanding our matrix model about the confining vacuum,
we find a cubic invariant for any $N \geq 3$.  
This result is not special to the parameters of our model: 
we show how if we expand a general matrix model about the point of
complete eigenvalue repulsion, then there is a cubic
invariant.  The presence of a cubic invariant, now in terms
of elements of the Lie algebra, instead of the Lie group,
implies that the deconfining phase transition is of first order not just
for $N=3$, but for $N \geq 4$.

The details of the transition in our simple matrix model appear to agree with
results from the lattice.  In particular, while the latent heat (scaled
appropriately by $N^2-1$) grows from $N = 3$ to large $N$, the order-disorder
interface tension, at $T_c$, is rather small for all $N$.  
In our model this is because there is a potential 
with a small barrier between two vacua which are relatively far apart.  

We also extend our analysis to include other gauge groups.  A particularly
interesting example is provided by the exceptional group $G(2)$
\cite{Holland:2002vk, Pepe:2006er, Cossu:2007dk, HoyosBadajoz:2007ds, Wellegehausen:2009rq, *Wellegehausen:2010ai, *Wellegehausen:2011sc,Caselle2012}
This group has a trivial center, so there is no order parameter for 
deconfinement.  Nevertheless, lattice simulations find that there
is a strongly first order phase transition between a deconfined phase
at high temperature, and a low temperature phase in which the expectation
values of Polyakov loops are very small.   That is, 
although there is no center symmetry, the low temperature phase still
appears to confine.

We do not find that the simplest matrix model reproduces the lattice data
for a $G(2)$ gauge group.
However, we show how to add terms in the effective matrix model to ensure
that the expectation value of Polyakov loops is small in the low
temperature phase.  Given our experience with $SU(N)$, for 
$G(2)$ we add terms which generate the complete repulsion of eigenvalues
in the low temperature phase.  Thus 
the expectation value of Polyakov loops are small not because
of a center symmetry, but because of eigenvalue repulsion.
This is also reflected in ther thermodynamic behavior.  
With the simplest choice of parameters in our model, we find
that the sharp maximum in $(e - 3p)/T^4$, found for $SU(N)$, does
not appear for a $G(2)$ gauge group.

Especially given the wealth of 
experimental results from heavy ion collisions at
ultrarelativistic energies, of course we wish to generalize this model to
theories with dynamical quarks, and in particular to QCD.  Before doing so,
however, we feel it is necessary to understand the transition in the pure glue
theory with some care.

The outline of the paper is as follows.  
In Sec. (\ref{sec:twoandthree}) we give an introduction to how our
matrix model works for the simplest case of two and three colors.  
We discuss the basic justification for our approach, which is a type of
large $N$ expansion.  We also discuss the quantities which the model
can compute.  Besides familiar quantities, such as
the pressure and the Polyakov loop, this includes
the 't Hooft loop, also known as the order-order interface tension.
We also compute the order-disorder interface tension at $T_c$ in our model.

There are of course other effective models which treat the theory
near $T_c$.  Besides the quasiparticle models discussed above
\cite{Peshier:1995ty,Castorina:2011ja}, these include:
$Z(N)$ effective theories 
\cite{Vuorinen:2006nz, *deForcrand:2008a, *Zhang:2011aa};
solutions of the functional renormalization group
\cite{Braun:2007bx, *Marhauser:2008fz, *Braun:2010cy}; and
Polyakov loop models \cite{Scavenius:2002ru, *Fukushima:2003fw, *Hell:2009by, *Fukushima:2010bq, *Buisseret:2011ms, *Horvatic:2010md, *Hell:2011ic, Layek:2005fn}.

We stress that we are not attempting to solve the theory near the
deconfining transition, but only to develop an effective theory.  As
such, we find it notable that our model, with only two parameters,
provides a good fit to two {\it functions} of temperature: both
to the pressure, and to the 't Hooft loop.  While our two parameters
are $N$-dependent, allowing this provides a good fit to $SU(N)$ for all $N$.

In Sec. (\ref{sec:effectivepotential})  
we discuss the type of matrix model which can be constructed for
arbitrary classical groups.  We emphasize the role which invariance
under the Weyl group plays, and the utility of understanding the concept
of the Weyl chamber.

In Sec. (\ref{sec:thermo_model}) we discuss models relevant to
$SU(N)$, up to those with two free parameters.  
We introduce a technical assumption, which we call the ansatz of
uniform eigenvalues, which allows us to compute many quantities 
analytically for arbitrary $N$.  
We also compare to lattice results on the interaction measure and
especially the latent heat.  

In Sec. (\ref{sec:interface}) we compute 
some of the interface tensions
which arise.  The order-disorder interface for
arbitrary $N$ is computed analytically under the uniform eigenvalue ansatz.
For two colors, we evaluate the 't Hooft loop analytically.
For three colors, the 't Hooft loop is computed numerically.

The numerical solution of the model for four to seven colors is given in
Sec. (\ref{sec:su45}).
We find that for the interaction measure and the Polyakov loop, that
the uniform eigenvalue ansatz works remarkably well for these
values of $N$.

In Sec. (\ref{sec:firstorder}) we demonstrate that matrix models
naturally explain why the deconfining transition is of first order
not just for three colors \cite{Svetitsky:1982gs}, but for
four or more.  

We consider matrix models for the $G(2)$ group in 
Sec. (\ref{sec:weyl}).  

A summary and conclusions are given in Sec. (\ref{sec:conclusion}).

\section{Outline of the method}
\label{sec:twoandthree}

In this section we give an elementary overview of how our matrix models
works for two and three colors.  Because we treat general gauge groups
later, here we shall concentrate on the 
assumptions implicit in our approach, and the
physical quantities which we can compute in our model.

\subsection{Two colors}

General results and lattice simulations show that the expectation value
of the Polyakov loop changes near $T_c$.  To model this, we take the
simplest ansatz which will generate such an expectation value, taking
$A_0$ to assume a constant, nonzero value.
By a gauge rotation we can take the background field for $A_0$.
to be diagonal.  For two colors there is only
one diagonal direction, along the Pauli matrix $\sigma_3$,
\beq
A_0 = \frac{\pi T}{g} \; q \; \sigma_3 \;\; , \;\;
\sigma_3 = 
\left(
\begin{array}{cc}
1 & 0  \\
0 & -1 \\
\end{array}
\right) \; .
\label{twocolor_ansatz}
\eeq
In this background field the Wilson line is
\beq
{\bf L}(\vec{x}) = {\bf P} \exp\left(i g \int^{1/T}_0 
A_0(\vec{x},\tau) \; d \tau \right)
\; = \;
\left(
\begin{array}{cc}
{\rm e}^{i \pi q} & 0    \\
0 & {\rm e}^{ - i \pi q} \\
\end{array}
\right)
 \; ,
\label{twocolor_lineA}
\eeq
and the Polyakov loop, in the fundamental representation, is
\beq
\ell = \frac{1}{2} \; \Tr \, {\bf L} = 
\cos(\pi q) \; .
\eeq

The usual perturbative vacuum is $q = 0$, ${\bf L} = {\bf 1}$, and
$\ell = 1$.  The $Z(2)$ transform of the usual vacuum is
$q = 1$, ${\bf L} = - {\bf 1}$, and $\ell = -1$.

It what follows it is convenient to restrict the variable $q$ to lie
in the region $q: 0 \rightarrow 1$.  If we do so, a $Z(2)$ tranformation
is given by
\beq
q \rightarrow 1 - q \;\; : \;\;
{\bf L} \rightarrow ( - ) \left(
\begin{array}{cc}
{\rm e}^{- i \pi q} & 0    \\
0 & {\rm e}^{ i \pi q} \\
\end{array}
\right)
 \;\;  ; \;\; \ell \rightarrow - \; \ell \; .
\label{twocolor_lineB}
\eeq
Notice that this Wilson line is only $-1$ times that in Eq.
(\ref{twocolor_lineA}) after allowing for permutation of the eigenvalues.

The confining vacuum is 
\beq
q_c = \frac{1}{2} \;\; : \;\;
{\bf L}_c = \left(
\begin{array}{cc}
i & 0    \\
0 & -i \\
\end{array}
\right)
 \;\;  , \;\; \ell_c = 0 \; .
\label{twocolor_conf}
\eeq

Given the known behavior of the Polyakov loop near $T_c$, 
this ansatz must characterize,
at least to some extent, the deconfining phase transition.  The essential
question of physics is the following.  For example,
in the confined phase does
$\ell_c$ vanish because it is dominated by $q_c$, or because
fluctuations, about various values of $q$, wash it out?

For an infinite number of colors, at any temperature 
the vacuum is dominated by a ``master field''.  At nonzero temperature
this master field must be related to the matrix $q$ above.
As is typical of large $N$, fluctuations in disconnected quantities,
such as the Polyakov loop, are suppressed by powers of $1/N^2$.  
To be sure, at a given temperature $T$, we can only deduce what the
value of $q$ is from measurements on the lattice.  These measurements
also give us no insight into the what effective theory determines
this $q$.

We now make the egregious assumption
that a large $N$ expansion is a good approximation
for all values of $N$, even for $N=2$.
As discussed in the Introduction,
Sec. (\ref{sec:introduction}), there are many similarities between
the transitions for small and large $N$.  A standard large $N$ expansion 
would imply computing at infinite $N$, and then expanding in
$1/N^2$.  Instead, we adopt a more expansive view,
and assume that we can expand about 
$q$'s appropriate to a given value of $N$.  
This can be considered as a type of ``generalized'' large $N$ expansion.
By expanding directly in the $q$'s appropriate to a given $N$, we are
directly incorporating some subset of corrections in $1/N^2$ more directly
than if we had followed the standard approach.  This is also natural,
since for either large or small $N$, we can only construct our effective
theory with the input of lattice data.  

This leaves open the question of how we could systematically develop a
procedure for computing corrections to our generalized large N expansion.
There will certainly be corrections in both $1/N$ and in powers of the
coupling constant, $g^2$.  We defer this analysis for now,
and proceed in developing an approximation to what is certainly lowest order.

The simplest thing to do is to compute the free energy in the presence of
the background field in Eq. (\ref{twocolor_ansatz}).  This is a standard
computation; see, {\it e.g.}, Sec. II of Ref. 
\cite{Bhattacharya:1990hk, *Bhattacharya:1992qb}.
It will be done for a general gauge group in Sec. 
(\ref{sec:effectivepotential}).  The result is
\beq
\V_{pt}(q) 
= \pi^2 \, T^4 \left( - \frac{1}{15} + \frac{4}{3} \; q^2 (1-q)^2 \right)
\; .
\label{two_color_pot}
\eeq
For $q = 0$ this is the free energy of an ideal gas of three, massless gluons.
This is degenerate with $q = 1$, which reflects the $Z(2)$ symmetry of the
pure glue theory.

This potential in $q$ can be used to compute the
't Hooft loop, or the order-order interface tension
\cite{Bhattacharya:1992qb,Giovannangeli:2002uv, *Giovannangeli:2004sg}. 
This is the action
for a state which tunnels from $q = 0$ at one end of a long spatial box,
to $q = 1$ at the other.  The computations are typical of
barrier penetration in one spatial dimension, 
and are given in Sec. (\ref{sec:interface}).  

Perturbatively, the confined state at $q = 1/2$ is an extremal point of the
potential, but a maximum.  To describe the transition to a confined
state, we have to add non-perturbative
terms to the effective Lagrangian to force the
vacuum to go from the perturbative vacua, at $q = 0$ and $1$, to $q = 1/2$.
After a little experimentation, the nature of these terms 
can be guessed.

Given the behavior of the interaction measure in Fig. (\ref{figwhot}),
we assume that any term is proportional to $T^2$.  It must also be $Z(2)$
symmetric.  Thus one such term is
\beq
\V_{npt} = - \; \frac{4 \pi^2}{3} \; c \; T^2 \, T_c^2 \; 
|{1\over 2}\Tr \; {\bf L}|^2 \;
= - \; \frac{4 \pi^2}{3} \; c \; T^2 T_c^2 \; \cos(\pi q)^2 \; .
\label{non_pert_first}
\eeq
To make up the mass dimensions, we use the critical temperature, $T_c$.
Since $T_c$ is a manifestly non-perturbative quantity, so is 
this potential.  

We then take the total potential as the sum of $\V_{pt}$ and $\V_{npt}$.
At high $T$, where $\V_{pt}$ dominates, the perturbative vacuum is favored.
Near  $T_c$, where $\V_{npt}$ becomes as important as $\V_{pt}$, the confined vacuum, $q = 1/2$, is.
It is easy to check that the deconfining transition occurs
for $c = -1/16$.  

The problem is that the transition is not of second order, but of first.  
That is, when $c = 1/16$, the vacua at $q = 0$ (or $1$) is degenerate with
the confined phase, $q = 1/2$, but there is a nonzero barrier between the
two.  That is, the theory stays in the perturbative QGP until $T_c$, when
it goes directly into the confined phase, with no semi-QGP in between.

This example illustrates a more general problem. 
For static fields, $A_0$ couples to the spatial degrees
of freedom $A_i$ as an adjoint scalar.  Thus when
$q$ and so $A_0$ are nonzero, there is an adjoint Higgs phase
\cite{Pisarski:2006hz, Unsal:2008ch, *Simic:2010sv}.
Thus in principle there could be {\it two} phase transitions.
Besides the usual 
deconfining transition at $T_c$, there could be a second transition,
at a temperature above $T_c$, when the theory first enters the adjoint Higgs
phase.  While possible, the lattice simulations give no indication of such
a second transition above $T_c$.  

To avoid this,
we add terms to the effective Lagrangian to ensure that
there is no chance of such a second 
transition developing.  
This is easy.
If we add a term which is {\it linear} in $q$ for small $q$, then such
a term will act to generate an expectation value for $q$ for any temperature.
That is, the theory is always in an adjoint Higgs phase.

Any term which we add must respect the $Z(2)$ symmetry.  Under $Z(2)$,
$ q \rightarrow 1 -q$.  This means that we cannot add a term 
which is just $\sim q $, but we can add a term $\sim q (1 - q)$,
\beq
\V_{npt} = - \; \frac{4 \pi^2}{15} \; T^2 \, T_c^2 \; 
 c_1 \; q (1 - q) \; .
\label{non_pert_twoA}
\eeq

This model was first proposed in Ref. \cite{Meisinger:2001cq}.
Meisinger and Ogilvie showed how it can arise from the expansion of
a massive field in a background field $A_0 \sim q$ 
\cite{Meisinger:2001fi, KorthalsAltes2012}.  Here we
emphasize that including such a term is not optional, but is essential to
avoid an unwanted phase transition above $T_c$.

To fit to the lattice data, we find it necessary to add two more terms
to the non-perturbative potential,
\beq
\V_{npt} = - \; \frac{4 \pi^2}{3} \; T^2 \, T_c^2 \; 
\left(  \frac{1}{5}c_1 \; q (1 - q) \; + \; c_2 \; q^2 (1 - q)^2 - c_3 \right) \; .
\label{non_pert_twoB}
\eeq

The term $\sim c_3$ is trivial as it does not affect the expectation value
of $q$. The term $\sim c_2$ is clearly allowed,
as it is identical to the perturbative
potential in Eq. (\ref{two_color_pot}).  

We end up with a model which appears to have
three free parameters.  However, 
we need to adjust the parameters so that the transition occurs at $T_c$.
Secondly, we need a constraint to fix the pressure in the confined phase.
At large $N$, the pressure in the confined phase
is $\sim 1$, relative to that $\sim N^2$ in the deconfined phase.
We adopt the simplest possible convention, and assume that
the pressure vanishes identically at $T_c^-$.  
In practice, it would be better to fit the pressure in the confined phase
to some sort of hadronic (that is, glueball) resonance gas.  
Because we don't do that here, we find that our model exhibits
unphysical behavior for the pressure below $T_c$.  This is entirely
an artifact of our overly simplistic assumptions.

Two conditions on three parameters then leaves one free parameter.
Surprisingly, we show later that fits with one free parameter do a remarkably
good job of fitting the pressure and the 't Hooft
loop, at least if one is not too near $T_c$.

The terms in $\V_{npt}$ are clearly not unique, as we can
add arbitrary
powers of $q(1-q)$.  Further, there is no reason why we could not add terms
which are not $\sim T^2$, but $T^0$, {\it etc.}.  In fact we shall have
to add a fourth term $\sim T^0 c_3'$ later, 
in order to fit the region near $T_c$ for $N \geq 3$.  This is then
a model with two free parameters.  With such a model, we can then fit the
region very near $T_c$, including the latent heat.

Even so, we find it striking that such simple models, with at most two
free parameters, can fit several functions of temperature.  
It satisfies one of the cardinal virtues of any good
mean field theory, which is simplicity.

(For two colors, in our theory the energy density
is negative in a narrow interval, to $\sim 1.01 T_c$.
This can be corrected by adding further terms to the potential.
However, we do not expect our model to describe the critical region,
near $T_c$, with precision.)

Viewing the $A_0$ as an adjoint Higgs
theory yields  the following. When 
$\langle q \rangle \neq 0$, there is a splitting of masses.
Since $A_0 \sim \sigma_3$, the off diagonal components develop
a mass $\sim q$, while the diagonal ones do not.  All components develop
an equal mass from Debye screening.
In our model, with Eqs. (\ref{non_pert_twoA}) and
(\ref{non_pert_twoB}), the theory is in
an adjoint Higgs phase for all $T > T_c$.  In practice, for the parameters
of the model, this condensate is {\it very} small 
except near $T_c$; above 
$\sim 1.2 \, T_c$, the condensate effectively vanishes. 

\subsection{Three or more colors}

For three colors the background field can lie in one of two directions,
\beq
A_0 = \frac{\pi T}{3 \, g} \; 
\left( q_3 \;  \lambda_3 + q_8 \; \lambda_8 \right) \;\; ; \;\;
\lambda_3 = {\rm diag}(1, -1,0) \;\; ; \;\;
\lambda_8 = {\rm diag}(1, 1,-2) \;\; . 
\label{ansatz_three}
\eeq
Except for overall constants, $\lambda_3$ and
$\lambda_8$ are the usual Gell-Mann matrices.

Unlike for two colors, the two directions now have different effects.
Moving along the $\lambda_8$ generates $Z(3)$ transformations:
${\bf L} = {\bf 1}$ when $q = 0$, and 
${\bf L} = \exp(2 \pi i/3) {\bf 1}$ when $q = 1$.  Moving along
$\lambda_3$ takes one to the confining vacuum, which is
\beq
{\bf L}_c = {\rm diag}\left( {\rm e}^{2 \pi i/3}, {\rm e}^{- 2 \pi i/3}, 1
\right) \;\; ; \;\; \ell_c = 0 \; .
\eeq

When $\langle q_3 \rangle \ll 1$, then the 't Hooft loop is determined
simply, by the path along $\lambda_8$.  
As discussed in Sec. (\ref{sec:interface}), though, near $T_c$, 
when $\langle q_3 \rangle$ is substantial, the path is along both
directions.

The generalization to arbitrary Lie groups is treated in the next section.
If we concentrate only upon the behavior of the pressure, then we only
need consider that path to the confining vacuum, and the problem is
relatively straightforward.  If we were to consider arbitrary 't Hooft
loops for $SU(N)$ when $N \geq 4$, though, it would be a much harder problem.

When the transition is of first order, as is generally true for most
gauge groups, then besides the order-order interface tension, there is,
at $T_c$, also an order-disorder interface tension.  This describes the
barrier between the deconfined, and the confined, phases at $T_c$.

\section{The effective potential for general gauge groups}
\label{sec:effectivepotential}

In this section we compute the perturbative effective potential 
for $SU(N)$.  Its general form, consistent with its symmetries, is discussed.  
We compute in sufficient generality that the result
be generalized to other classical groups, or to $G(2)$,
in Sec. (\ref{sec:weyl}).
Orthogonal group groups are not simply connected.  To avoid lattice
artifacts due to $\Pi_1(SO(N))$ being non-trivial, lattice simulations
are usually done using the spin representation.

For $SU(N)$, $\bf q$ is a traceless, diagonal $N \times N$ matrix.
The Wilson line is
\bea
{\bf L}
= \exp\left(2 \pi i \; {\bf q} \right)
=\left (\begin{array}{cccc}{\rm e}^{2 \pi i q_1}&0&\cdots&0\\
0&{\rm e}^{2 \pi i q_2}&\cdots&0\\
\vdots&\vdots&\ddots&\vdots\\
0&0& \cdots&{\rm e}^{2 \pi i q_N}
\end{array}\right) \; .
\label{eq:polloop1}
\eea

We wish to compute the effective potential $\V({\bf q})$.
This can be done starting from  the field theory path integral .  Do the path integral keeping 
the eigenvalues of the Polyakov  fixed to the value $\exp(i2\pi{\bf q})$.
To do this in a manifestly gauge invariant way
we take  traces of powers the Polyakov loop.  We need as many of these powers as there are
independent eigenvalues, the rank $r$ of the group:
\beq
\exp(-V_3\; \V({\bf q})/T)=
\int {\cal D}A \; \Pi_{n=1}^r \; \d
\bigg(\Tr \; {\rm e}^{2 \pi i \, n \, {\bf q}}
- \Tr \; \overline{P(A_0)^n}\bigg)\exp(-S(A)) \; .
\la{eq:pathinthistogram}
\eeq
The average over the spatial volume $V_3$ is denoted by a bar.
This path integral is up to a normalization the probability that a given
configuration ${\bf q}$ of phases occurs in the system.
As is well known, this constrained path integral 
is in the large volume limit ($V_3\rightarrow\infty$) the traditional
free energy as a function of the quantum average of the loop.
In perturbation theory this path integral has been evaluated to order $g^3$. 

To lowest order one computes about a constant background field,
\beq
{\bf A}_0 = \frac{2 \pi T}{g} \; \bf q \; .
\label{fund_ansatz}
\eeq
The basic variables of our matrix model are the eigenvalues of
the Wilson line, which are gauge invariant.  
To leading order in the coupling constant $g^2$ these are given by expanding
about the background field in Eq. (\ref{fund_ansatz}).  Since
in general $A_0$ is gauge dependent, though, it is not surprising
to find that the relationship between the background $A_0$, and
the eigenvalues of the Wilson line, is more complicated
beyond leading order in $\sim g^2$
~\cite{Belyaev:1991gh, KorthalsAltes:1993ca, Bhattacharya:1992qb, Giovannangeli:2002uv, Giovannangeli:2004sg}.

For $SU(N)$ the only constraint on the eigenvalues is given by unimodularity:
\beq
q_1+q_2+\cdots+q_N=0.
\la{eq:unimod}
\eeq
The number of independent $q$'s in Eq. (\ref{eq:polloop1}) is $r$, the
rank of the group; for $SU(N)$, $r = N-1$.

What is the general form of the effective potential 
$\V({\bf q})$ which we can take?
The trace of the Wilson line, in an arbitrary representation $\cal R$,
${\bf L}_{\cal R}$, is gauge invariant.
By the character expansion, in the sum we can take only single
traces, ${\rm tr} \, {\bf L}_{\cal R}$,  if
arbitrary representations are included
\cite{Dumitru:2003hp, Dumitru:2004gd, Oswald:2005vr}.

In practice we find it convenient to take
traces of powers of loops, as
\beq
\V({\bf q})= T^4 \sum_R \sum_{n\ge 1} w^R_n(t)
\Tr_R\left({\bf L}^n +  \left({\bf L}^\dagger\right)^n \right) \; .
\la{eq:generalformpot}
\eeq
The weights $w_n^R(t)$ are taken real. 
As we see later, using infinite sums, as in Eq. (\ref{eq:generalformpot}),
allows us to write the our (matrix)
mean field theory in an especially simple manner.

At high temperature, to one loop order the potential is as
in Eq. (\ref{eq:generalformpot}), where only the adjoint representation
appears,
\beq
w^{adj}_n\sim{1\over{n^4}}.
\la{eq:pertwn}
\eeq

The traces in Eq. (\ref{eq:generalformpot}) involve the
identities
\beq
\sum_{n= 1}^\infty \; {1\over{n^{2p}}} \; \cos(2\pi n x)=(-)^{p-1}
{(2\pi)^{2p}\over {2(2p)!}} \; \widetilde{B}_{2p}(x)\; ,
\la{eq:bernoulliwinding}
\eeq
where $\widetilde{B}_{2p}(x)$ is 
a Bernoulli polynomial ~\cite{rizhikgradshtein}.

For $p=1$ and $p=2$ we define 
\bea
B_2(x)&=&x(1-|x|),~\mbox{mod}~1 \; , \nonumber\\
B_4(x)&=&x^2(1-|x|)^2,~\mbox{mod}~1 \; .
\label{define_B}
\eea
We make the unconventional choice of defining 
$B_k(x) = \widetilde{B}_k(x) - \widetilde{B}_k(0)$, so
that our $B_k$'s vanish at the origin, $B_2(0) = B_4(0) = 0$.
Outside of the range $|x| \leq 1$, 
they are defined to be periodic in $x$, modulo one.  This
reflects the fact that the $q_i$'s
are periodic variables.  

The quantities $B_2$ and $B_4$ are the building blocks of our 
matrix model.  Note that after the infinite summation over loops,
that we have a quartic polynomial in the eigenvalues of the Wilson line, the
$q$'s.  We then need to make a judicious
choice for the weights $w^{adj}_n$.   

The $SU(N)$ groups have a additional global
symmetry, the center group symmetry $Z(N)$. This means
that the potential is the same for ${\bf L}$ 
and for ${\rm e}^{2 \pi i k/N}{\bf L},~k=1,\cdots,N-1$.
This symmetry limits the representations $ \cal R$ to those 
having $N$-ality zero, such as the adjoint.

There is no such requirement in 
the absence of a center group symmetry, as occurs for the 
group $G(2)$.  Thus
besides the adjoint representation, which is a ${\bf 14}$,
we can also include the fundamental representation, which is a 
${\bf 7}$.  This is useful in Sec. (\ref{sec:weyl}).

\subsection{General computation to one loop order}
\la{sec:oneloopeffpotallgroups}
 
Our effective potential, $\V_{tot}({\bf q})$, is constructed from two
quantities: a perturbative potential, 
$\V_{pt}(\bf q)$, in which the only mass scale is the
temperature, and a non-perturbative 
term $\V_{non}(\bf q)$, which involves both a non-perturbative mass scale
and the temperature.

The perturbative potential is computed to one loop order order using
the steepest descent method, applied to Eq. (\ref{eq:pathinthistogram}):
\beq
\V_{pt}({\bf q})\; = \; T\; \Tr\log \det(-D^2({\bf q}))/V_3 \; .
\la{eq:oneloopdet}
\eeq
The trace is over all momentum and colour degrees of freedom. Spin
degrees of freedom are already summed over.
The gauge covariant d'Alembertian $D^2({\bf q})$ is 
\beq
D_\m({\bf q})=\partial_\m \; + \; 2 \pi i\; \d_{\m,0}\; [ \, {\bf q}, \; .
\la{eq:covariantderiv}
\eeq  

The color algebra can be diagonalized by using the Cartan basis.
This is comprised of $N-1$ diagonal matrices, the Cartan generators,
$\vec H=H_1,\ldots{.....,}H_{r}$, and
$N^2 - N$ off-diagonal matrices, the $E_{\a}$. Their
commutation relations define the root vectors $\vec\a$ in Cartan space:
\bea 
[\vec H,E_{\a}]&=&\vec\a ~E_{\a} \; , \\
~[E_{\a},E_{-\a}]&=&\vec\a \cdot \vec H\; , \\
~[E_\a,E_\b]&=&(\vec\a+\vec\b)E_{\a+\b} \; ,
~\mbox{if  $\vec\a+\vec\b$  is a root.}
 \la{eq:commrel}
\eea
We normalize all generators as
\beq
\Tr \left( H_iH_j \right)
={1\over 2}\; \d_{ij}
\; \; , \; \; \Tr \left( E_\a E_\b \right)\; 
= \; {1\over 2} \; \d_{\a,-\b} \; .
\la{eq:generatornorm}
\eeq

The roots have a length proportional to the normalization of the 
matrices $\vec H$. But  the combination:
\beq
H_\a={\vec \a.\vec H\over{{\vec\a}^2}} 
\eeq
does not depend on normalization. The commutation relations 
Eq. (\ref{eq:commrel}) tell us that the triplet
$\hat E_{\pm\a}=E_{\pm\a}/|\vec\a|$ and $ H_\a$ form a 
$SU(2)$ algebra:
 \beq
~[ H_\a, \hat E_{\pm\a}]=\pm \hat E_\a \; , \;  
[\hat E_\a,\hat E_{-\a}]=H_\a \; .
 \la{eq:su2ha}
 \eeq
As a diagonal matrix, the ${\bf q}$ can be rewritten
in terms of the $\vec H$,
$
{\bf q} \; =\; \vec{\q} \cdot \vec{H} 
$.
In contrast to the $q_i$, the $r$ components of $\vec{\q}$ are independent
quantities.
 
The covariant derivative (\ref{eq:covariantderiv}) acquires
${\bf q}$-dependence 
by acting with the commutator term on gauge field fluctuations
proportional to $E_\a$. 
For a fixed root $\a$, 
the d'Alembertian $-D^2({\bf q})$ becomes
 \beq
 -D^2({\bf q}) \; E_\a
=\; \left( \left(2\pi T \left( n +\vec\a \cdot \vec{\q}\, \right)\right)^2 
+\vec{p}^{\; 2}\right) E_\a \; ;
\eeq
the Matsubara frequency is $2 \pi T n$, where $n$ is an integer.

Finally, integrating over the spatial momenta, and summing over the
$n$'s and the $\a$, gives the one loop perturbative potential
\beq
 \V_{pt}(\bfq) \; = \; -
\frac{(N^2 -1) \pi^2}{45} \; T^4 \; + \; {2\pi^2\over 3} \; T^4 \; \sum_{\a}
B_4\left( \vec{\a} \cdot \vec{q} \, \right)
\; ,
\la{eq:potroot}
\eeq
where $B_4$ is given in Eq. (\ref{define_B}).

The arguments $\vec\a \cdot \vec{q}$ can be rewritten as
$
\vec\a \cdot \vec{\q}\; =\; 2\; \Tr\; 
\left( \vec{\a} \cdot \vec{H} \; {\bf q} \right)
$.
Below we write this argument explicitly for the 
four types of classical groups, using 
standard group theory~\cite{georgi}.  
We split the 
$\vec\a$ into 
$
\vec\a \rightarrow \vec{\a}^-, \vec {\a}^+ ~\mbox{and}~\vec\b.
$.
Definitions of these quantities, and 
a detailed analysis, will appear separately \cite{KorthalsAltes2012}; here
we simply present the results.

For $SU(N)$ the argument becomes
\beq
\vec\a^-_{ij} \cdot \vec{\q}
\; = \; 
2\; \Tr\; \vec\a^-_{ij} \cdot \vec H \; {\bf q} \; = \; q_i \; - \; q_j \; ,
~1\le i<j\le N \; .
\la{eq:sunargument}
\eeq 
For the orthogonal groups $SO(2N)$ these arguments involve both
differences and sums of the $q_i$, 
$2N(N-1)$ in total, living in the Cartan algebra of $N$ dimensions:
\beq
\vec\a^-_{ij}.\vec{\q} = q_i-q_j \;\; ; \;\;
\vec\a^+_{ij}.\vec{\q} = q_i+q_j,~1\le i<j\le N,\\
\label{eq:so2nrootprojq}
\eeq
together with roots of the opposite sign.

For $SO(2N+1)$ groups the dimension of the Cartan subalgebra is 
the same as that of $SO(2N)$.  Apart from the  $2N(N-1)$ roots involving 
$\a_{ij}^{\pm}.\vec{\q}=q_i\pm q_j$,
as in Eq. (\ref{eq:so2nrootprojq}) also $2N$ short roots $\b_i$ leading to
\beq
\pm \vec\b_i.\vec{\q}=\pm q_i,~1\le i\le N.
\la{eq:so2nplus1shortrootproj}
\eeq

For the symplectic groups $Sp(2N)$ 
(rank $N$ and dimension equal to that of $SO(2N+1)$) 
the arguments are obtained simply by 
leaving the $2N(N-1)$ projections 
$\a^{\pm}_{ij}.\vec q=q_i\pm q_j$ the same, but 
changing the $2N$ short roots into long roots.  This gives
\beq
 \pm \vec\b_i.\vec{\q}= \pm 2q_i,~1\le i\le N. 
\la{eq:sympllongrootproj}
\eeq
  
A comment is in order.  Transforming the generic root $\vec r$ into its dual,
$\vec r*=r/\vec r^{\;2}$,
leaves the $\vec\a^{\pm}$ invariant, while
the short roots $\vec\b$ transform into the long roots, and vice versa.
The root systems of $SU(N)$  and of $SO(2N)$ are invariant 
under this transformation.
The root systems of $SO(2N+1)$ and of $Sp(2N)$ 
transform into one another, 
Eqs. (\ref{eq:so2nplus1shortrootproj}) and (\ref{eq:sympllongrootproj}). 
The duality between the roots of  $Sp(2N)$ and $SO(2N+1)$ 
implies a duality between the potentials for $\bf q$;
see, also, the discussion following Eq. (\ref{eq:vksu7so7}) 
\cite{KorthalsAltes2012}.
  
The root system of the exceptional group $G(2)$ is 
dealt with in Sec. (\ref{sec:weyl}). There we will use 
the projections in Eqs. (\ref{eq:sunargument}), 
(\ref{eq:so2nrootprojq}) and 
(\ref{eq:so2nplus1shortrootproj}) for $SU(7)$ and $SO(7)$.

\subsection{Weyl groups and Weyl chambers}
\la{sec:weylgroupchamgener}
 
An interesting aspect of Eq. (\ref{eq:potroot}) is that it
is a sum over {\it all} roots of the gauge group.  This guarantees
that the potential is invariant under the symmetries of the roots, which
comprise the Weyl group.  
This invariance 
is especially useful in generalizing our potential to 
other gauge groups, such as $G(2)$ in Sec. (\ref{sec:weyl}).

Weyl transformations are generated by the the
reflection of the simple root $\vec\a$ into the mirror, $M_{\b}$, 
which is orthogonal
to the simple root $\vec{\g}$.  This reflection produces another root $\vec{w}_\g(\vec{\a})$:
\beq
 \vec w_\g(\vec\a) \; = \; \vec\a \; - \; 2
\; {\vec\a \cdot \vec\g\over{\vec\g^2}} \; \vec \g \; ,
\eeq
where
\beq
 2 \; \frac{\vec\a \cdot \vec\g}{\vec\g^2} \; = \; m \; ,
\la{eq:roots}
 \eeq
and $m$ is an integer.  Eq. (\ref{eq:roots}) is 
invariant under the interchange of $\a$ and
$\g$, although the integer $m$ may change. 
Together these conditions imply that the 
the roots lie on a lattice.  The only
possible angles between adjacent roots are:  
$30^o$, $45^o$, $60^o$, and $90^o$; 
the relative lengths between $|\vec \a|$ and  $|\vec\g|$
can be $1$, $\sqrt{2}$ , or $\sqrt{3}$.

To use representations other than the adjoint, 
we need their weight vectors, $\vec v$. Eq. (\ref{eq:roots})
is invariant if $\vec v$ is replaced by $\vec \a$.

Classical groups have root systems with at most two different lengths:
\bi
\item for $SU(N)$ and $SO(2N)$ all roots $\vec\a^{\pm}$ are equal.  
\item $SO(2N+1)$ has $2N$ short roots  $\vec\b$ and
$N(2N-1)$ long roots $\vec\a^{\pm}$, $\sqrt{2}$ longer.
\item The root system for $Sp(2N)$ is like that
for $SO(2N+1)$, except that the short and long roots are interchanged.
\item $G(2)$ has six short roots and six long roots, of relative length $\sqrt{3}$.  This root system is obtained 
by a simple projection of the $SO(7)$ root system discussed in section (\ref{sec:weyl}).
\ei

The Weyl group is a set of orthogonal transformations and therefore leaves the 
length of the roots invariant. For any classical group,
by Eq. (\ref{eq:roots}) there are at most two lengths involved;
for $SU(N)$ and $SO(2N)$, there is only one length. 

For $SU(N)$ there are $N-1$ independent reflections.  These generate
a finite group, which is the Weyl group $W$.
The Weyl group of $SU(N)$ is the permutation group of the $N$
fundamental indices.  Thus the order of the Weyl group is $d_W = N!$. 

Lastly we introduce the concept of the Weyl chamber, $\W$.
The Weyl chamber has as its walls the mirrors
$M_\a$, which are perpendicular to the root $\a$; here
$\a$ runs through the $r$ simple roots that span the Cartan algebra.
No element of the Weyl group leaves the Weyl chamber invariant, as the 
resulting $d_W$ Weyl chambers fill all of the Cartan space.

Let us return to Eq. (\ref{eq:potroot}), the perturbative potential
for the ${\bf q}$'s.  If we wish to consider more general potentials,
we need to require that they are invariant under the Weyl group.
This is constructed  by exploiting the separate invariance
of the  roots   $\a^{\pm}$ and the  roots   $\b$.  Instead 
of giving long and short roots the same weight,
we can take the linear combination:
\beq
a\; \sum_\a\bigg(B_4(\vec\a^{\pm} \cdot \vec \q))+
a_2 \; B_4(2\vec\a^{\pm} \cdot \vec \q)+...\bigg)
\; +b
\sum_\b\bigg( \; B_4(\vec\b \cdot \vec \q)+b_2\;
B_4(2\vec\b \cdot \vec \q)+ \ldots \bigg).
\la{eq:generalweylinv}
\eeq
Each term in this sum is invariant under the Weyl group.
This is essential in using our approach in gauge theories other than
$SU(N)$, like $G(2)$, in Sec. (\ref{sec:weyl}).  
  
Once we know the potential Eq. (\ref{eq:generalweylinv}) inside a Weyl
chamber, we can determine it everywhere in the Cartan space by using
Weyl transformations, and the periodicity of the Bernoulli polynomials.
The Weyl symmetry is a property of the Lie algebra.
Additional symmetries, such as $Z(N)$ for a $SU(N)$ gauge group,
arise from global properties of the gauge group.

\subsection{The Weyl chamber of $SU(N)$}
\label{sec:weyl_chamber}

For $SU(N)$, there is an 
alternate basis for the $\vec{H}$'s which is useful
in what follows.  Consider the diagonal matrices 
$\Y_k$, where $k = 1,2\ldots N-1$:
\beq
{\bf Y}_k
= \;
{1\over  N}\; \mbox{diag}(k,\ldots k, k-N,\ldots,k-N).
\la{eq:hypercharges}
\eeq
There are $N-k$ entries $k$, and $k$ entries $k-N$, so $\Y_k$
has zero trace; we call them hypercharges.
The ${\bf Y}_k$ are orthogonal to the simple roots $H_{i,i+1}$:
\beq
\Tr\left({\bf Y}_k \; H_{i, i+1}\right) \; = \; {1\over 2} \; \d_{ik} \; .
\label{eq:ortho}
\eeq
They also obey
\beq
\Tr \left( {\bf Y}_k \, {\bf Y}_l \right) \; = \;
{1\over N}(N~\mbox{min(k,l)}- \, k\, l)\ge 0 \; .
\eeq
Consequently, the angle between two hypercharges is also less than $\pi/2$.

The ${\bf Y}_k$ are useful because they serve as generators
of elements of $Z(N)$:
\beq
\exp\left( 2 \pi i \, {\bf Y}_k \right) 
\; = \; \exp\left( {2\pi i k \over N} \right){ \bf 1} \; .
\la{eq:znelnts}
\eeq

Further, the ${\bf Y}_k$ are 
the edges of the Weyl chamber of $SU(N)$.
To see this, take 
the set of $N-2$ $\Y_k$ matrices, excluding $\Y_i$ . 
Because of Eq. (\ref{eq:ortho}), this set forms the
Weyl mirror $M_{i, i+1}$, which is orthogonal to the root $\a_{i,i+1}$.

Now form a polyhedron whose $N-1$ edges are given
by the $\Y_k$. The rest of the
edges are given by drawing edges between the endpoints of the  hypercharges,
Eq. (\ref{eq:hypercharges}). Then the $N-1$ Weyl mirrors $M_{i,i+1}$ are the
faces of this polyhedron. This polyhedron is the Weyl chamber.  
Note that by Eq. (\ref{eq:znelnts}),
the vertices of the Weyl chamber are the points  corresponding to
the $N-1$ elements of the centergroup $Z(N)$.

Consider the average of all of the hypercharges;
we call this average the barycenter,
${\bf Y}_c$, of the Weyl chamber:
\beq
{\bf Y}_c \; \equiv\; {1\over N}\; \left(
{\bf Y}_1 \, +\, {\bf Y}_2 \, + \cdots {\bf Y}_{N-1} \right) \; .
\label{barycenter_def}
\eeq
The corresponding element of the Lie group is
\bea
{\bf L}_c \; =\; 
\exp\left(2 \pi i \; {\bf Y}_c\right) = 
\left( \begin{array}{cccc}{\rm e}^{\pi i(N-1)/N}&0&\cdots&0\\
 0&{\rm e}^{\pi i (N-3)/N}&\cdots&0\\
 \vdots&\vdots&\ddots&\vdots\\
0&0& \cdots&{\rm e}^{- \pi i (N-1)/N}
 \end{array}\right) \; .
 \la{eq:pzero}
 \eea

The confining vacuum is precisely the barycenter ${\bf Y}_c$.
This is clear from the $Z(N)$ symmetry: since by Eq. (\ref{eq:znelnts})
the hypercharges generate elements of $Z(N)$,
the barycenter, as the average of all of the hypercharges,
is automatically $Z(N)$ invariant.  
This implies, and it can be checked, that the appropriate traces of
${\bf L}_c$ vanish:
\beq
{\rm tr} \; {\bf L}_c^k = 0 \; , \; k = 1\ldots N-1 
\;\;\; ; \;\;\;
{\rm tr} \; {\bf L}_c^{N} = {\bf 1} \; .
\eeq
From the explicit form of Eq. (\ref{eq:pzero}), we also see
that the eigenvalues are equally distributed about the unit circle,
with a spacing 
$2 \pi/N$.  That is, in the confined vacuum there is an uniform
repulsion of eigenvalues.  

\begin{figure}
\includegraphics{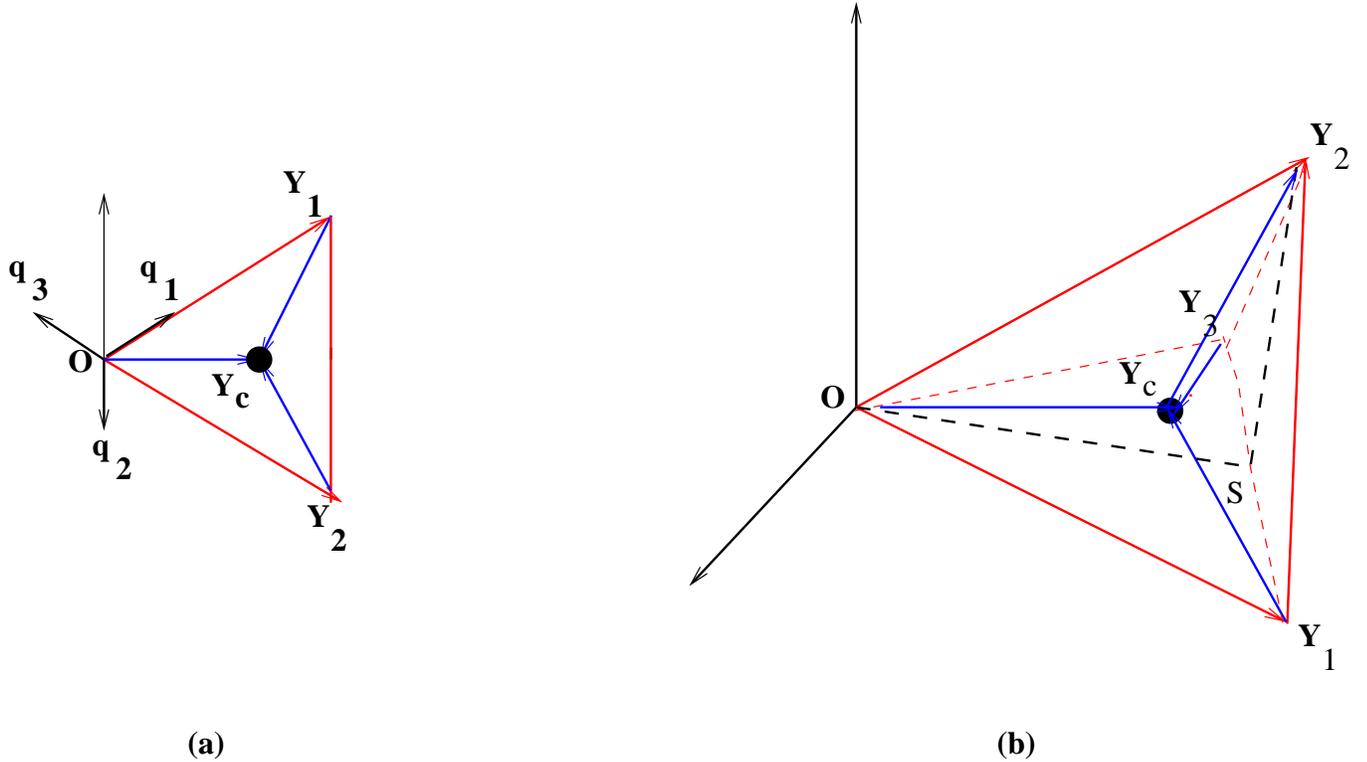}
\vspace*{-0.61cm}
\caption{\label{fig:su34} The Weyl chamber for three (a) and
four (b) colors. See also fig. (\ref{fig:su4realtraceplane}). }
\label{fig:weyl34}
\end{figure}

The Weyl chamber is illustrated in Fig. (\ref{fig:su34})
for three and four colors.  This figure is useful later in
Sec. (\ref{sec:firstorder}) in understanding why the deconfining
transition is of first order for four or more
colors.  

For three colors the Weyl chamber is an
equilateral triangle, with corners $\bf O$, ${\bf Y}_1$, 
and ${\bf Y}_2$.   
$Z(3)$ invariance of the potential divides the Weyl chamber into
three equivalent triangles, with the invariant barycenter $\Y_c$ in common.
Note that the loop ${\bf L}$ is real
along the line ${\bf O}{\bf Y}_c$.
By a global $Z(3)$ rotation we can required that the minina lie along
this line for any temperature.  

For four colors, 
the Weyl chamber is a tetrahedron, with corners $\bf O$, ${\bf Y}_1$, 
${\bf Y}_2$ and ${\bf Y_3}$. The four faces of the tetrahedron
are congruent triangles: sides ${\bf O} {\bf Y}_2$ and 
${\bf Y}_1 {\bf Y}_3$ have length $1$, while the other four sides have
length $\sqrt{3}/2$. 
The barycenter, ${\bf Y}_c$, is common to the
four  $Z(4)$ equivalent tetrahedrons defined by the four faces 
$\bf O Y_1 \bf Y_2$, {\it etc.}. 
We also indicate the path from the perturbative vacuum to the confining
vacuum, ${\bf O}{\bf Y}_c$, and  
${\bf O S}= {\bf Y}_2 {\bf S}=1/\sqrt{2}$.  
For four colors the loop ${\rm tr} \, {\bf L}$ is real
in the plane ${\bf O}{\bf S}{\bf Y}_2$, spanned by 
${\bf O}{\bf Y}_{13}$ and ${\bf O} {\bf Y}_2$,
and the point ${\bf S}$, ${\bf S}=(1/2)({\bf Y}_1+{\bf Y}_2)$.

In Fig. (\ref{fig:weyl34}) the hypercharges ${\bf Y_k}$ are
fixed numerical matrices. The absolute length of the roots,
however, is convention dependent, and so we do not show them.

\section{One and two parameter models for $SU(N)$}
\label{sec:thermo_model}

\subsection{Possible Potentials}

To model the transition, we assume that the total potential,
\beq
\V_{tot}(\bfq)=\V_{pt}(\bfq)+\V_{non}(\bfq) \; ,
\eeq
is a sum of perturbative and non-perturabtive contributions.

To one loop order, the perturbative potential for $\bfq$
was computed in Eq. (\ref{eq:potroot}), and involves
\beq
V_2(\bfq) \; = \; {1\over2} \; 
\sum_\a B_4 \left( \vec\a \cdot \vec{\q} \right) \; ,
\eeq
with $B_4(x)$ given in Eq. (\ref{define_B}).  Because of the
sum over the roots $\alpha$, this potential is invariant under the Weyl
group, and we can require the $\bfq$'s to lie in the Weyl chamber.

Given our experience with two colors in Sec. (\ref{sec:twoandthree}),
it is easy to guess possible forms for the non-perturbative potential.
To avoid a second phase transition, above $T_c$, 
it is necessary to add a term which is linear in the $\bfq$'s for
small $q$, as in Eq. (\ref{non_pert_twoA}).  The generalization of
this term for $SU(N)$ is
\beq
V_1(\bfq) \; = \; {1\over 2}\; \sum_\a B_2(\vec\a \cdot \vec{\q}) \; .
\label{linear_gen}
\eeq
where $B_2$ is given in Eq. (\ref{define_B}).
Summation over all roots in Eq. (\ref{linear_gen}) ensures that
the result is invariant under the Weyl group.

The simplest assumption is to assume that any non-perturbative term
is proportional to $\sim T^2 T_c^2$.  Thus we start by taking the
non-perturbative potential to be 
\beq
\V_{\rm non}(\bfq) \; = \; - \; {4\pi^2 \over 3} \;
T^2 \, T_c^2 \; \left( \frac{1}{5}\,
c_1 \; V_1(\bfq)\, + \, c_2\; V_2(\bfq) 
- \frac{(N^2 -1)}{60} \; c_3 \right) \; .
\label{vtotnpt}
\eeq
We shall show that with this model, we cannot explain the latent
heat.  Thus we generalize
the model slightly, and let $c_3$ be temperature dependent,
\beq
c_3(T) \; = \; c_3(\infty) + 
\left(c_3(T_c) - c_3(\infty)\right) \; \frac{T_c^2}{T^2}  \; .
\label{define_c3t}
\eeq
Given how $c_3(t)$ enters the
potential, this is equivalent to introducing 
MIT ``bag'' constant for the theory, whose value is
\beq
B = + \; \frac{\pi^2 (N^2 - 1)}{45} 
\; \left(c_3(1) - c_3(\infty) \right) \; T_c^4 \; .
\label{MIT_bag}
\eeq
In trying to fit to the lattice data, we also tried adding terms
$\sim T_c^4$ times both $B_2(\bfq)$ and $B_4(\bfq)$.  Surprisingly,
we found that the simplest possibility, Eq. (\ref{define_c3t}),
did the best job of fitting the lattice data.

We comment that charge conjugation is a symmetry of $SU(N)$.
This is generated by ${\bf A}_0 \rightarrow - {\bf A}_0$, or
$\bfq \rightarrow - \bfq$.  Requiring each $q_i$ to lie in the region
between $0$ and $1$, this is equivalent to 
$\bfq \rightarrow {\bf 1} - \bfq$.  This is why the powers of
$q (1-q)$ enter in the $B_{2 n}$.

We turn to parametrizing the path between the perturbative vacuum,
$\bfq = {\bf 0}$, and the confining vacuum, 
${\bf Y}_c$.  By a global $Z(N)$ rotation, we can assume that the
Wilson line ${\bf L} = {\bf 1}$ at high temperature, and that
the trace of the Wilson line remains real for all temperatures.

For an even number of colors, we generalize the
two color solution of Eq. (\ref{twocolor_ansatz}): for $SU(2M)$,
we take $M$ pairs of eigenvalues, $\pm q_i$, $i = 1\ldots M$.  
For an odd number of colors, $N = 2M + 1$,
we take one eigenvalue to vanish, leaving again $M$ pairs $\pm q_i$.
Thus the stationary point of the potential
$SU(2M)$ or $SU(2M+1)$ involves $M$
independent variables.  

The simplest possible path is a straight line
from the origin to $\Y_c$:
\beq
{\bf q}(s) \; = \; s \; {\bf Y}_c \; ,  ~~ 0 \le \; s \; \le 1 \; .
\la{eq:ansatz1}
\eeq 
where
\beq
q_j(s) \; = \; {N-2j + 1\over{2N}} \; s \; .
\la{eq:ansatz2}
\eeq
We stress that this is an {\it ansatz}.  It applies
for two or three colors, but is {\it not} a solution for four
or more colors.  This ansatz assumes that the $M$ eigenvalues have
constant spacing: for $2 M$ colors, by
reordering the eigenvalues we have $q_j = j q_1$, $j = 1 \ldots M$,
with the other $M$ eigenvalues given by $ - q_j$.
In the limit of infinite $N$, this ansatz given a uniform eigenvalue
density to some maximum. 
We thus refer to Eq. (\ref{eq:ansatz2}) 
as the {\it uniform eigenvalue ansatz}.

The advantage of making the uniform eigenvalue 
ansatz is that it is then easy to compute
{\it analytically} for arbitrary $N$. 
For finite $N > 3$,
the exact solution must be determined numerically.
We have done so for four to seven colors
in Sec. (\ref{sec:su45}).
For these values of $N$, we find that the difference between 
the uniform eigenvalue ansatz and the exact solution is 
remarkably small.  
The differences are naturally greatest at $T_c$.
For $N=4$ to $7$,
for all thermodynamic quantities
and for the expectation value of the Polyakov loop,
even at $T_c$ 
the difference between the uniform eigenvalue ansatz and the exact
solution is less than $\sim 1\%$.  This
difference is within the width of the curves 
in the figures given below.  Thus we do not present these (coincident)
curves, and give results only for the eigenvalues themselves in
Sec. (\ref{sec:su45}).

The difference between the constant eigenvalue ansatz and the exact
solution increases with the number of colors.
Results for larger numbers of colors
will be given separately \cite{Pisarski2012}.  At
infinite $N$, the model can be solved analytically.
At $T_c$, expectation value of the Polyakov loop
is $1/2$ for the exact solution, versus $2/\pi \sim 0.64$
versus for the ansatz, Eq. (\ref{critical_ell}). 
Rescaled as in Eq. (\ref{rescaled_latent_heat}),
the latent heat of the exact
solution is $1/\pi^2 \sim 0.10$, versus $\sim 0.16$ with the ansatz, 
Eq. (\ref{latent_heat_results}).

Given the close numerical coincidence between the uniform eigenvalue
ansatz and the exact solution, especially for
moderate values of $N$, we find it most useful to investigate
this ansatz in great detail.
We do so in the remainder of this section.  

\subsection{Evaluating the potential under the uniform eigenvalue ansatz}
\label{sec:evaluation}

From Eq. (\ref{eq:sunargument}), 
\beq
\vec\alpha^-_{ij} \cdot \vec{\q} = q_i - q_j
\; =\; \frac{i-j}{N} \; s \; .
\la{ccinvtline}
\eeq
The perturbative vacuum is given by $s = 0$, 
${\bf L} = {\bf 1}$, while 
the confined vacuum is $s = 1$, where ${\bf L} = {\bf L}_c$,
Eq. (\ref{eq:pzero}).
The variable $s$ is also convenient because the effective potential is
$\sim N^2$, times a potential of $s$.  The coefficients of this potential
have a smooth limit at $N \rightarrow \infty$.

We need to compute
\beq
V_k(\bfq) \; = \; \sum_{1\le i<j \le N} \; B_{2k}(q_i-q_j) \; .
\eeq
These potentials involve the sums
\beq
{\cal S}_n = \sum_{j = 1}^N \sum_{k = 1}^N \; |j - k|^n = 
2 \; \sum_{i = 1}^N \sum_{j = 1}^{j-1} \; (j - k)^n = 
2 \; \sum_{j = 1}^N \sum_{k = 1}^{j-1} \; k^n \; .
\eeq
The last identity follows by relabeling $j - k \rightarrow k$.

We need the first four sums,
$$
{\cal S}_1 = \frac{1}{3} \; N (N^2 - 1) \; \;\; ; \;\;\;
{\cal S}_2 = \frac{1}{6} \; N^2 (N^2 - 1) \; ;
$$
\beq
{\cal S}_3 = \frac{1}{30} \; N (N^2 - 1) (3 N^2 - 2) \;\;\; ; \;\;\;
{\cal S}_4 = \frac{1}{30} \; N^2 (N^2 - 1) (2N^2 - 3) \; .
\label{sums}
\eeq

Since the only mass scale in our model
is set by the critical temperature $T_c$, we introduce
the dimensionless ratio
\beq
t = \frac{T}{T_c} \; .
\eeq
It is convenient to redefine the potential as
 \beq
 \V_{tot}(s,t) = \frac{\pi^2 (N^2 - 1)}{45} \; T_c^4 \; t^2 \; 
\left(t^2 - c_2 \right) {\cal W}(s,t) \; ,
\label{definition_vtot}
\eeq
where
$$
{\cal W}(s,t) = \frac{1}{t^2 - c_2}\left(
- \, t^2 - 2 c_1 \left( s - \frac{s^2}{2}\right)+ c_3(t) \right) 
$$
\beq
+ 5s^2-6 \left(1-\frac{2}{3N^2}\right)s^3
\, +\, 2 \, \left( 1- \frac{3}{2N^2}\right)s^4 \; .
\la{eq:potccline}
\eeq
The first term, $-t^2/(t^2-c_2)$, is the ideal gas term.
That $\sim c_1$ is from $B_2(\bfq)$ is from the non-perturbative
potential, as is the constant $\sim c_3$.  
The quartic potential in $s$ arises from $B_4(\bfq)$,
in both perturbative and non-perturbative terms.

To simplify the expressions, we let $c_3$ independent of temperature;
in Sec. ({\ref{bag_constant}) we show that
it is trivial to incorporate.

Fixing the parameters is done as for two colors.
We start with a model with three parameters, $c_1$, $c_2$, and
$c_3$.  To destabilize the perturbative vacuum, $c_1$ must be positive.
As it stands, $T_c$ is just a mass parameter.  One of the
parameters, say $c_1$, can be fixed by requiring that $T_c$
is the temperature for the phase transition.

We also need a condition to fix the value of the pressure at $T_c$.
We make the somewhat
unphysical choice that the pressure vanishes at the transition,
which implies
\beq
\V_{tot}(s = 1, t =1) = 0 \; .
\label{zero_pressure}
\eeq
This is used to determine $c_3$.

That leaves $c_2$ as one free parameter.  We shall solve
the model for arbitrary values of $c_2$, and determine its
value by comparison to the lattice data in the next section.
The value of $c_2$ is tuned to ensure there is a sharp peak in the
interaction measure, as seen in numerical simulations.

Given the $s$-dependence of the potential, it is useful to
introduce the parameter
\beq
z(t) \; =\; \frac{c_1}{t^2-c_2} \; .
\label{eq:variables}
\eeq  
We also introduce
\beq
r\, =\, 1\, -\, s \; .
\eeq
In this variable, $r = 0$ is the confining vacuum, and $r=1$ is
the perturbative vacuum.

The value of the potential in the confined vacuum is
\beq
{\cal W}(0,t) \; = \; \frac{1}{N^2}
+ \frac{(- c_1 - c_2 + c_3)}{t^2 - c_2} \; .
\eeq
The $r$-dependent terms in the potential are
\beq
{\cal W}(r,t) - {\cal W}(0,t)
= \; - \left( 1  +  {6\over N^2} -  z(t)\right)\; r^2  - 
2 \left(1- {4\over N^2}\right)  r^3  + 
\left(2 -  {3\over{N^2}}\right)r^4 \; .
 \la{eq:wpot}
\eeq

Before proceeding to the details of the solution, we make a general
remark, which we expand upon later.  
In Eq. (\ref{eq:wpot}) we have reduced our model
to a standard mean field theory, with terms which are quadratic, cubic,
and quartic in $r$.  The term linear in $r$ vanishes because the
confining vacuum is necessarily extremal in $r$.  
This follows because 
${\bf Y}_c$ is the barycenter of the Weyl chamber.  

When $N=2$, the term cubic in $r$ vanishes, and the model has
a second order phase transition.  

For three or more colors, though, the cubic term is nonzero.
By standard mean field theory, this implies 
that the deconfining transition is of {\it first}
for three {\it or more} colors. 
Clearly, the coefficient of the
$r^3$ term in Eq. (\ref{eq:wpot}) is special to our model.
We argue in in Sec. (\ref{sec:firstorder}), though, that generally
the term cubic in $r$ is nonzero.  That is, the first order transition
for $N \geq 3$ is not an accident of the particular form of our model.

The vacuum at a given temperature, $r_0(t)$, is given by
requiring that it is an extremum of the potential,
\beq
\left. \frac{\partial }{\partial r} \; \V_{tot}(r,t) \; \right|_{r = r_0(t)}
\; = \; 0 \; .
\label{eq_W}
\eeq
The vacuum at a temperature $T = t \, T_c$ is the minimum, $r_0(t)$.
Note that $r_0(t)$ is temperature dependent solely because $z(t)$ is.

The pressure is minus the value of the potential at this minimum,
\beq
p(T) = - \; \V_{tot}(r_0(t),t) \; .
\eeq

The equation of motion 
gives a quadratic equation for $r$ which is easily solved. The solutions
are $r = 0$, and 
 \beq
 r_{0\pm}(t)=
{1\over{8(1-{3/(2N^2)})}}\left(
3 \left( 1\, -\, {4\over{N^2}}\right) \pm 
\sqrt{25-16 \left(1-{3\over{2N^2}}\right)z(t)}\right) \; .
 \la{eq:minima}
 \eeq
As $t \rightarrow \infty$, $z(t) \rightarrow 0$, and one can see that
$r_{0+}$ corresponds to the  minimum in the deconfined phase.
We discuss the role which the other root, $r_{0-}$, plays in
Sec. (\ref{sec:overheating}).

\subsubsection{Behavior at $T_c$}
\label{sec:behavior_Tc}

This gives us $r_{0+}(t)$ as a function of $z(t)$, but it does not determine
the value of $z(t)$ at some temperature, such as $z(1)$.

To determine this, we first compute the value of $r$ at the critical
temperature, $r_c = r_0(1^+)$.  This can be done by
a trick.  
Remember that we require that the pressure vanishes in the confined
phase, Eq. (\ref{zero_pressure}).  
Consequently, 
whether the transition is of first or second order, 
at $T_c$ the pressure of
the deconfined phase must then equal that in the confined phase, and
so vanish.  This gives two conditions:
\beq
{\cal W}(0,1) = 0 \;\; ; \;\;
{\cal W}(r_c, 1) = 0 \; .
\label{second_condition}
\eeq
By manipulating these two conditions,
the terms involving $z(1)$ can be eliminated.  Doing so immediately gives
the value of $r_c$,
\beq
r_c \; = \; {N^2-4\over{2N^2-3}} \; .
\la{eq:critr}
\eeq

At the critical temperature the potential has a simple form,
\beq
{\cal W}(r,1) = \left( 2 - \frac{3}{N^2} \right) \;
r^2 \left( r - r_c\right)^2  \; .
\label{eq_crit_pot}
\eeq
This is the standard potential expected in mean field theory.
For two colors, $r_c = 0$, and at $T_c$ there is a purely quartic
potential, $\sim r^4$.  For three or more colors, $r_c \neq 0$,
and the potential has two degenerate minima, 
at $r=0$ and $r = r_c$, as is typical of a first order transition.

The value of $r_c$ increases with $N$,
\beq
r_c(3) = \frac{1}{3} \;\; ; \;\;
r_c(4) = \frac{12}{29} \;\; ; \; \;
r_c(6) = \frac{32}{69} \;\; ; \;\; 
r_c(\infty) = \frac{1}{2} \; . 
\label{critical_r}
\eeq
The value of the Polyakov loop at $T_c^+$ equals
\beq
\ell_c = \frac{1}{N} \; \frac{\sin(N \delta)}{\sin(\delta)} \; \; ; \;\;
\delta \; = \; \frac{(N^2 + 1) \pi}{N (2 N^2 - 3) } \; .
\la{eq:jumploop}
\eeq
Explicitly,
\beq
\ell_c(3) = .449... \;\; ; \;\;
\ell_c(4) = .542... \;\; ; \;\;
\ell_c(6) = .597.... \;\; ; \;\;
\ell_c(\infty) = \frac{2}{\pi} = .637... \; .
\label{critical_ell}
\eeq
Since the Polyakov loop vanishes at $T_c^-$, these values are the
discontinuity in the loop at $T_c$.

The value of $r_c$ and $\ell_c$ for an infinite number of colors has a
simple interpretation in terms of the eigenvalue density, which is a
function of an angle $\theta = 2 \pi j/N$.
At infinite $N$, $\theta$ is a continuous variable, 
from $-\pi$ to $\pi$.  
In the perturbative vacuum, the eigenvalue density is a delta function
at $\theta = 0$.  
In the confining vacuum, the eigenvalue density is constant, over the
entire circle, from $-\pi$ to $+ \pi$.  
Under the uniform eigenvalue ansatz, at
$T_c^+$, the eigenvalue density is nonzero only over half the unit 
circle, from $-\pi/2$ to $+ \pi/2$.
We stress that the eigenvalue density for the exact solution at
infinite $N$ is not constant.

The value of $z_c = z(1)$ is found to be
\beq
z_c \; = \; \frac{(N^2+1)(3 N^2-2)}{N^2 (2 N^2-3)} \; .
\la{eq:c1fixesc2}
\eeq
As a function of the number of colors,
\beq
z_c(2) = \frac{5}{2} \;\; ; \;\;
z_c(3) = \frac{50}{27} \;\; ; \;\;
z_c(4) = \frac{391}{232} \;\; ; \; \;
z_c(6) = \frac{1,961}{1,242} \;\; ; \;\; 
z_c(\infty) = \frac{3}{2} \; . 
\eeq

Note that both $r_c$, Eq. (\ref{eq:critr}) 
and $z_c$, Eq. (\ref{eq:c1fixesc2}), are independent of the parameter $c_2$.  
It can be shown that this is not special to
the uniform eigenvalue ansatz, but is also a property of the exact
solution of our model for any $N$.

Given the definition of $z(t)$, Eq. (\ref{eq:variables}),
this determines $c_1$,
\beq
c_1 = z_c \; ( 1-c_2 ) \; .
\label{eq:determinec1}
\eeq  

Lastly, we can use the condition ${\cal W}(0,1) = 0$ to determine $c_3$,
\beq
c_3 = 1 +  \left( z_c - 1 - \frac{1}{N^2}\right) \left( 1 - c_2 \right) \; .
\label{eq:determinec3}
\eeq
Using this value, 
\beq
{\cal W}(0,t) = \frac{1}{N^2} \left( \frac{t^2 - 1}{t^2 - c_2} \right) \; .
\eeq

The behavior of the pressure in the confined phase is deserving of
comment.  From Eq. (\ref{definition_vtot}) the pressure of the confined
phase, where $s = 1$ and $r = 0$, is
\beq
p(T) \; = \; - \; \V_{tot}(1,t) \; = \;
- \; \frac{\pi^2}{45} \; 
\left( 1 - \frac{1}{N^2} \right) \left( T^4  \; - \; T_c^2 \, T^2 \right) \; .
\label{pressure_confined_phase}
\eeq

At large $N$, the pressure in the deconfined phase is $\sim N^2$,
while that in the confined phase is $\sim 1$.  This is satisfied
by Eq. (\ref{pressure_confined_phase}), but 
as discussed following Eq. (\ref{non_pert_twoB}), below 
$T_c$ we should match to a 
hadronic resonance gas.  
This is a gas of massive glueballs, and so
will be a series of Boltzmann factors.  If there are many massive
glueballs, such as from a Hagedron spectrum, the temperature dependence
can be more involved, involving 
powers of $T_H - T$, where $T_H$ is the Hagedorn temperature.

This is not what Eq. (\ref{pressure_confined_phase}) represents, though.
Rather, it reflects the limitations of an incomplete large $N$
approximation, where such a matching to a hadronic resonance
gas has not been done.  
Eq. (\ref{pressure_confined_phase})
includes a {\it negative} pressure from two, massless
degrees of freedom, $- T^4$, and a positive term $\sim + T^2 T_c^2$.  
In the confined phase, these contributions are manifestly
unphysical; for example, while the pressure is positive below
$T_c$, the entropy is negative.

This shows that our model is applicable only in the deconfined
phase, for $T \geq T_c$.  Since it is explicitly motivated by
an expansion in large $N$, using it in the confined phase, which involves
corrections $\sim 1/N^2$, is dubious. 

\subsubsection{Latent heat}
\label{sec:latent_heat_model}

In this subsection we derive
the interaction measure and so the latent heat.
The interaction measure is related to the energy density, $e(T)$,
and the pressure, $p(T)$, as
\beq
\D(T)= \frac{e - 3p}{T^4} = T \frac{\partial}{\partial T}
\left(\frac{p}{T^4} \right)
= \; - \; T\; \frac{\partial}{\partial T} \frac{\V_{tot}(r_0(t),t)}{T^4} \; .
\eeq
The temperature derivative acts both upon explicit and implicit
temperature dependence.  The explicit $T$ dependence is from
the overall factor of $T^4$ in the perturbative potential, and $T^2$
in the non-perturbative potential.  Clearly only the latter contributes.
There is also the implicit dependence of the solution, $r_0$, with temperature.
Since $r_0$ is a solution of the equation of motion, 
Eq. (\ref{eq_W}), this contribution vanishes.  Thus
the interaction measure depends only upon the non-perturbative potential
at the minimum,
\beq
\D(T) \; = \; 2 \; \frac{\V_{non}(r_0(t),t)}{T^2}  
= - \; 2 \; \left( \frac{p(T) + \V_{pt}(r_0(t),t)}{T^4} \right) \; .
\la{eq:deltapressurepot}
\eeq

The latent heat is the jump in the energy density at $T_c$.  By
construction, we assume that both the pressure and the energy density
vanish in the confined phase.  The latent heat is then
$-2$ times the perturbative potential at $T_c$.  This equals
\beq
\frac{e(T_c)}{T_c^4} \; = \; {\, \pi^2\over{15}} \; (N^2 - 1) \; f(N) \; .
\label{latent_model}
\eeq
where
\beq
f(N) \; =\; z_c r_c^2 - \frac{1}{N^2}
= \; \frac{(3N^8-31N^6+74N^4-22N^2-5)}{N^2(2N^2-3)^3} \; .
\label{latent2}
\eeq
As for other quantities at $T_c$, in our model the latent heat is independent
of the parameter $c_2$.  

Note that while there is an overall factor of $N^2 - 1$ in the latent heat,
the function $f(N)$ increases markedly as $N$ does.
Its value is $5/54 \sim .09$ for three colors, to 
$3/8 \sim .375$ for an infinite number of colors.
We comment upon this later when we compare to the lattice data in 
Sec. (\ref{sec:latent_heat}).

\subsubsection{Nonzero ``bag'' constant}
\label{bag_constant}

The effective potential can be generalized to include terms other
than those $\sim T^2$.  The simplest is to include terms $\sim T^0$,
as in Eq. (\ref{define_c3t}).

In the previous
section, we found it convenient to consider $c_2$ as the
single free parameter,
with $c_1$ determined by Eq. (\ref{eq:determinec1}) and $c_3$
by Eq. (\ref{eq:determinec3}).  Note that both are determined by
the behavior of the potential at the critical temperature, $t = 1$.

Since the term proportional to $c_3$ is independent of $r$,
we can generalize the previous solution immediately.
Again $c_1$ is fixed by Eq. (\ref{eq:determinec1}).
The quantity $c_3(1)$ is fixed by
Eq. (\ref{eq:determinec3}).  The leaves $c_3(\infty)$ as a free
parameter, with along with $c_2$, gives a model with two free parameters.

When $c_3(1) \neq c_3(\infty)$, there is an additional contribution
to the interaction measure,
\beq
B = \; \frac{\pi^2 (N^2 - 1)}{45} \; \frac{(c_3(1) - c_3(\infty))}{T^4} \; .
\eeq
We comment that in order to fit the thermodynamics of $SU(N)$, that
we need $c_3(1) > c_3(\infty)$.  By
Eq. (\ref{MIT_bag}), this corresponds to a positive sign
for the MIT bag model, $B > 0$.  This is physical, as the confining
vacuum has negative pressure.
Given the other terms in our model $\sim T^2$, 
though, probably not too much should be made of this.

\subsubsection{Over and under heating}
\label{sec:overheating}

We can also use the potential to compute the temperature for over
and under heating.  This is related to the behavior of the other
root, $r_{0-}(t)$, in Eq. (\ref{eq:minima}).

Over heating is the following.  
Suppose one increases the temperature from $T_c$.  If the theory is
originally in the confined vacuum, $r=0$, then in thermal equilibrium,
it tunnels to the deconfined
phase, at $r_c$.  If we raise the temperature sufficiently 
quickly, however, it will stay at $r=0$
until the quadratic term in $r$, about $r = 0$, vanishes; 
at this point, it must roll down the potential to $r_0 \neq 0$.
This is the temperature for over heating, $t_{oh}$.
From Eq. (\ref{eq:wpot}), the mass squared for $r$
vanishes at a temperature
\beq
z_{oh} = z(t_{oh}) \; = \; 1 \; +\; \frac{6}{N^2} \; .
\la{eq:overheating}
\eeq 
Using the value of $z_c$,  we can then compute the ratio of the
overheating temperature to $T_c$,
\beq
t_{oh}^2 \; = \; \left(\frac{T_{oh}}{T_c} \right)^2 \; = \;
{z_c(N)\over{z_{oh}(N)}}\left( 1 -c_2\right) + c_2 \; .
\eeq
The ratio $z_c/z_{oh}=1$ for $N=2$;
$z_c/z_{oh} = 10/9$ when $N=3$, and increases monotonically with $N$,
$= 3/2$ at $N = \infty$.  

For two colors, then, $T_{oh} = T_c$.
That is correct, over heating cannot occur for a second order
transition.

For three or more colors, $T_{oh} > T_c$.  Unlike other quantities
in our model at $T_c$, however, this ratio {\it does} 
depend upon the parameter $c_2$.
When $c_2$ vanishes, $(T_{oh}/T_c)^2 = z_{oh}/z_c$.  
For three colors, for example, $T_{oh}/T_c = \sqrt{10/9}$.  
As we make $c_2 \rightarrow 1$ (remember it must be less than one),
we find that $T_{oh} \rightarrow T_c$, independent of $N$.

That is, when $c_2$ is near one, the width of the 
transition region {\it narrower} than for $c_2 = 0 $.
We shall see in the next section that the behavior of the interaction
measure is even steeper than the behavior of $T_{oh}/T_c$ indicates.
Nevertheless, it gives us some intuition as to why we find
it necessary to choose a value of $c_2$ near one, at least for small $N$.

At very high temperature, one can check that the other root
of Eq. (\ref{eq:minima}), $r_{0-}$, is negative.  As the temperature
decreases, $r_{0-}(t)$ moves towards the origin.  At the
temperature for overheating, this root coincides
with the origin, $r_{0-}(t_{oh}) = 0$.  This is why
the mass squared for $r$ vanishes at $r = 0$ at $t_{oh}$.

As the temperature decreases below $t_{oh}$, $r_{0-}(t)$ represents
a maximum in the potential, between $r = 0$ and $r_{0+}$.
This is true at the critical temperature, where by
Eq. (\ref{eq_crit_pot}), $r_{0-}(1) = r_c/2$.

As the temperature is lowered below $T_c$, the point for
$r_{0+}(t)$ represents a relative minimum, which is unstable to tunneling
to the absolute minimum at $r = 0$.  At the temperature for under cooling,
the two minimum coincide, $r_{0+}(t_{uh}) = r_{0-}(t_{uh})$.  This gives
\beq
z_{uc} = \frac{ 25 \, N^2}{8 (2 N^2 - 3)} \;  ,
\eeq
or
\beq
t_{uh}^2 \; = \; \left(\frac{T_{uc} }{T_c} \right)^2
= \frac{z_c(N)}{ z_{uc}(N) } \left( 1 - c_2 \right) + c_2 \; .
\eeq
At the temperature for underheating, there is no barrier for the theory
at $r_{0}$ to roll down to the absolute minimum at $r = 0$.  
The qualitative behavior is the same as for over heating, except that
the variation with $N$ is much weaker.  

\subsection{Comparison between lattice data and the uniform eigenvalue
ansatz}
\subsubsection{One parameter model}
\label{sec:one_parameter}

In this section we review the results for the zero parameter model
of Meisinger, Miller, and Ogilvie \cite{Meisinger:2001cq}, and
the one parameter model which we analyzed before
\cite{Dumitru:2010mj}.  This is done for completeness, and to make
clear why it is necessary to generalize the model further.

We remark that in this paper the constant's $c_i$ differ from those
in Ref. \cite{Dumitru:2010mj}.  If we denote $\widetilde{c}_i$ by
those in Ref. \cite{Dumitru:2010mj}, then they are related to those
in the present work by
\beq
\widetilde{c}_1 = - \; \frac{2 \pi^2}{15} \; c_1 \;\; ; \;\;
\widetilde{c}_2 = - \; \frac{2 \pi^2}{3} \; c_2 \;\; ; \;\;
\widetilde{c}_3 = \; \frac{\pi^2 (N^2 - 1)}{45 N^2} \; c_3 \; .
\eeq
The change in notation was made to make the results more transparent.
In particular, the point where $c_2 = 1$ is special.
There are terms $\sim \V_2(\bfq)$ both in the non-perturbative
potential, $\sim - c_2 \, T^2 T_c^2 \, \V_2(\bfq)$, Eq. (\ref{vtotnpt}), 
and in the perturbative potential,
$\sim T^4 \, \V_2(\bfq)$, Eq. (\ref{eq:potroot}).  
When $c_2 = 1$, these terms cancel identically at $T_c$.  Because
of the lattice data, at least for small $N$ 
we are driven to a point close to $c_2 = 1$.

In Fig. (\ref{fig:nc3_compare}) we show the results for
the interaction measure, 
$\Delta(T) = (e - 3p)/(8T^4)$, for the zero parameter
model, $c_2 = 0$, and our optimal fit for the one parameter model,
$c_2 = 0.8297$.   Note that here and henceforth, we rescale
the interaction measure by the number of perturbative gluons, $N^2 -1$.

As is clear from the figure, there is sharp discrepancy between
the model with $c_2 = 0$ and $c_2 = 0.8297$.  With the zero
parameter model, the peak in the interaction measure is off by
about $\sim 50\%$.  By introducing $c_2$, we can fit this to within
a few percent.  To do so, we have to take a value very near one.

The difference between the models is
only clear once one plots the interaction measure.  If one were to
plot the pressure or energy density, scaled by $T^4$, it would
be difficult to see the difference between the two models.

We remark, however, that this behavior is similar to what is seen in
an analysis of the Schwinger-Dyson equations by Braun, Gies, and Pawlowski,
Ref. \cite{Braun:2007bx}.  The numerical values for, {\it e.g.}, 
$\langle r \rangle$ do not agree, but in both cases, the region in
which the condensate is non-zero is unexpectedly small.  

For completeness, 
in Fig.~(\ref{fig:SUNLoop}) we also plot the expectation values $\ell_0$ of
Polyakov loops in the fundamental representation of SU(N) for $N=3$,
$4$, $6$ and $64$ for $T>T_c$; they vanish when
$T < T_c^-$. At $T_c^+$, the
expectation values agree with Eq. (\ref{critical_ell}). As for
three colors, the expectation value of the Polyakov loop approaches unity
quickly as $T$ increases.

\begin{figure}
\includegraphics[width=10cm]{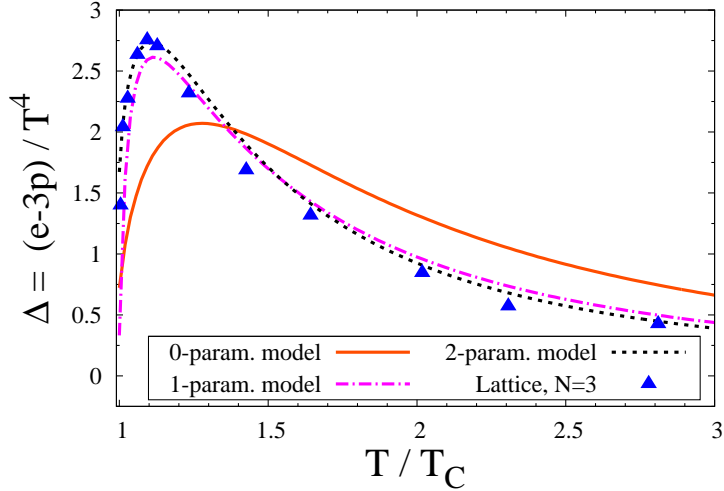}
\caption{\label{fig:nc3_compare}A comparison of the interaction
measure, $(e-3p)/(8 T^4)$ 
for three colors in the models with zero \cite{Meisinger:2001cq}, 
one \cite{Dumitru:2010mj}, and two parameters, versus 
lattice measurements.}
\end{figure}

\subsubsection{Latent heat}
\label{sec:latent_heat}

Nevertheless, the one parameter model has serious problems near the
critical temperature.
For three or more colors, the transition is of first order, which is
parameterized by the latent heat.
We introduce a dimensionless measure of the latent heat,
rescaling it by both $T_c^4$ and the number of
perturbative gluons \cite{Datta:2010sq}:
\beq
L(N) \; = \; \frac{e(T_c^+)}{(N^2 - 1) \, T_c^4} \; .
\label{rescaled_latent_heat}
\eeq

On the lattice, $L(N)$ has been measured for $N=3$ by
\cite{Beinlich:1996xg,Datta:2010sq}, and
$N=4$, $6$, and $8$ \cite{Lucini:2005vg,Datta:2010sq}.  
Datta and Gupta \cite{Datta:2010sq} give a simple
analytic form,
\beq
L(N) \; = \; 0.388 \; - \; \frac{1.61}{N^2} \; \; .
\label{DG_latent}
\eeq
We summarize these results in a table,
\begin{equation}
\begin{array}{ccccc}
N \;\;\; & \;\;\; {\rm Ref.} \, \mbox{\cite{Beinlich:1996xg}} \;\;\; & \;\;\; {\rm Ref.} \, \mbox{\cite{Lucini:2005vg}} \;\;\; & \;\;\; {\rm Ref.} \, \mbox{\cite{Datta:2010sq}} \;\;\; & \;\;\; {\rm Model} \\
3  \;\;\; &  \;\;\; .175  \;\;\; & \;\;\;    \;\;\; & \;\;\; .209 \;\;\; & \;\;\; .041 \\
4  \;\;\; & \;\;\;    \;\;\; &  \;\;\; .311  \;\;\; & \;\;\; .287   \;\;\; & \;\;\; .099\\
6  \;\;\; & \;\;\;    \;\;\; &  \;\;\; .349  \;\;\; & \;\;\; .342   \;\;\; & \;\;\; .137\\
8  \;\;\; & \;\;\;    \;\;\; &  \;\;\; .321  \;\;\; & \;\;\; .363   \;\;\; & \;\;\; .149\\
\infty  \;\;\; & \;\;\;    \;\;\; &  \;\;\; .344 \;\;\; & \;\;\; .388 \;\;\; & \;\;\; .165 \\
\end{array}
\label{latent_heat_results}
\end{equation}
The model results are those for the one parameter theory
of the previous subsection, Eqs. (\ref{latent_model}) and
(\ref{latent2}).  Remember that these values are independent of $c_2$.

Thus the lattice data shows that $L$, while a quantity of order one,
does increase from $\sim 0.2$ for three colors, to $\sim 0.36$ for
an infinite number.  Our model exhibits a similar increase, but
the latent heat is too small by about a factor of five for three colors,
and a factor of two for an infinite number.
Given the simple nature of your model, if we wish to describe
the latent heat, we have to generalize the model further.

\subsubsection{Two parameter model}
\label{sec:two_parameter}

To ameliorate this problem, we adopt the temperature dependent parameter
$c_3(t)$, Eq. (\ref{define_c3t}).  Before doing so, we stress that
since only $c_3$ is temperature dependent, that many results are unchanged.
This is because while the temperature dependence of the potential changes,
the $r$-dependent potential does not.  Thus the following quantites
are unchanged: the value of $r_c^+$, Eq. (\ref{eq:critr});
the corresponding value of the  Polyakov loop
at $T_c^+$, Eq. (\ref{eq:jumploop});
and the form of the potential in $r$ at $T_c$, Eq. (\ref{eq_crit_pot}).

With two parameters, there is some freedom in how they can be chosen.
For three colors, Fig. (\ref{figwhot}) shows that
the rescaled conformal anomaly,
$\widetilde{\Delta}(T)$ in Eq. (\ref{rescaled_conf_anomaly}),
is nearly constant from $1.2 \, T_c$ to $2.0 \, T_c$.
Although the data for $N=4$ and $6$ \cite{Datta:2010sq} is much noisier than
that for $N=3$ \cite{Umeda:2008bd}, 
again $\widetilde{\Delta}(T)$ appears to be constant over 
a similar range in temperature.  This data
also shows that the value which $\widetilde{\Delta}(T)$ attains
for $T > 1.2 \, T_c$ is approximately independent of $N$.

From our solution, $\langle r \rangle \neq 0$ in a narrow region,
below $1.2 \, T_c$.  Above this temperature, the behavior of
$\widetilde{\Delta}(T)$ is controlled entirely by the constant $c_3(\infty)$.
Thus we take the same value of $c_3(\infty)$ for all $N$.
For three colors, the best value of this parameter is 
\beq
c_3(\infty) = 0.95 \; .
\eeq

We then determined the remaining parameter, $c_2$, by fitting to the
latent heat.  To be definite, we used Eq. (\ref{DG_latent})
of Ref. \cite{Datta:2010sq}.  The results are
\begin{equation}
\begin{array}{ccccc}
N\;\;\;\;& \;\;\;\; c_2\;\;\;\;  & \;\;\;\; c_1\;\;\; \; & \;\;\;\; c_3(1) \;\;\;\;  & \;\;\;\; B^{1/4} (MeV)  \\ 
3\;\;\;\;& \;\;\; 0.552 \;\;\;  & \;\;\; 0.830\;\;\;  & \;\;\; 1.332 \;\;\;  & \;\;\; 244 \; \\ 
4 \;\;\;\;& \;\;\; 0.391 \;\;\;  & \;\;\; 1.026\;\;\;  & \;\;\; 1.379 \;\;\;  & \;\;\;  294 \;  \\
6\;\;\;\;& \;\;\; 0.236 \;\;\;  & \;\;\; 1.205\;\;\;  & \;\;\; 1.421\;\;\;  &  \;\;\;  372 \;  \\
64\;\;\;\;& \; \;\;\; 0.081 \;\;\;  & \;\;\; 1.379\;\;\;  & \;\;\; 1.460\;\;\;  & \;\;\;  1,249 \; 
\end{array}
\label{values_two_parameter}
\end{equation}
Note that for three colors, the value of $c_2$ in the two
parameter model, $0.5517$, is significantly smaller 
than for the one parameter model, 
$c_2 \sim 0.8297$, Sec. (\ref{sec:one_parameter}) 
\cite{Dumitru:2010mj}.  While a relatively large change in parameters,
since they were determined in rather different ways, this change is
not that surprising.

Fitting to the latent heat, 
$c_2$ becomes small as $N$ increases.
For infinite $N$, the value of $c_2$ is very close to zero.  We
have no insight as to why this is true, but it is certainly indicated by
the lattice data and the form of the model.

We find that in all cases, $c_3(1) > c_3(\infty)$, so that
the MIT ``bag'' constant, $B$, as computed from
Eq. (\ref{MIT_bag}), is positive.  
The numerical value of $B$ was determined by taking $T_c = 270$~MeV.
The $B$ constant increases with $N$;
since $c_3(1)$ is relatively insensitive to $N$,
this mainly reflects the definition,
$B \sim N^2$ in Eq. (\ref{MIT_bag}).

A positive sign for the bag constant is 
in contrast to alternate models
of the semi-QGP, such as that of
Begun, Gorenstein, and Mogilevsky {\cite{Begun:2010eh}.
In their model,
the pressure is a power series in $T^4$, $T^2$, and $T^0$, with
fixed coefficients, and no dynamical fields.  
They find that in order to fit interaction measure, that the
bag constant must be negative.  Their model
is equivalent to ours above $\sim 1.2 \, T_c$, where
$\langle r \rangle \approx 0$, but not closer to $T_c$,
where $\langle r \rangle \neq 0$.  Indeed, 
having fixed $c_3(\infty)$, by Eq. (\ref{MIT_bag}) 
the bag constant follows from $c_3(1)$.
Fitting to the latent heat gives 
$c_3(1) > c_2(\infty)$, and thus
$B > 0$.  We do not have a general argument as to why 
the lattice data requires $c_3(1) > c_3(\infty)$, and so $B > 0$.

The value of the bag constant appears sensible, although this is
for the pure glue theory.  We stress again, however,
that since we have terms $\sim T^2$ in the potential, not too much should
be made of the value of the coefficient $\sim T^0$, or the sign of 
the bag constant, $B$.  See, in particular, our comments at the end of
sec. (\ref{sec:behavior_Tc}).  

\begin{figure}
\includegraphics[width=10cm]{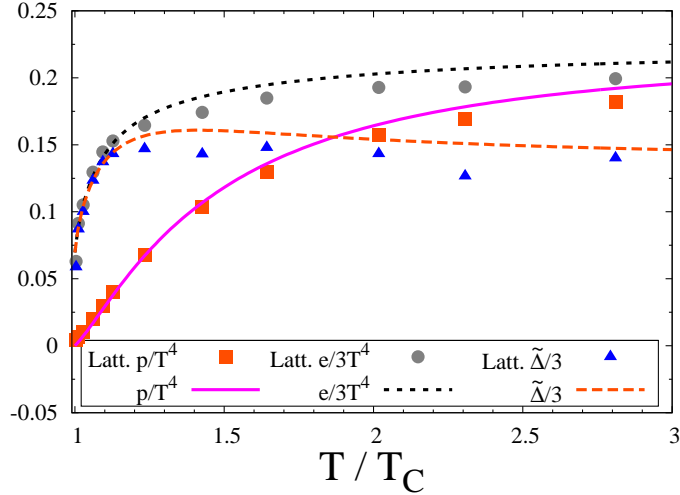}
\caption{\label{fig:SU3Thermo}Thermodynamics of SU(3): pressure
$p/T^4$, energy density $e/(3 T^4)$, and the interaction measure
times $T^2/T_c^2$, 
$\widetilde{\Delta}$ in
Eq. (\ref{rescaled_conf_anomaly}). All quantities are also scaled by $1/8$.}
\end{figure}

\begin{figure}
\includegraphics[width=10cm]{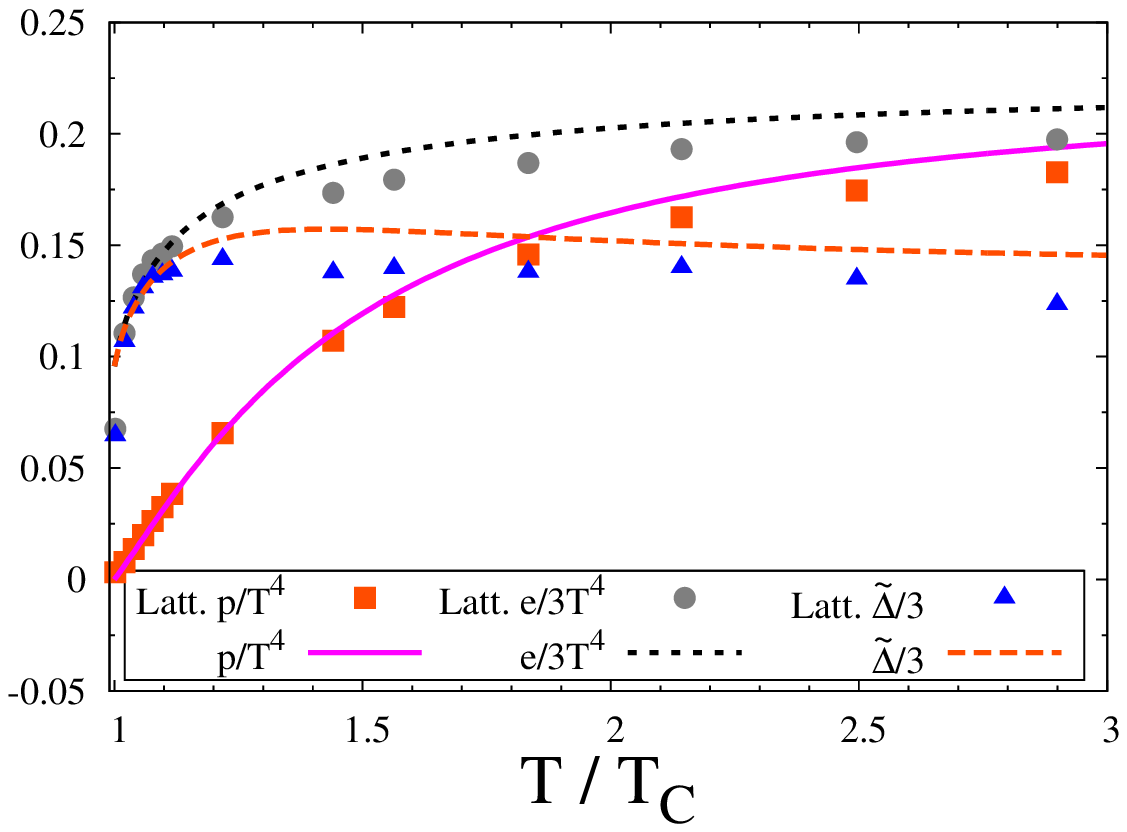}
\caption{\label{fig:SU4Thermo}Thermodynamics of SU(4)
under the uniform eigenvalue ansatz: pressure
  $p/T^4$, energy density $e/(3 T^4)$, and the interaction measure
times $T^2/T_c^2$, $\widetilde{\Delta}$ in
Eq. (\ref{rescaled_conf_anomaly}).
All quantities are also scaled by $1/15$.}
\end{figure}

\begin{figure}
\includegraphics[width=10cm]{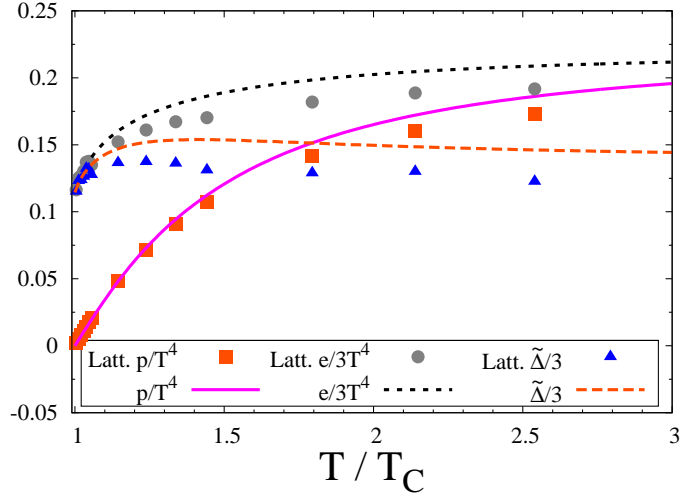}
\caption{\label{fig:SU6Thermo}Thermodynamics of SU(6)
under the uniform eigenvalue ansatz: pressure
  $p/T^4$, energy density $e/(3 T^4)$, and the interaction measure
times $T^2/T_c^2$, 
$\widetilde{\Delta}$ in
Eq. (\ref{rescaled_conf_anomaly}). All quantities are also scaled by $1/35$.}
\end{figure}

We show the results for thermodynamic quantities in the following figures:
Fig. (\ref{fig:SU3Thermo}) for three colors,
Fig. (\ref{fig:SU4Thermo}) for four colors,
and Fig. (\ref{fig:SU6Thermo}) for six colors.
Remember that for four and six colors, this is under the uniform
eigenvalue ansatz.  However, a detailed comparison to the exact solutions
in Sec. (\ref{sec:su45}) shows that the difference
between the uniform eigenvalue ansatz, and the exact solution, is small,
less than $\sim 1\%$.

In each figure, we show the pressure, $p/T^4$, one third the energy density,
$e/(3 T^4)$, and one third the interaction measure rescaled by
$T^2/T_c^2$, 
$\widetilde{\Delta}(T)$ in Eq. (\ref{rescaled_conf_anomaly}).
All quantities are also scaled by the number
of perturbative gluons, $N^2 - 1$.  

Overall, the model appears to reproduce the
lattice data reasonably well, especially near the
transition. Deviations are visible at higher temperatures.
In the conclusion, Sec. (\ref{sec:conclusion}) we discuss how
this might be improved.

\begin{figure}
\includegraphics[width=10cm]{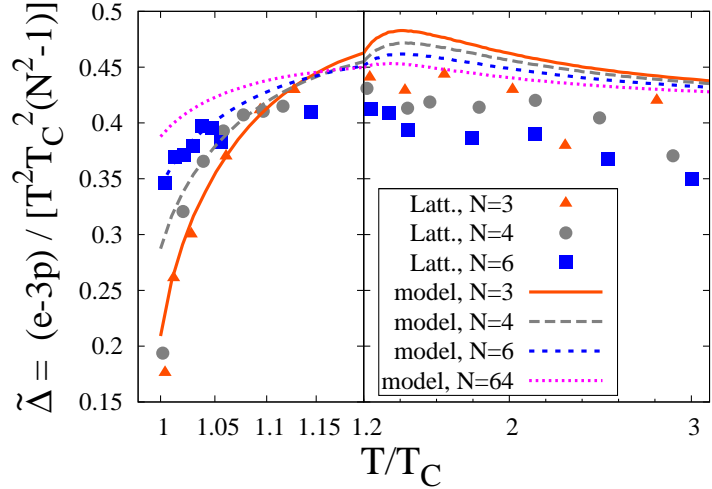}
\caption{\label{fig:SUNDelta}Rescaled
interaction measure, 
$\widetilde{\Delta}=(e-3p)/((N^2-1) T^2 T_c^2)$, 
Eq.\ (\ref{rescaled_conf_anomaly}), for different values
of $N$ from our model, and the lattice.  For $N= 4$, $6$, and $64$
we make the uniform eigenvalue ansatz.
We plot the regions $T_c \rightarrow 1.2 \, T_c$ and 
$1.2 T_c \rightarrow 3.0 \, T_c$ on different abscissa scales;
all curves, and their derivatives, are smooth across $1.2 T_c$.}
\end{figure}

Fig.~(\ref{fig:SUNDelta}) shows the interaction measure of $SU(N)$ for
various $N$. The panel on the left zooms into the region near $T_c$,
from $T_c$ to $1.2 \, T_c$.  Because of the increase in the latent heat
with $N$, $\widetilde{\Delta}(T)$ increases slightly with $N$. 
Since we fit one parameter in our model to the latent heat, our model
agrees well in this region.  The panel on the right shows the region
from $1.2 T_c$ to $3.0 \, T_c$; where
the agreement between the model and the data is not quite as good.
We discuss in Sec. (\ref{sec:conclusion}) how this
might be improved.  We stress, however, that by multiplying 
the interaction measure by $T^2/T_c^2$, to form $\widetilde{\Delta}(T)$,
that we are greatly magnifying the errors in any possible fit.

As discussed before,
chooosing $c_2$ to be near one
makes the width of the transition region 
narrower.  In Fig. (\ref{fig:nc3_compare_loop})
we show the result for the Polyakov loop between the models with
zero, one, and two parameters.  
The width of the transition is broadest for zero parameters, with
$c_2 = 0$, followed by that with two parameters, 
where $c_2 = 0.5517$, and then by that with one parameter,
$c_2 = 0.8297$.

We also plot the results for the renormalized loop from
lattice simulations.  The results for the
Polyakov loop in our model differ sharply from those obtained from the
lattice.  We do not understand the reason for this discrepancy.

Lastly, we show the results for the Polyakov loop, in our model,
from different numbers of colors in Fig. (\ref{fig:SUNLoop}).  They
differ near $T_c$, below $\sim 1.05 \, T_c$, but above this temperature,
they are all rather close to one another.  This is what one expects from
the conformal anomaly, which up to an overall factor of $N^2 -1 $,
scales similarly for all $N$.

\begin{figure}
\includegraphics[width=10cm]{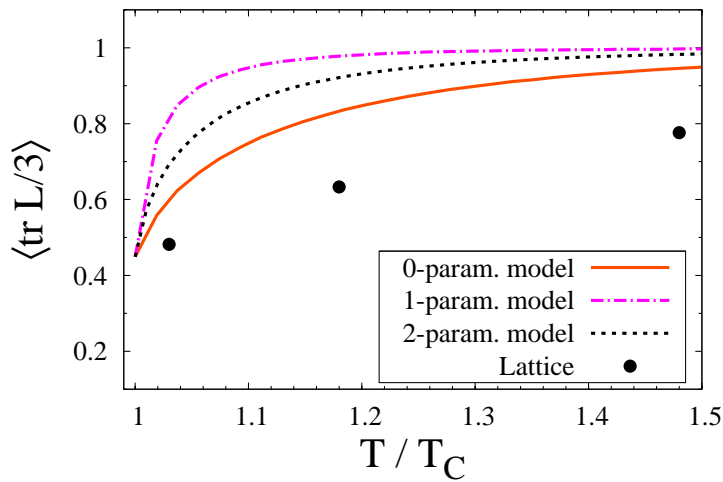}
\caption{\label{fig:nc3_compare_loop}A comparison of the Polyakov loop
for three colors in models with zero, one, and two parameters, 
and from lattice measurements.}
\end{figure}

\begin{figure}
\includegraphics[width=10cm]{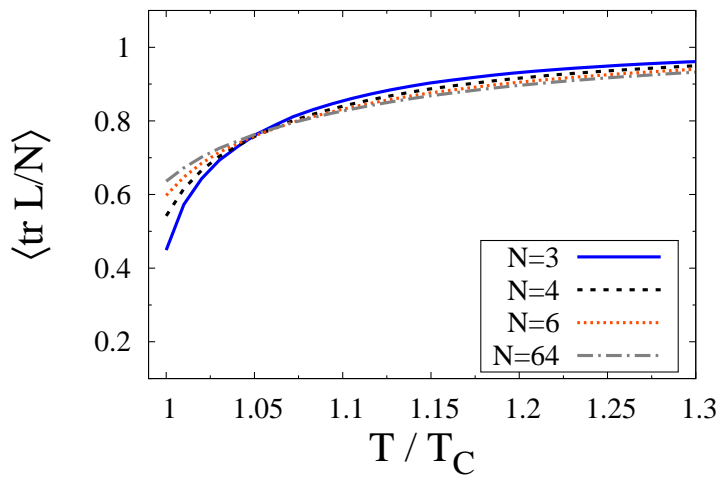}
\caption{\label{fig:SUNLoop}Expectation values of the fundamental
  Polyakov loop in our model for SU(N);
for $N= 4$, $6$, and $64$ we make the uniform eigenvalue ansatz.
 }
\end{figure}

\newcommand{\qi}{q_\alpha}
\newcommand{\qd}{q_\beta}
\section{Interface tensions}
\label{sec:interface}

An interface tension is computed as follows.  Put the system in a box
which is long in one spatial direction, say the $z$ direction.  Let
the system be in one vacuum at one end of the box, and a degenerate
but inequivalent vacuum at the other end of the box.
The theory is in a vacuum state
at both ends, but not in between, so forming this interface costs
action.  This action is proportional to the transverse volume, $V_\text{tr}$,
with the coefficient defined to be the interface tension.

Above $T_c$, one can have the theory in one $Z(N)$ vacua at one end of the
box, and a different $Z(N)$ vacua at the other.  This is known as the
order-order interface tension.  It is equivalent to a 't Hooft loop
in the deconfined phase \cite{KorthalsAltes:1999xb, *KorthalsAltes:2000gs}.
We only compute here for two and three colors, where there is only one
't Hooft loop.  For four or more colors, there is more than one 't Hooft
loop, which we comment upon in the Conclusions, Sec. (\ref{sec:conclusion}).

If the transition is of first order, as it is for $N \geq 3$, then precisely
at $T_c$, the confined vacuum, with $r = 0$, is degenerate with the
deconfined state, at $r = r_c$.  This can then be used to define the
order-disorder interface tension.

The computation of interface tensions in the semi-QGP
is close, but not identical, to that in the
perturbative QGP \cite{Bhattacharya:1990hk, *Bhattacharya:1992qb,Giovannangeli:2002uv, *Giovannangeli:2004sg}.
In the effective action $S$, in addition to the potential
$\mathcal{V}_\text{tot}({\bf q})$ we need a kinetic term,
\begin{equation}
\begin{split}
S=\int d\tau \; d^3x
\; \left(\mathcal{T}_\text{kin}(\bm{q})+ \mathcal{V}_\text{tot}(\bm{q}) \right).
\end{split}
\end{equation}
In general, the kinetic term can be of the form
\begin{equation}
\begin{split}
\mathcal{T}_\text{kin}(\bm{q})
= \frac{1}{2}
\; \sum_{a,b=1}^N G^{ab}(\bm{q}) \; \partial_i q_a(x) \partial_i q_b(x)
\; ,
\end{split}
\end{equation}
where $G^{ab}(\bm{q})$ is a ``metric'' depending on $\bm{q}$.
Such a nontrivial metric arises in computing interface tensions at
next to leading order in the coupling constant
\cite{Bhattacharya:1992qb,Giovannangeli:2002uv, *Giovannangeli:2004sg}

At leading order, however, we can use the form of the kinetic term
at tree level,
\begin{equation}
\begin{split}
\mathcal{T}_\text{kin}(\bm{q})=\frac{1}{2}\mathrm{tr}F_{\mu\nu}^2
= \frac{4\, \pi^2 \, T^2}{g^2}\; \sum_{a=1}^N\left(\frac{d q_a}{dz}\right)^2 \;
.
\end{split}
\label{eq:Tkin}
\end{equation}
We have assumed that $q_a$ is a function only of the long spatial direction,
$z$, of the interface.

Assume that the vacua at the two ends of the box,
$z = -L$ and $z = +L$, are
$\varOmega_i$ and $\varOmega_f$.  These correspond to the two
minima, $\bar{\bm{q}}_i$ and $\bar{\bm{q}}_f$.
We take the spatial length $L \rightarrow \infty$.

The interface tension is related to the shortest path
between $\bar{\bm{q}}_i$ and $\bar{\bm{q}}_f$.
This path obeys the following equation of motion
\begin{equation}
\frac{8\pi^2 T^2}{g^2} \; \frac{d^2q_a}{dz^2} =
\frac{d\mathcal{V}_\text{tot}}{dq_a}\,,
\end{equation}
with the boundary condition
$\bm{q}(-L)=\bar{\bm{q}}_i$ and $\bm{q}(L)=\bar{\bm{q}}_f$.
Multiplying $dq_a/dz$, and integrating over $z$,
we obtain an ``energy'' density,
\begin{equation}
\mathcal{E} \; =\;
\frac{4\pi^2 T^2}{g^2}\; \sum_{a=1}^N
\left(\frac{dq_a}{dz}\right)^2-\mathcal{V}_\text{tot}(\bm{q}) .
\end{equation}
This quantity is independent of $z$, so its value can be taken from either
end.   The kinetic term does not contribute at $z = \pm L$, so the energy
is given by the potential in vacuum,
$\mathcal{E}_\text{vac}= - \, \mathcal{V}_\text{tot}(\bar{\bm{q}}_i)$.

Using the conservation of ``energy'', we define
\begin{equation}
\delta \mathcal{V}_\text{tot}\equiv
\mathcal{V}_\text{tot}(\bm{q}) + \; \mathcal{E}_\text{vac} \; =\;
\frac{4\pi^2 T^2}{g^2}\sum_{a=1}^N\left(\frac{dq_a}{dz}\right)^2.
\label{eq:energyConsevation}
\end{equation}
The effective action becomes
\begin{equation}
S \; = \; V_\text{tr}\int dz
\left(\mathcal{T}_\text{kin}(\bm{q})+\mathcal{V}_\text{tot}(\bm{q}) \right)
\; = \; S_0 \; + \;
2 V_\text{tr}\int dz \; \delta \mathcal{V}_\text{tot}(\bm{q})
\,,
\end{equation}
where $S_0= - 2 V_\text{tr}L \mathcal{E}_\text{vac}$.

The interface tension is then
\begin{equation}
\alpha \; =\;
\frac{1}{V_\text{tr}}\frac{S-S_0}{T}= \frac{2}{T}\;  \int dz \; \delta \mathcal{V}_\text{tot}
\; .
\end{equation}
This is a general form of the interface tension.  The potential depends
upon the problem at hand.

\subsection{The order-disorder interface tension under the
uniform eigenvalue ansatz}

The order disorder interface is the simplest to consider.  We work at
the critical temperature, so we are tunneling from the deconfined state
at $r_c$, to the confined vacuum, at $r=0$, and compute
under the uniform eigenvalue ansatz.

From Eqs. (\ref{fund_ansatz}) and (\ref{eq:ansatz2}),
the kinetic term becomes
\beq
\mathrm{tr} \left( \frac{ \partial A_0}{\partial z} \right)^2
= \frac{\pi^2 \; (N^2 -1 ) \; T^2}{3 g^2 N} \;
\left( \frac{ d r(z)}{d z} \right)^2 \; ,
\label{kinetic_order}
\eeq
where we now allow $r$ to be a function of the spatial direction $z$.

At the critical temperature, the potential is
\beq
{\cal V}_\text{tot}(r,1) = \frac{\pi^2 \; (N^2 -1 ) \; T_c^4}{45}
\left( 1 - c_2 \right) \; {\cal W}(r,1) \; ,
\eeq
where the potential ${\cal W}(r,1)$ is given by Eq. (\ref{eq_crit_pot}).
We now rescale the coordinate $z$ as
\beq
\widetilde{z} = \sqrt{\frac{(1 - c_2) g^2 N}{15}} \; T_c \; z \; ,
\label{rescaling_z_oo}
\eeq
so the action becomes
\beq
S = V_{\text{tr}} \;  \frac{T_c^3}{\sqrt{g^2 N}}
\; \frac{\pi^2 \; (N^2 - 1)}{3 \sqrt{15}} \; \sqrt{1 - c_2} \;
\int \; d\widetilde{z} \;
\left( \left( \frac{d r}{d \widetilde{z}} \right)^2
+ \; {\cal W}(r,1) \right) \; .
\eeq
Using the conservation of ``energy'', the integral becomes
\beq
2 \int \; d\widetilde{z} \; {\cal W}(r,1) \; = \;
2 \; \int^{r_c}_0 \; dr \; \sqrt{{\cal W}(r,1)}
\; = \; \frac{\sqrt{2 N^2 - 3}}{3 N} \; r_c^3  \; ,
\eeq
where $r_c$ is given in Eq. (\ref{eq:critr}).  Using this value,
the order disorder interface tension is given by
\beq
\alpha_{o-d} = \frac{T_c^2}{\sqrt{g^2 N}} \; \frac{\pi^2}{3^{5/2} \, 5^{1/2}}
\; \frac{(N^2 - 1)\,(N^2 - 4)^3}{N \, (2N^2-3)^{5/2} }
\; \sqrt{1 - c_2} \; .
\label{order_disorder}
\eeq
As for the order-order interface tension, this is proportional not to
$\sim 1/g^2$, but to $\sim 1/\sqrt{g^2}$, because the potential for
$r$ is generated at one loop order.  It is proportional to $N^2$ at
large $N$, which is typical of the free energy in the deconfined phase.

To compare to the lattice, we need to assume a value for the coupling.
We take $\alpha_s(T_c) = 0.3$, simply to get an idea of the numbers,
and in particular, of the $N$-dependence.

On the lattice, there are results for three colors
by Beinlich, Peikert, and Karsch \cite{Beinlich:1996xg}
and by Lucini, Teper, and Wenger \cite{Lucini:2005vg}.
For four or more colors, there are results from Ref. \cite{Lucini:2005vg}.

We summarize the comparsion for order-disorder interface tension,
$\alpha_{o-d}/T_c^2$, in the following table:
\begin{equation}
\begin{array}{ccccc}
N \;\;\; & \;\;\; {\rm Ref.} \, \mbox{\cite{Beinlich:1996xg}} \;\;\; & \;\;\; {\rm Ref.} \, \mbox{\cite{Lucini:2005vg}} \;\;\; & \;\;\; 1 \; \text{parameter model} \;\;\; & \;\;\; 2 \; \text{parameter model} \\
3  \;\;\; &  \;\;\; .0155  \;\;\; & \;\;\; .0194   \;\;\; & \;\;\; .014 \;\;\; & \;\;\; .022 \\
4  \;\;\; & \;\;\;    \;\;\; &  \;\;\; .121  \;\;\; & \;\;\; .049   \;\;\; & \;\;\; .093\\
6  \;\;\; & \;\;\;    \;\;\; &  \;\;\; .394  \;\;\; & \;\;\; .167   \;\;\; & \;\;\; .35\\
\infty  \;\;\; & \;\;\;    \;\;\; &  \;\;\; .0138 N^2 \;\;\; & \;\;\; .006 N^2 \;\;\; & \;\;\; .0139 N^2 \\
\end{array}
\end{equation}
Taken at face value, the values of the order-disorder interface tension
computed in the two parameter model agree remarkably well with those
from the lattice.

This agreement could well be fortuitous.
We have only included the result to leading order in the coupling
constant, $g^2$.  For the order-disorder interface tension, it is
known that including at least corrections to $\sim g^2$
\cite{Bhattacharya:1992qb}, and to $\sim g^3$
\cite{Giovannangeli:2002uv,Giovannangeli:2004sg} are essential to
obtain agreement with lattice results, even at temperatures $\sim 10 T_c$,
well into the perturbative QGP \cite{deForcrand:2005pb}.

Even so, the fit to the latent heat indicates a lower value of $c_2$ in
the two parameter model than in the one parameter model,
Eq. (\ref{values_two_parameter}).  Because
of the overall factor of $\sqrt{1 - c_2}$ in the order-disorder interface
tension, Eq. (\ref{order_disorder}), the results do support a value of
$c_2$ which decreases as $N$ increases.

We conclude this section with a general comment.  While the order-disorder
interface tension is of order $N^2$ at large $N$, in fact the coefficient,
from either lattice simulations or our two parameter model, is {\it extremely}
small, $\alpha_{o-d}/(N^2 \, T_c^2) \approx 0.014 $.

This is in sharp contrast to the latent heat, which is, properly normalized,
a number of order one.
For an ideal gas of $N^2-1$
gluons, the energy density is $(N^2-1) T^4$ times a pure number,
$\pi^2/15 \sim .66$.
At infinite $N$, by Eq. (\ref{DG_latent}) the energy density
at $T_c$ is $N^2 T_c^4$
times $\sim .39$; thus the latent
heat is almost $60\%$ the energy density of an
ideal gas.  This is well known from
three colors: the energy density very quickly approaches that of an ideal
gas, close to $T_c$.

Our model provides a qualitative explanation for why
the latent heat is large, but the order-disorder interface tension
is small.  The latent heat
is given by the ``jump'' in the order parameter, essentially by $r_c$.
For any $N \geq 3$, this is not a small number;
at infinite $N$, $r_c = \frac{1}{2}$, Eq. (\ref{critical_r}).  Note that
since $r$ is a number between $0$ and $1$, we can speak of its magnitude
without qualification.

In contrast, the order-disorder interface tension is given by the probability
for tunneling through the barrier at $T_c$.  Even if the jump in $r$ is large,
the probability not to tunnel through the barrier can be small, if the
height of the barrier is very small.
That is, it is a {\it shallow} potential.

Such a shallow potential is exhibited by our model.
For simplicity, consider the potential at infinite $N$.
At the critical temperature, by Eq. (\ref{eq_crit_pot}) the potential is
\beq
{\cal W}(r,1)_{N=\infty} = 2 \;
r^2 \left( r - \frac{1}{2}\right)^2  \; .
\label{eq_crit_pot_inf_N}
\eeq
The confined vacuum is at $r = 0$, while at $T_c$,
the deconfined vacuum is at $r = 1/2$.
Between $r=0$ and $r = 1/2$, the maximum occurs at
$r = 1/4$.  At this maximum, the value of the potential is
${\cal W}(r=1/4,1)_{N=\infty} = 1/128$.
This is certainly a barrier, but to gauge its height, consider the value of
the potential at the perturbative vacuum, $r=1$, which is
${\cal W}(r=1,1)_{N=\infty} = 1/2$.  That is, the potential
at the maximum of the barrier is smaller by a factor of $64$
than what might have been expected.  This is why the order-disorder
interface tension is so much smaller than expected, because it is
a broad, but very shallow, potential in $r$.

Needless to say, this also illustrates the virtue of computing quantities,
such as the order-disorder interface tension: it gives one insight into
more detailed properties of the theory than available from just bulk
thermodynamics.

\subsection{The order-order interface tension for two colors}

We next turn to the order-order interface tension, or
't Hooft loop, for two colors.  This is simple because there is only one
direction in the Weyl chamber, along the Pauli matrix $\sigma_3$.
Thus moving to a confining phase, or tunneling from one $Z(2)$ vacua
to another, occurs along the same direction.  This greatly simplifies
the computation, and allows us to compute analytically, as
in the previous section.

For two colors, one can show from Eq. (\ref{eq:wpot})
that the potential for $r$ has a simple form,
\beq
{\cal W}(r,t) - {\cal W}(0,t)
\; = \; 5 \; \left(
- \; \frac{m(t)^2}{2} \, r^2 \; + \; \frac{1}{4} \, r^4 \right) \; \; ; \;\;
m(t)^2 \; = \; \frac{t^2 -1}{t^2 - c_2} \; .
\eeq
The value of the potential at $r = 0$, ${\cal W}(0,t)$, enters only into
the vacuum energy, $\mathcal{E}_{\text{vac}}$, and can be ignored.

This potential behaves as expected: the mass for $r$ vanishes at $t = 1$,
so the transition is of second order.  In the deconfined phase, $t > 1$,
there are two degenerate vacua, at $r(t) = \pm \, m(t)$.
The $\pm$ represents the two
degenerate $Z(2)$ vacua in the theory.  At a given $t > 1$, 
we want to determine
the tunneling probability between the $r = - m(t)$ and $r = + m(t)$.

The kinetic term is as in Eq. (\ref{kinetic_order}).  We rescale
the position as
\beq
\widetilde{z} = \sqrt{\frac{2 g^2}{3}}  \; \sqrt{t^2 - c_2} \;  T_c\;z \; ,
\eeq
so that the action becomes
\beq
\frac{1}{\sqrt{g^2}} \frac{\pi^2 \, T^2\,T_c}{\sqrt{6}} \;
\sqrt{t^2 - c_2} \;
\int \; d \widetilde{z}
\left(
\left( \frac{dr}{d\widetilde{z}} \right)^2 +
\frac{1}{5} \left( {\cal W}(r,t) - {\cal W}(0,t) \right)
\right) \; .
\eeq
By changing $r \rightarrow r - m(t)$, the potential, and the related integral,
is precisely that of the previous section.
The result for the order-order interface tension is
\beq
\alpha_{o-o} = \frac{4 \, \pi^2 \,}{3 \, \sqrt{6\, g^2}}
\; T^2 \;
\frac{ (t^2 - 1)^{3/2}}{t (t^2 - c_2 )} \; ;
\eeq
remember $t = T/T_c$.

As $T \rightarrow T_c$, the order-order interface tension vanishes,
as $\alpha_{o-o} \sim (T - T_c)^{3/2}$.  By universality, this
interface tension should vanish as $\alpha_{o-o} \sim (T-T_c)^{2 \nu}$,
where $2 \nu \sim 1.26$; lattice results
by de Forcrand, D'Elia, and Pepe \cite{deForcrand:2000fi}
find $2 \nu \sim 1.32$.  Our result is a type of mean field theory, though,
and so we certainly do not expect our result to correctly describe the
critical region, which is very near $T_c$.

\subsection{The order-order interface tension for three colors}

For three colors, the path to the confined vacuum is along the ${\bf Y_c}$
direction, while that for the 't Hooft
loop is along the ${\bf Y}_1$ direction in Fig.~(\ref{fig:path}).
Thus we have to determine a path
in two dimensions.  This problem can be solved numerically, but is not
amenable to analytic solution.

We choose a vacuum in the semi-QGP as $\bar{\bm{q}}_i$.
\begin{equation}
\bar{\bm{q}}_i =( \bar{q}_i, 0,-\bar{q}_i) \;\;\; , \;\;\;
\bar{q}_i=
\frac{1}{4}\left(1-
\sqrt{1
- \frac{80}{81}\left(\frac{1-c_2}{t^2-c_2}\right)}\right) \; .
\end{equation}
We wish to find a path which tunnels from this point to a $Z(3)$ transform
of this point, $\bar{\bm{q}}_f$,
\begin{equation}
\bar{\bm{q}}_f =
(\frac{2}{3}-\bar{q}_i ,
-\frac{1}{3}+\bar{q}_i,
-\frac{1}{3})
\; .
\end{equation}
We parameterize this path as
\begin{equation}
\begin{split}
\bm{q}= \bar{\bm{q}}_i+(
\frac{2\qi}{3},
 \qd-\frac{\qi}{3},
-\qd-\frac{\qi}{3}) \; .
\end{split}
\end{equation}

The potential which governs this tunneling is
\begin{equation}
\begin{split}
\delta \mathcal{V}_\text{tot}(q_1,q_2,q_3)
=\mathcal{N}_{\mathcal{V}_3}
\mathcal{V}_\text{norm}(\qd,\qi) \; ,
\end{split}
\end{equation}
where
\begin{equation}
\begin{split}
\mathcal{N}_{\mathcal{V}_3}=\frac{8\pi^2}{3}T_c^4t^2(t^2-c_2) \; .
\end{split}
\end{equation}
Without loss of generality, we can choose the variables to satisfy
$0\le \qi <1$, $0\le \qd<\qi$, and $\qd<1-\qi$.  The potential becomes
\begin{equation}
\begin{split}
\mathcal{V}_\text{norm}(\qd,\qi)&= \qi^2 (\qi - \bar{q}_f)^2 +
2 \qi (\qi - \bar{q}_f) (1 - \bar{q}_f) \qd \\
&\quad+
  (6 \qi(\qi -   \bar{q}_f) + \bar{q}_f (2 + \bar{q}_f)) \qd^2
- 2 (1 + 3 \bar{q}_f) \qd^3 + 9 \qd^4 \; ,
\end{split}
\end{equation}
where $(\qd,\qi)=(0,\bar{q}_f) \equiv (0,1-3\bar{q}_i) $ corresponds to $\bar{\bm{q}}_f$.

The kinetic term is
\begin{equation}
\mathcal{T}_\text{kin}=\mathcal{N}_{\text{cl}3}
\Bigl[\Bigl(\frac{d\qi}{dz}\Bigr)^2+3\Bigl(\frac{d\qd}{dz}\Bigr)^2    
\Bigr]\, ,
\end{equation}
where $\mathcal{N}_{\text{cl}3}={8}\pi^2T^2/{(3g^2)}$.

The interface tension becomes
\begin{equation}
\begin{split}
\alpha_{o-o}&=2 \sqrt{\mathcal{N}_{\mathcal{V}_3}\mathcal{N}_{\text{cl}3}}
\int d \widetilde{z} \mathcal{V}_\text{norm}(\qi,\qd)\\
&= \frac{16}{3g}\pi^2T^2\sqrt{1-\tilde{c}_2t^{-2}}\;
\int d \widetilde{z} \mathcal{V}_\text{norm}(\qi,\qd) \; ,
\end{split}
\end{equation}
where $\widetilde{z} = g \; T_c \; \sqrt{t^2 - c_2} \; z \; $.

For two colors, the path is in one direction, and so we can use energy
conservation to determine the action to tunnel, even without an explicit
form of the solution.  For three colors, the path is in two directions,
and so energy conservation does not, by itself, determine the solution,
nor its action.

Thus we need to explicit determine the path which tunnels between the two
degenerate vacua.  This satisfies the equation of motion,
\begin{align}
\frac{d^2\qi}{d \widetilde{z}^2} =  \frac{1}{2}\frac{d\mathcal{V}_\text{norm}}{d\qi}\,,\qquad
\frac{d^2\qd}{d \widetilde{z}^2} =  \frac{1}{6}\frac{d\mathcal{V}_\text{norm}}{d\qd}\,.
\end{align}

The boundary conditions which the solution obeys is
\begin{align}
\qi(-\infty)&=0,\qquad \qi(\infty)=\bar{q}_f, 
\qquad\qd(-\infty)=0,\qquad \qd(\infty)=0\,. \label{eq:boudaryqd}
\end{align}
These boundary conditions do not uniquely determine the solution, because
we need to specify the turning point.  We require
that the turning point occurs at the middle of the interface, $\widetilde{z} = 0$, so
that
\begin{align}
\qi(0) = \frac{ \bar{q}_f}{2} \, .
\end{align}
This is natural, since
the potential is symmetric under $\qi\leftrightarrow(\bar{q}_f-\qi)$.
This also implies that at the turning point,
\begin{equation}
\frac{d\qd(0)}{d \widetilde{z}}=0 \,.
\end{equation}
At this point, the derivative of $\qi$ is obtained by
energy conservation, Eq.~(\ref{eq:energyConsevation}):
\begin{equation}
\frac{d\qi(0)}{d \widetilde{z}} =
\sqrt{{\mathcal{V}}_\text{norm}(\qd(0),\qi(0))} \,.
\label{eq:boundaryCondition}
\end{equation}
In the numerical computations,
we use Eq.~(\ref{eq:boundaryCondition}) as a boundary condition
instead of Eq.~(\ref{eq:boudaryqd}).

In the perturbative QGP, $\bar{q}_f=1$ is the vacuum,
so that as $t\to\infty$, the derivative
$d \mathcal{V}_\text{tot}/d\qd=0$ vanishes at
$\qd=0$.  In this case, the path with a straight line is the
solution of the equation of motion.  This is a path along ${\bf Y}_1$
\cite{Bhattacharya:1990hk,Bhattacharya:1992qb}.

In the semi-QGP, the straight line is no longer
the solution of the equation of motion.  That is, the path is along
both ${\bf Y}_1$ and ${\bf Y}_2$.  In this case, it is necessary to
solve the equations numerically.

In Fig.~(\ref{fig:path}) we show the path
$(\qi(z), \qd(z))$ in the plane of $\qi$ and $\qd$.
This shows that as $T$ approaches $T_c^+$,
the tunnelling path passes closer and closer to the
$SU(3)$ confining vacuum, where both ${\rm tr} {\bf L}$
and ${\rm tr}{\bf L}^2$ vanish.
For three colors, the fact that $|\ell|$ is small at $T_c^+$ 
in the middle of the
interface has been observed previously within 
a Polyakov loop model~\cite{Layek:2005fn}.
We also note that the interface tension has been analyzed in linear
models \cite{Vuorinen:2006nz, *deForcrand:2008a, *Zhang:2011aa}.

In Fig.~(\ref{fig:interface}),  we give the result for the
order-order interface tension, $\alpha_\text{o-o}$.
We remark that while in principle $\langle q \rangle \neq 0$ for all
temperatures, in practice this is very small except close to $T_c$.  This
means that to a good approximation, we can take a path along ${\bf Y}_1$
when $T > 1.2 \, T_c$.
We also note that in plotting the interface tension in our model, versus
that from the lattice \cite{deForcrand:2005rg}, that we have included
a perturbative correction $\sim g^2$.  This correction was computed in the
perturbative QGP, and so should be recomputed for the semi-QGP.  Except
for $T < 1.2 \, T_c$, however, it can be shown that this correction is
correct \cite{Dumitru2013}.

\begin{figure}
\includegraphics[width=0.40\textwidth]{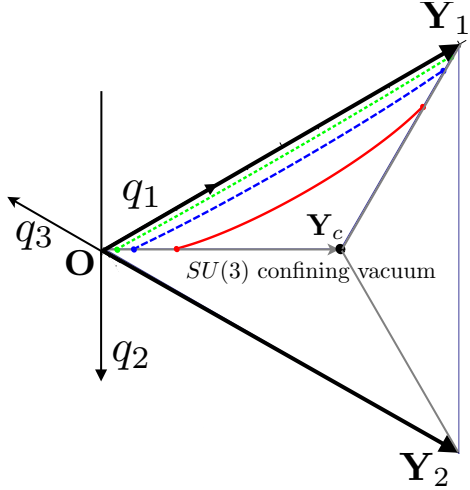}
\caption{The classical path of the interface
tension for $SU(3)$ pure gauge theory.}
\label{fig:path}
\end{figure}

\begin{figure}
\includegraphics{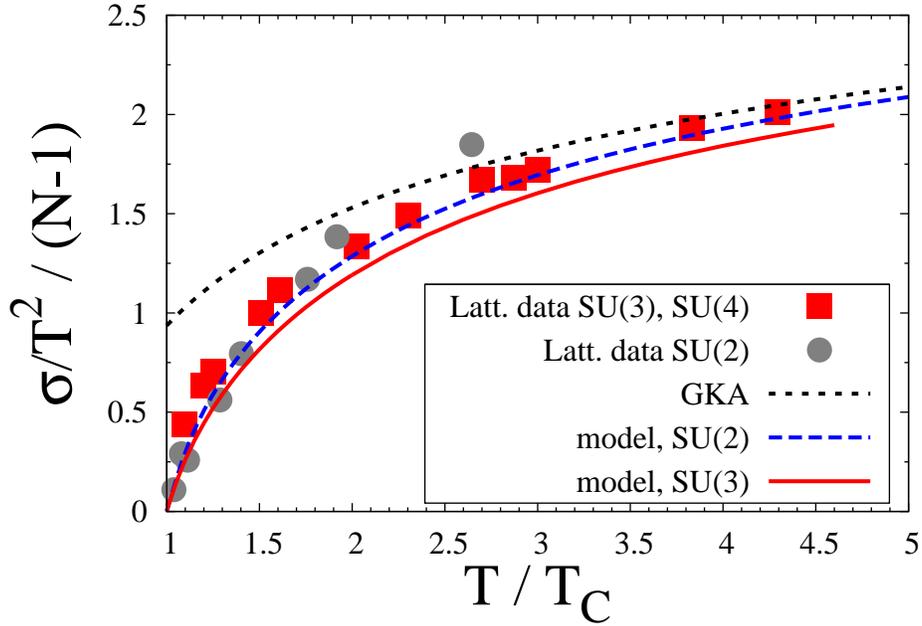}
\caption{ The interface tension for $SU(3)$ pure gauge theory.}
\label{fig:interface}
\end{figure}

We conclude by discussing the problem of order-order interface tensions
for four or more colors.
In the complete QGP, the path for order-order
interface tension is along one of the hypercharge
directions, the ${\bf Y}_k$ of Eq. (\ref{eq:hypercharges})
\cite{Giovannangeli:2002uv,Giovannangeli:2004sg}.
The example of three colors shows, however, that near $T_c$,
one will have to compute a path in the full space of $N -1$ dimensions.
This is an interesting but nontrivial exercise in
minimization, which we defer for now.  We comment further about
these interface tensions in the Conclusions, Sec. (\ref{sec:conclusion}).

\section{Numerical solution of the model for four to seven colors}
\la{sec:su45}

Before describing our numerical results, we first
make an elementary but useful remark.
As discussed preceeding Eq. (\ref{eq:ansatz1}), 
by a global $Z(N)$ rotation we can require that the expectation value
of the Polyakov loop is real.  If $N = 2M$ or $2M + 1$, this means
that the solution involves $M$ degrees of freedom.
At infinitely high temperature
all ${\bf q}$'s vanish.  As the temperature decreases the 
${\bf q}$'s move along
a curve in this $M$-dimensional space, until at
the critical temperature they end up at a point
${\bf q}^+_c$.  At this point, the value of the potential
equals that in the confined vacuum, where
${\bf q}^-_c = {\bf Y}_c$, Eq. (\ref{barycenter_def}).  This assumes
that the transition is of first order, so that
${\bf q}^+_c \neq {\bf q}^-_c$; 
we justify this later in Sec. (\ref{sec:firstorder}).

For finite $N > 3$,
the path in the $M$-dimensional space 
can be determined numerically.  The total potential in our model is given by
\beq
{\cal V}_{tot}({ \bf q})
= f_0(T) + f_1(T) \; V_1({\bf q}) 
+ f_2(T) \; V_2({\bf q})  \; .
\label{general_potential_Vs}
\eeq
We have chosen very specific forms for the functions $f_0(T)$, 
$f_1(T)$, and $f_2(T)$, 
but the following conclusion is independent of their specific form.

The solution of the model at a given temperature,
${\bf q}_0(T)$, is given by requiring that the total potential
is stationary with respect to the ${\bf q}$'s,
\beq
\left.
\frac{\partial {\cal V}_{tot}({\bf q})}{\partial {\bf q}} 
\right|_{{\bf q} = {\bf q}_0(T)} = 
f_1(T) \; \frac{\partial}{\partial {\bf q}} 
\left( V_1({\bf q}) + \frac{f_2(T)}{f_1(T)} \;  V_2({\bf q}) 
\right)_{{\bf q} = {\bf q}_0(T)} = 0  \; ,
\label{general_solution}
\eeq
which is the generalization of Eq. (\ref{eq_W}).  

Because the ${\bf q}$-dependence only enters through two functions,
$V_1({\bf q})$ and $V_2({\bf q})$, trivially
this equation only involves only the ratio $f_2(T)/f_1(T)$.  Whatever
the specific form of these two functions, at a given temperature this
is a pure number.  

This implies that 
in the $M$-dimensional space of the ${\bf q}$'s,
that the path is {\it independent} of the choice of these two functions.
Further, it follows immediately that the endpoint of the path,
${\bf q}_c^+$, is also {\it independent} of this choice of these two functions.

That the endpoint is independent of these functions was observed
previously for the uniform eigenvalue ansatz.  There, 
it was found that the properties of the solution at $T_c$ were independent
of the one, free parameter, $c_2$.

Needless to say, different choices of the $f_i(T)$ functions {\it do} give
different physics.  While the path in ${\bf q}$-space is independent
of the $f_i$'s, thermodynamic quantities also involve derivatives with
respect to temperature, and these change as $f_1(T)$ and $f_2(T)$ do.
That is, the temperature dependence of how one proceeds along this
fixed path depends upon the choice of these functions.
In the uniform eigenvalue
ansatz, this is what changing the parameter $c_2$ does: it shifts the
overall scale of $T/T_c$ in a nonlinear fashion.

\begin{figure}[htbp]
\includegraphics[width=10cm]{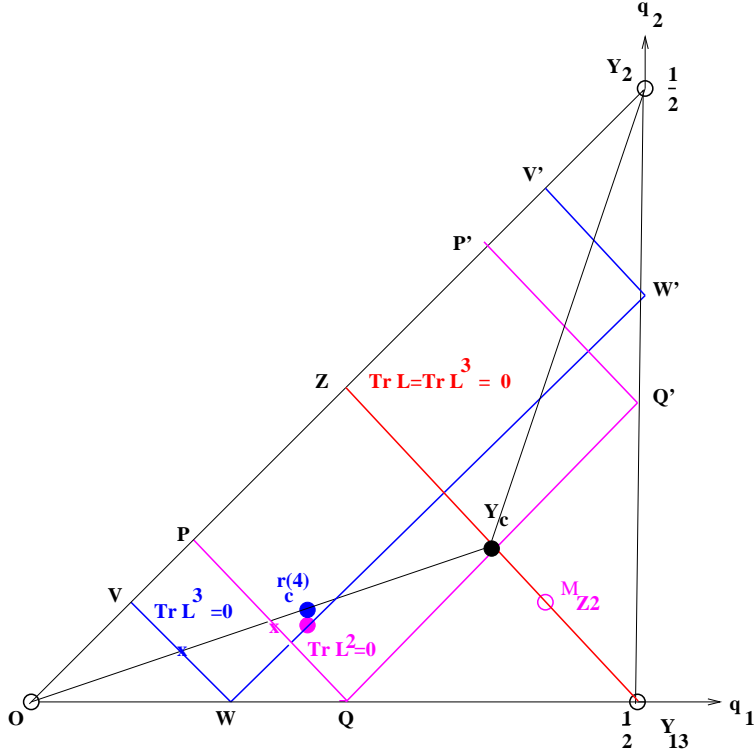}
\caption{\label{fig:su4realtraceplane} 
For four colors, the region of the Weyl chamber where the expectation
value of the Polyakov loop is real.}
\end{figure}

\subsection{Four and five colors}

We now turn to the case of four colors.  In Fig.
(\ref{fig:su4realtraceplane}) we illustrate the
region of the Weyl chamber in which the expectation value of the
Polyakov loop is real.  This is the plane spanned by 
the hypercharge $\Y_2$ and by $\Y_{13}=(\Y_1+\Y_3)/2$.

The confining vacuum is given by the barycenter, ${\bf Y}_c$.
The curves where 
${\rm tr} \, {\bf L}^p$ vanish, for $p=1$, $2$, and $3$, are also indicated.
The uniform eigenvalue ansatz is the line ${\bf O}{\bf Y}_c$.
In this ansatz, the point ${\bf q}_c^+$ is given by a blue point;
that for the exact solution is given by a magenta cross.

Under a $Z(2)$ transformation, 
the triangle ${\bf O} {\bf Y}_2 {\bf Y}_{13}$ maps onto itself:
${\bf O}$ and ${\bf Y}_2$ interchange with one another,
while the lines where ${\rm tr}\, {\bf L}={\rm tr}\,{\bf L^3}$ 
vanish are left invariant. 

The numerical solution for the trajectory of 
$SU(4)$ is plotted in 
Fig. (\ref{fig:pathsu4}). 
There are two positive eigenvalues, $q_1$ and $q_2$.  In the uniform
eigenvalue ansatz, $q_2 = 3 q_1$.  The left panel shows 
the path in the plane of $q_1$ and $q_2$; visually,
this is obviously very close to 
a straight line.  The left panel shows the values of $q_1$ and
$q_2/3$; the amount by which $q_1 \neq q_2/3$ indicates the
deviation from the uniform eigenvalue ansatz. 
Even at $T_c$, this deviation is very small.
As discussed previously, for thermodynamic quantities, and the
expectation value of the Polyakov loop,
the results are within the width of the curves in Figs.
(\ref{fig:SU4Thermo}) and (\ref{fig:SUNLoop}), respectively.

In $Z(4)$ spin systems there is a possibility to have mixed phases between
ordered and disordered phases where $Z(4)$ is broken but
$Z(2)$ is unbroken. For $SU(4)$ such a $Z(2)$ invariant phase
would have $\Tr\,{\bf L }=\Tr\,{\bf L}^3=0$ and $\Tr\,{\bf L}^2\neq 0$. 
For example, the point along this line where ${\rm tr}\,{\bf L}^2 = 1$
is indicated by the point $M_{Z2}$.  We find no evidence for
such a $Z(2)$ phase at any temperature.
This agrees with numerical simulations on the lattice~\cite{Lucini:2002ku} .

\begin{figure}[htbp]
\includegraphics[width=0.48\textwidth]{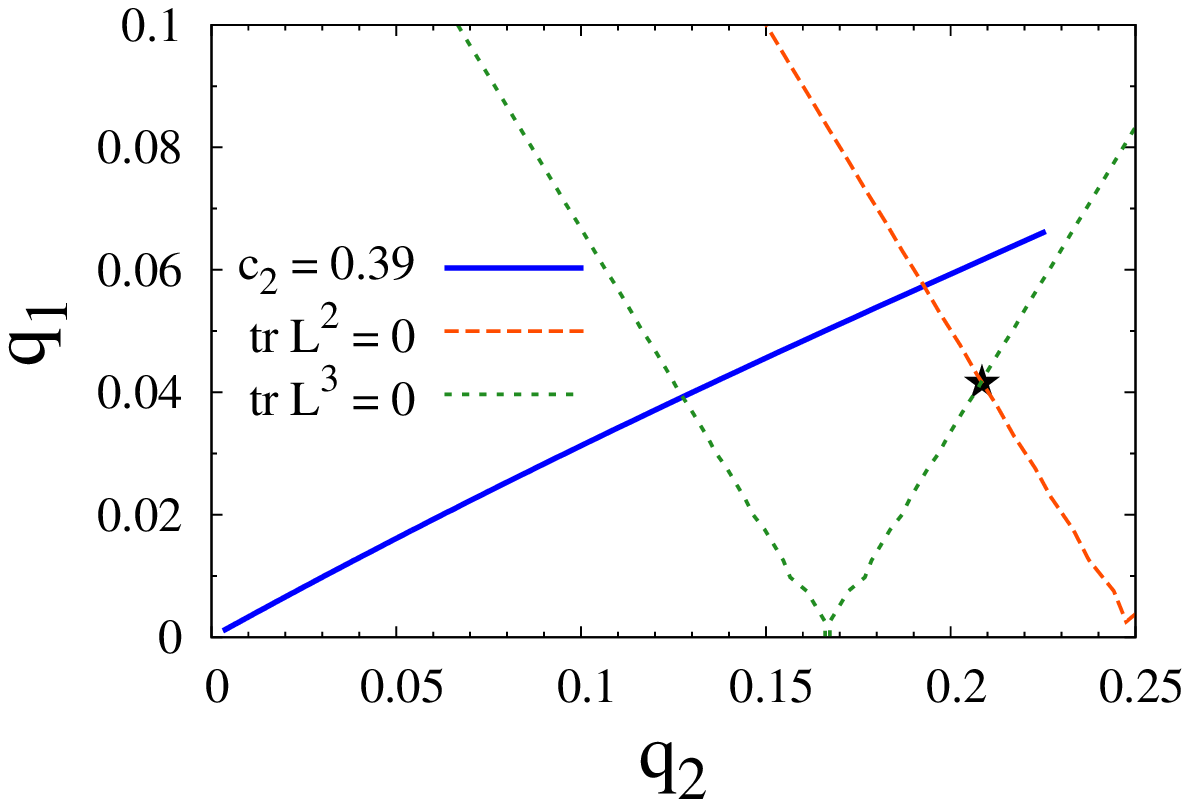}
\includegraphics[width=0.48\textwidth]{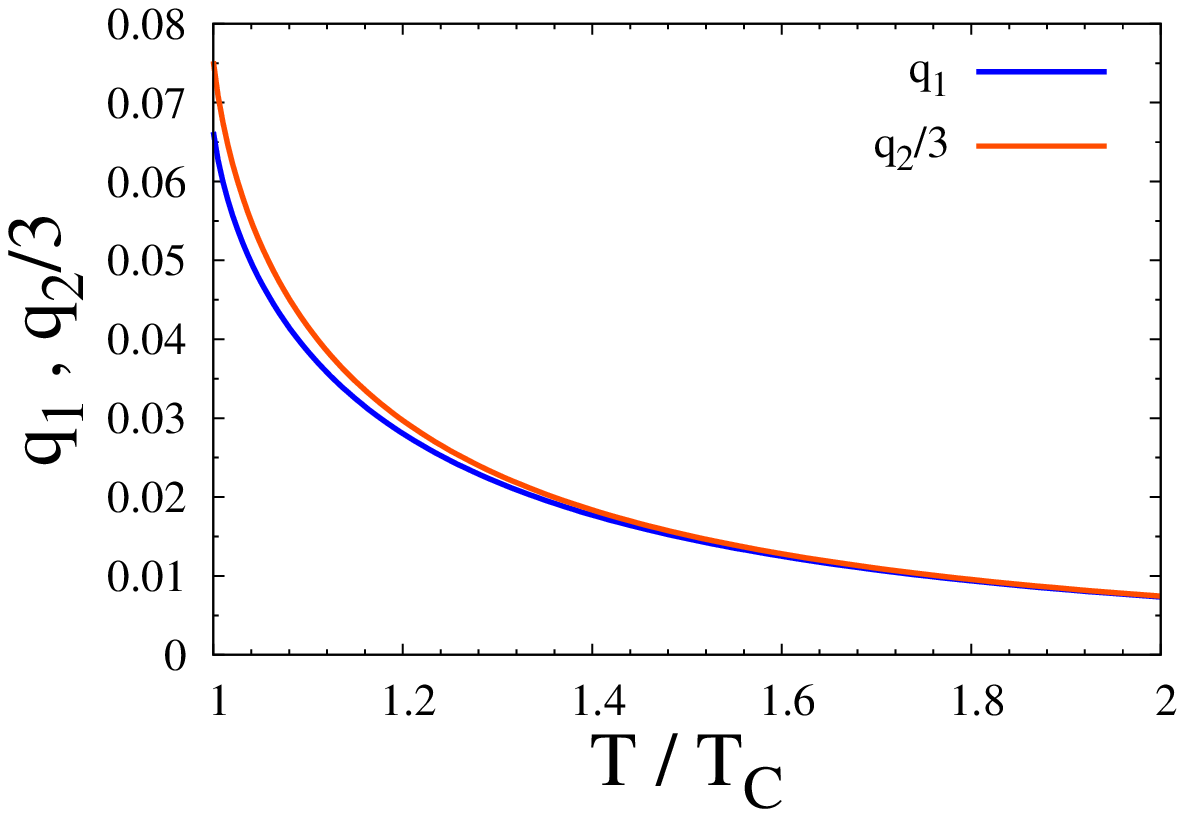}
\caption{\label{fig:pathsu4} The numerical solution for four colors,
with $c_2 = 0.39$.  The left panel is in the $q_1$-$q_2$ plane,
while the right panel gives $q_2/3$ and $q_1$. 
}
\end{figure}

For five colors the plane where the Polyakov loop is real is shown in Fig.
(\ref{fig:su5realtraceplane}); it is spanned by the line
$\Y_{14}=(\Y_1+\Y_4)/2$ and $\Y_{23}=(\Y_2+\Y_3)/2$.  
In this plane
${\rm tr}\,{\bf L}$ is real and we expect the trajectory of the minima to
again nearly coincide with the straight line {\em ansatz} from the
origin to $\Y_c$, as for four colors. This is indeed the
case as is evident from \fig (\ref{fig:pathsu5}). Only near $T_c$ do
the eigenvalues of the Wilson line deviate from the straight line.
Here, too, we found that the thermodynamic functions
with those obtained above with the straight line {\em ansatz}.

\begin{figure}[htbp]
\includegraphics[width=10cm]{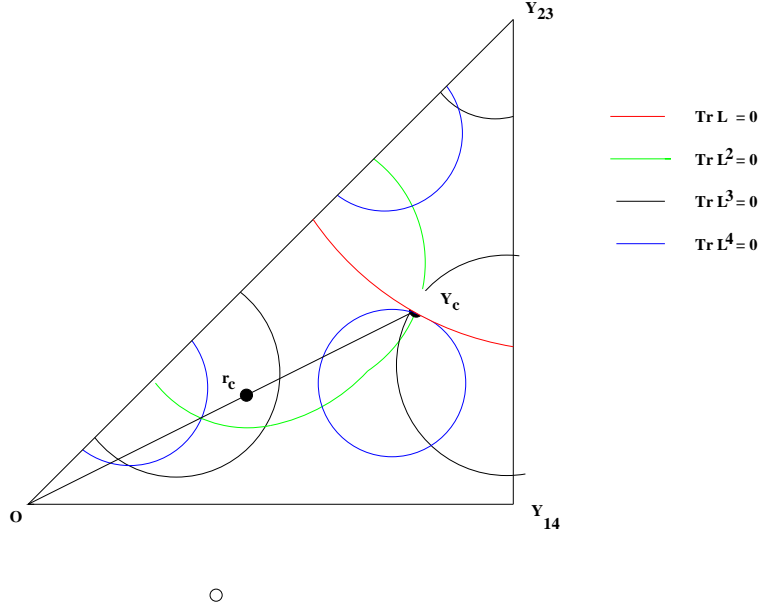}
\caption{\label{fig:su5realtraceplane} For five colors, the plane
where the Polyakov loop is real.}
\end{figure}

\begin{figure}[htbp]
\includegraphics[width=0.48\textwidth]{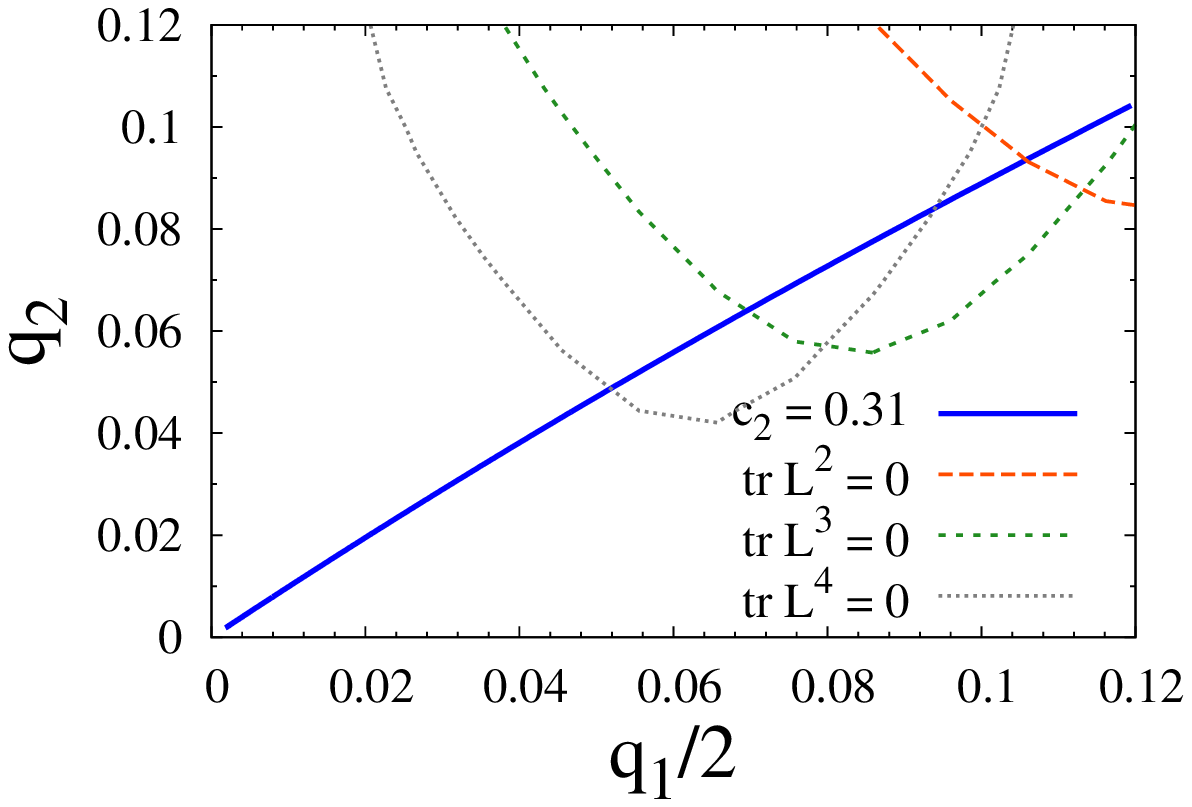}
\includegraphics[width=0.48\textwidth]{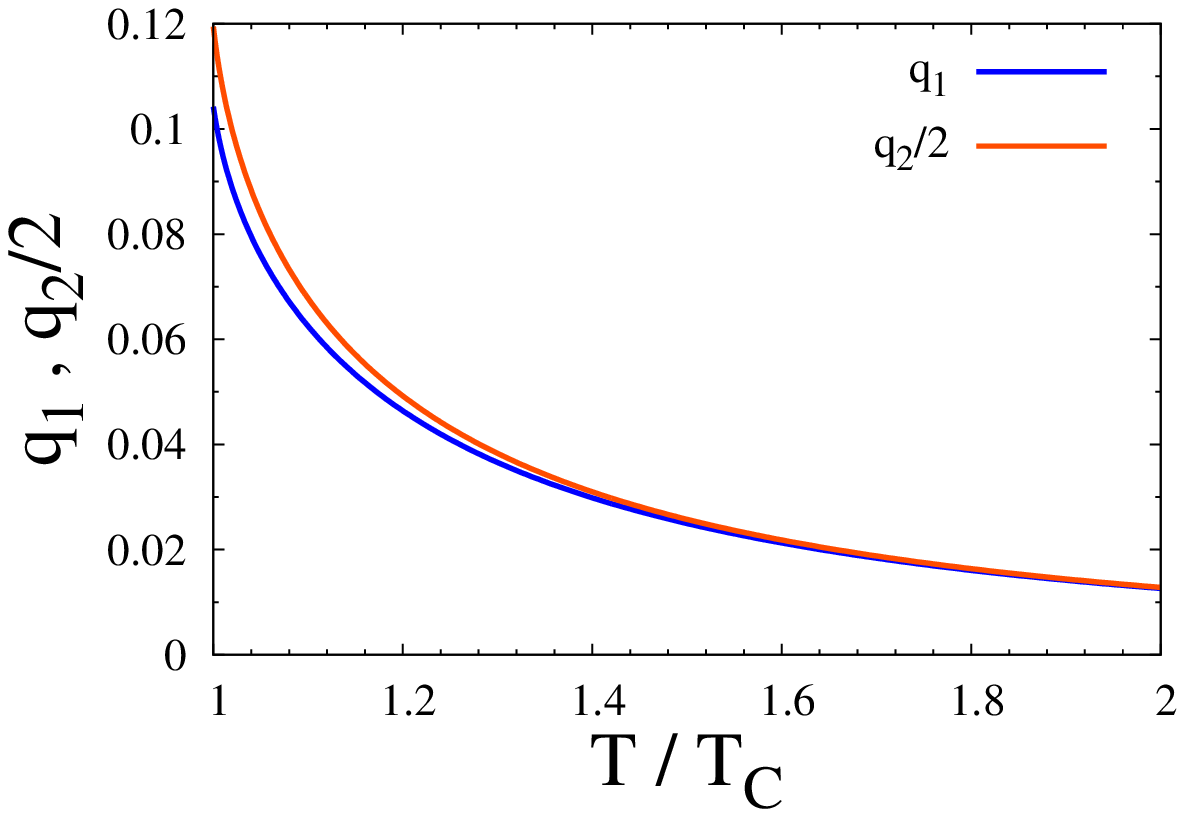}
\caption{\label{fig:pathsu5} The numerical solution for five colors,
with $c_2 = 0.31$.  The left panel is in the $q_1$-$q_2$ plane,
while the right panel gives $q_2/2$ and $q_1$. 
}
\end{figure}

\subsection{Six and seven colors}

In this case the space where the Polyakov loop is 
real is three dimensional.  

For six colors we find again that the exact solution is close to
that of the uniform eigenvalue ansatz.  It is also possible
to have intermediate phases in which there is a global $Z(2)$ or
$Z(3)$ symmetry.  We find no evidence of such partially deconfined
phases.

We comment that the symplectic group $Sp(6)$ is the pseudo-real part of
$SU(6)$; by exhanging the short and long roots of
$SO(7)$, this is th dual of
$Sp(6)$~\cite{Holland:2003kg}. As already remarked in section
(\ref{sec:effectivepotential}), to one loop order the perturbative
potentials ${\cal V}_{pert}({\bf q})$ 
are therefore related by duality.  If the non-perturbative potentials
are assumed to be dual as well, then the deconfining
phase transitions are the same \cite{KorthalsAltes2012}.
This can be tested through numerical simulations on the lattice.

In Fig. (\ref{fig:pathsu7}) we show the path for the exact solution.
In the uniform eigenvalue ansatz, $q_1/2 = q_2/3 = q_3/4$.  We find
that the deviation from this is small except close to $T_c$.
The results for seven colors are interesting because of their relevance for
the confining states in $SO(7)$ and $G(2)$ groups, discussed in Sec.
({\ref{sec:weyl}}).

\begin{figure}
\includegraphics[width=15cm]{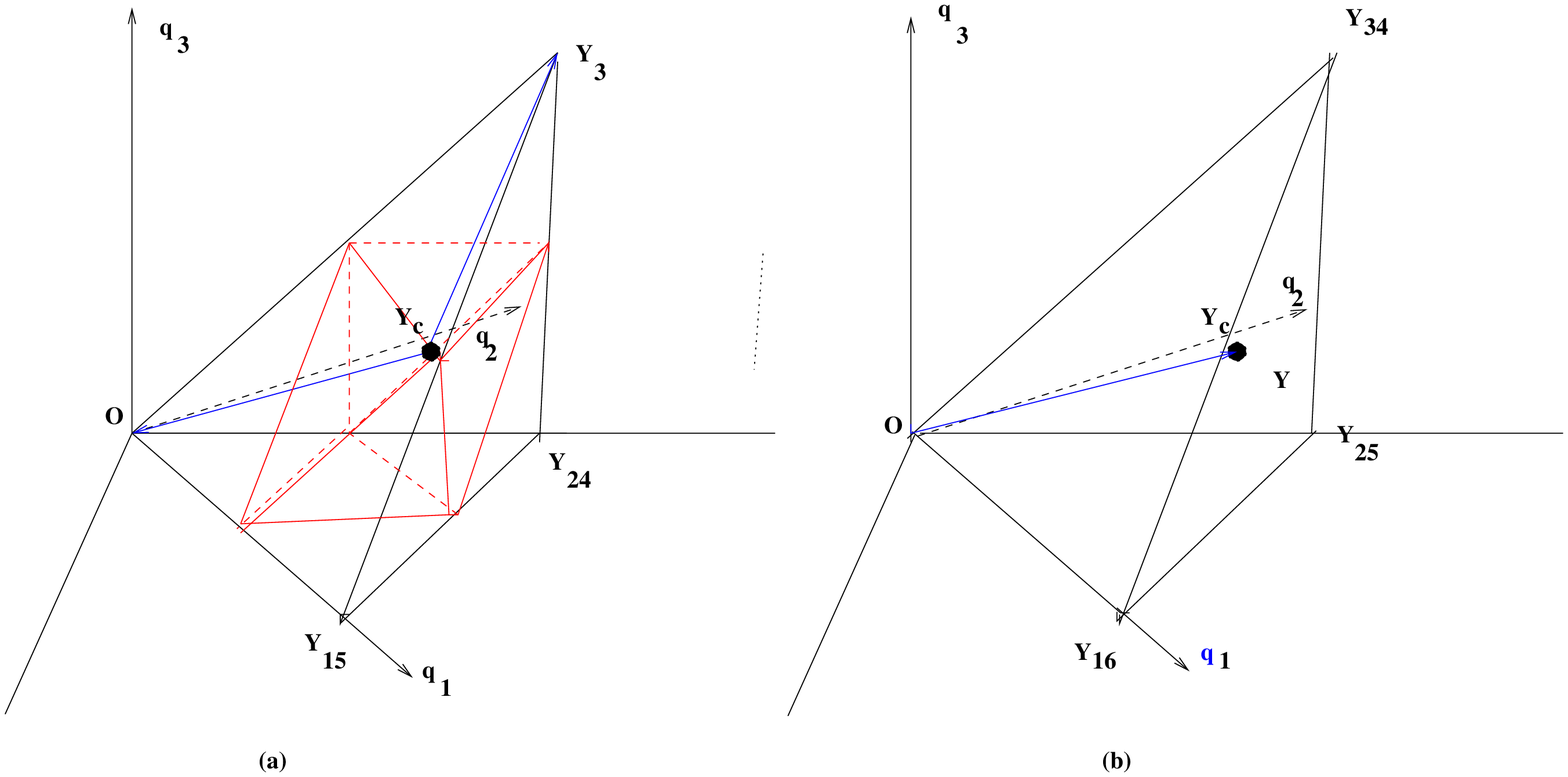}
\caption{\label{fig:su67realtrace} 
The three dimensional subspaces
where the Polyakov loop is real: left for six
colors, and right for seven.  
The planes where ${\rm tr} \, {\bf L}^n$ vanish are given in the left
panel.}
\end{figure}

\begin{figure}[htbp]
\includegraphics[width=0.48\textwidth]{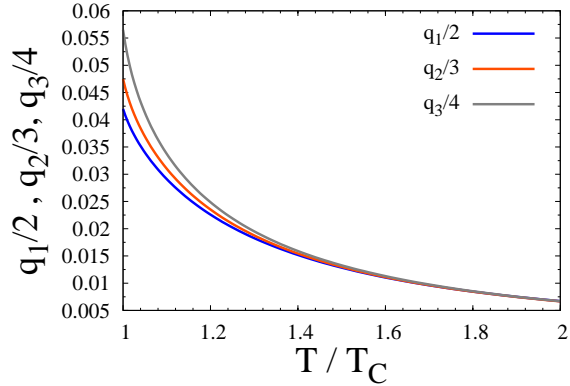}
\caption{\label{fig:pathsu7} The numerical solution for seven colors,
with $c_2 = 0.23$.  The deviation from $q_1/2 \neq q_2/3 \neq q_3/4$
indicates the deviation from the uniform eigenvalue ansatz.
}
\end{figure}

\section{Why the deconfining phase transition is of first order
for four or more colors}
\label{sec:firstorder}

In this section we discuss why a first order transition is expected
generically in matrix models when $N \geq 3$.

We first review the standard argument for why the transition is of
first order for three colors \cite{Svetitsky:1982gs}.  If $\ell$ is
the loop in the fundamental represetation, we consider a general
potential invariant under $Z(N)$ transformations,
\beq
\ell \rightarrow {\rm e}^{2 \pi i/N} \; \ell \; .
\eeq
The corresponding potential includes two terms.  First, there is
\beq
V_{O(2)}(\ell) \; = \; m^2 \; |\ell|^2 \; + \; \lambda_{O(2)} \; 
\left( |\ell|^2 \right)^2 \; ,
\label{O2}
\eeq
which is invariant under $O(2)$ transformations.  Second, there is
\beq
V_{Z(N)}(\ell) \; = \; \lambda_{Z(N)} \left( \ell^N \; + \;
\left( \ell^* \right)^N \right) \; .
\label{ZN}
\eeq
which is invariant only under $Z(N)$.

For three colors, Eq. (\ref{ZN}) is a cubic invariant, and is,
in the sense of the renormalization group, a relevant
operator.   By standard mean
field analysis, the transition is of first order.  We note that there is
a qualification: if the coupling $\lambda_{Z(N)}$ vanishes, then the
transition could be of second order (if $\lambda_{O(2)}> 0$).  There is
no symmetry reason why $\lambda_{Z(N)}$ should vanish, though, and so one
expects a first order transition.

For four colors, Eq. (\ref{ZN}) is a quartic term, and so marginal.  For
five or more colors, it is of pentic or higher order, and so an irrelevant
operator.  Then the deconfining transition is of second order when
$\lambda_{O(2)} > 0$, and of first order $\lambda_{O(2)} < 0$.  In the
latter case, a positive term $\sim (|\ell|^2)^3$ stabilizes the potential.

Why, then, is the deconfining transition of first order for any $N \geq 3$?
This follows in mean field theory from a matrix model.  From
Eq. (\ref{eq:wpot}) of Sec. (\ref{sec:effectivepotential}), 
in our particular model there is a cubic term in $r$ about
the confining vacuum, $r = 0$.  As discussed there, in mean field
theory this implies that the deconfining transition is of first order.

To see that this is not an accident, consider the general form of an
effective potential.  We express our matrix model in terms of the $q_i$,
and then $r$, but in general we can construct any function of the $q_i$
as a series in powers of ${\rm tr}\, {\bf L}$, ${\rm tr} \, {\bf L}^2$, and
so on, up to ${\rm tr}\, {\bf L}^{N-1}$.  Each term must be
invariant under $Z(N)$ transformations, so the simplest possible terms
include
\beq
|{\rm tr} \, {\bf L}|^2 \;\;\; ; \;\;\;
|{\rm tr} \, {\bf L}^2|^2 \;\;\; ; \;\;\;
|{\rm tr} \, {\bf L}^3|^2 \;\;\; ; \;\;\;
\left( {\rm tr} \, {\bf L} \right) 
\left( {\rm tr} \, {\bf L}^3  \right) +
{\rm c.c.} \;\; + \; \ldots\ \; .
\label{possible_matrix_terms}
\eeq
There is clearly an infinite series of such terms.  The first three terms
are invariant under $O(2)$; the last, under $Z(4)$.  
This multiplicity of terms is in contrast to a 
loop model, which only involves powers of the loop in the
fundamental representation, ${\rm tr} \, {\bf L}$.

Now we perform an elementary computation.
Take the basic ansatz for $A_0$, Eq. (\ref{fund_ansatz}).
We do not assume the uniform eigenvalue ansatz, Eq. (\ref{eq:ansatz1}),
but consider a general path, parametrized as
\beq
{\bf q} = (q_1, q_2, -q_2, -q_1) 
\;\;\; ; \;\;\;
q_1 = \frac{1}{8} (1 - \widetilde{x} + \widetilde{y}) 
\;\;\; ; \;\;\;
q_2 = \frac{1}{8} (3 - \widetilde{x}  - \widetilde{y}) \; .
\eeq
The confined vacuum is $\widetilde{x}  = \widetilde{y} = 0$.  
For the straight line path,
$\widetilde{x}  = 2 \widetilde{y}$, 
but this parametrization is convenient, so that the
expansion in Eq. (\ref{expand_matrix_terms}) has a simple form.

We can then easily compute the expansion of these quantities about
the confining vacuum.  We introduce $x = \pi \widetilde{x}$ and 
$y = \pi \widetilde{y}$,
$$
\left|\frac{1}{4} \; {\rm tr} \, {\bf L}\right|^2 \; = \;
\frac{1}{2} \, x^2 \; + \; \frac{1}{4}
x^2 \; y +  \ldots \; , 
$$
$$
\left|\frac{1}{4} \; {\rm tr} \, {\bf L}^2\right|^2 \; = \;
4 \, y^2 \; 
- \; x^2 \, y^2
- \frac{1}{3} \, y^4 + \ldots \; , 
$$
\beq
\left|\frac{1}{4} \; {\rm tr} \, {\bf L}^3 \right|^2 \; = \;
\frac{9}{2} \, x^2 
\; - \; \frac{27}{4} \, x^2 \, y 
+ \ldots \; .
\label{expand_matrix_terms}
\eeq
All of these loops vanish in the confined vacuum, $x=y=0$;
as the confined vacuum is the barycenter of the Weyl chamber,
the terms linear in $x$ and $y$ do as well, Sec. (\ref{sec:weyl_chamber}).
These loops then begin with terms quadratic in $x$ and $y$.
What is of relevance here is the existence of terms cubic in 
$x$ and $y$, which are always $\sim x^2 \, y$.

The general pattern is clear.  For most terms, such as
$|{\rm tr} \, {\bf L}|^2$ and $|{\rm tr} \, {\bf L}^3|^2$, 
{\it etc.}, have terms cubic in $x$
and $y$.  This is true for any term which involves 
${\rm tr} \, {\bf L}^n$, for odd $n$.  Thus, if there are {\it any} such
terms in the
effective potential, then 
in mean field theory the transition is of first order.
Note that having two fields doesn't alter the conclusion: the point is
that one cannot obtained a ``flat'' potential, 
typical of a second order transition.
The variables $x$ and $y$ are useful because then the quadratic
terms are diagonal.

Note that in a loop model, terms such as 
$|\ell|^2 \sim |{\rm tr}\, {\bf L}|^2$ 
are even in $\ell$, and do not give a first order transition.  A first
order transition only follows in a matrix model, where one expands 
about the barycenter of the Weyl chamber.

As illustrated by Eq. (\ref{expand_matrix_terms}), if a term
only involves ${\rm tr} \, {\bf L}^n$, for even $n$, then there is no
term cubic in $x$ and $y$, and the transition can be of second order.
This does not invalidate the expectation of a first
order transition.  If the effective Lagrangian involves
only even powers of ${\rm tr} \, {\bf L}^n$, then the global symmetry of
the theory is not $Z(4)$, but $Z(2)$.  The global symmetry is then not
like that expected for four colors, but for two colors, where 
the transition is of second order.
However, as for the analysis of three colors in a loop model, there is no
reason to expect such an accidental $Z(2)$ symmetry to occur.  
Indeed, for this to happen in a matrix model,
an infinite number of terms would have to vanish.

The above analysis can be generalized for any $N \geq 3$.  In general,
there are cubic terms in the expansion of $|{\rm tr} \, {\bf L}^n|^2$,
except for those $n$ and $N$ where there may be a residual $Z(n)$ symmetry
in $Z(N)$, as for $n=2$ and $N=4$.

Why the deconfining phase transition is of first order
for $N \geq 3$ can also be understood geometrically from 
the Weyl chamber for three and four colors in Fig. (\ref{fig:su34}).  

For three colors the Weyl
chamber is an equilateral triangle, bounded by the vectors
${\bf Y}_1$ and ${\bf Y}_2$.  The confined vacuum, ${\bf Y}_c$, is the
barycenter of this triangle.  The path from the perturbative vacuum, to
the confined vacuum, is a straight line, ${\bf O} {\bf Y}_c$.  

Consider expanding an effective potential about ${\bf Y}_c$, along
this straight line.  Even without computation, it is evident that the
region to the right is smaller than that to the left.  Thus if we
expand in $r$, where $r = 0$ is the confined vacuum, we would expect
terms of cubic order, and so a first order transition.

For four colors, instead of considering the full three dimensional space
of the Weyl chamber, we can limit ourselves to the plane where the Polyakov
loop is real, Fig. (\ref{fig:su4realtraceplane}).
Now the path from the perturbative vacuum to the
confined vacuum is not a straight line, ${\bf O} {\bf Y}_c$,
but slightly bent.  This is irrelevant, though: 
the point is that if we expand any effective potential about
${\bf Y}_c$, simply from the shape of the Weyl chamber, it is clearly
not symmetric in any manner.  Thus we expect the expansion
of an effective potential, about ${\bf Y}_c$, to expand the lack of
such symmetry.  This is the reason for the cubic terms in 
Eq. (\ref{expand_matrix_terms}).

The same argument applies for higher $N$.
When $N = 2M$, the part of the Weyl chamber in which the Polyakov loop
is real is $M$-dimensional, and so it becomes more difficult to draw.
Even so, the lack of symmetry about the confined vacuum for six colors
can also be seen from Fig. (\ref{fig:su67realtrace}).
\section{A $G(2)$ gauge group and the law of maximal eigenvalue repulsion}
\la{sec:weyl}

\subsection{Motivation}

For $SU(N)$ groups the Polyakov loop vanishes below $T_c$
and there is a strict definition of the
deconfining transition temperature.
That the Polyakov loop vanishes is a consequence of the center of $SU(N)$,
and the associated global $Z(N)$ symmetry.
The confined vacuum, ${\bf Y}_c$, is also the point where
there is maximal repulsion of the eigenvalues, Eq. (\ref{eq:pzero}).
We shall see in this section that it may be more useful to think
of confinement as arising not from the center
symmetry {\it per se}, but from eigenvalue repulsion.

To this end, we consider groups without a center.
Such groups have been considered before, starting with 
$SO(3)$ \cite{Langfeld:2000sc,deForcrand:2002vs,Greensite:2006sm}. 
The group $SO(3) = SU(2)/Z(2)$ and it has $SU(2)$ as a 
two-fold covering group.  
Consequently, the first homotopy group is
nontrivial, $\Pi_1(SO(3)) =Z(2)$.  
This is true for all $SO(2N+1)$ groups: they 
have a two-fold covering representation, $Spin(2N+1)$,
which has a center of $Z(2)$, and are doubly connected.

Like the odd dimensional rotation groups,
$G(2)$ has a trivial center.  Unlike $SO(2N+1)$ groups, though,
$G(2)$ is simply connected.  For this reason, it 
is especially interesting to consider
~\cite{Langfeld:2000sc,deForcrand:2002vs,Holland:2002vk,Cossu:2007dk,Pepe:2006er,Holland:2003kg,Greensite:2006sm}. 
In fact, $G(2)$ is a subgroup of $SO(7)$.
This means that the eight dimensional spin representation of  
$SO(7)$ has no subgroup which corresponds to a two-fold covering of $G(2)$. 

A feature common between $SO(3)$ and $G(2)$ is that Polyakov loops in
the fundamental representation can be screened dynamically.
In $SO(3)$, two fundamentals screen each other,
$
{\bf 3}\otimes{\bf 3}={\bf 1}+{\bf 3}+{\bf 5}
$.
In $G(2)$, a fundamental field, in the ${\bf 7}$ representation,
can be screened by three adjoint fields, in the ${\bf 14}$:
$
{\bf14}\otimes {\bf14}\otimes{\bf 14}={\bf 1}+{\bf 7}+....,
$.

For both $SO(3)$ and $G(2)$, then,
the Polyakov loop in the fundamental representation
is always screened. This is unlike $SU(N)$, where the fundamental field
cannot be screened by the $Z(N)$ symmetry.  It appears similar to
$QCD$ with dynamical, light quarks.
Consequently, we might expect there is no deconfining
phase transition, but perhaps just a cross-over.

Nevertheless, lattice simulations for the $G(2)$ group
find that there appears to be a first order transition
\cite{Holland:2002vk, Pepe:2006er, Cossu:2007dk, HoyosBadajoz:2007ds, Wellegehausen:2009rq, *Wellegehausen:2010ai, *Wellegehausen:2011sc,Caselle2012}.
Even more strikingly, the expectation value of the Polyakov 
loop in the fundamental representation appears to be {\it very} small
below the temperature for this first order transition.  
Thus we are led to consider how to construct effective potentials which give 
confinement, and a deconfining phase transition, in the
absence of any center symmetry.

The simplest approach was suggested following
Eq. (\ref{eq:generalformpot}).  
For $SU(N)$, the perturbative potential is a power series in the adjoint
loop.  The $Z(N)$ symmetry allows arbitrary powers of
the $Z(N)$ neutral adjoint loop to appear,
but forbids terms linear in the fundamental loop.  Since $G(2)$
has no center symmetry, though, in the nonperturbative potential
there is nothing to forbid us from adding a term a term where in Eq. 
(\ref{eq:generalformpot}) we sum over powers of the {\it fundamental}
loop, the ${\bf 7}$.  If the coefficient of such a 
term is large and of the right sign, one can drive the theory to
the confined phase at low temperatures.

Another approach is to observe that 
$
SU(7)\supset SO(7) \supset G(2) 
$.
This implies that we can construct a non-perturbative potential 
from both $SU(7)$  and $G(2)$ potentials.
Of course we must restrict the $SU(7)$ 
potentials to the two dimensional Cartan space of $G(2)$.
An easy exercise shows the confining vacuum for $SU(7)$, ${\bf Y}_c(7)$,
lies lies in the Cartan space of $G(2)$.  Thus by 
carefully adjusting the parameters of the
potential, we can ensure that the 
system lies in the $SU(7)$ confining vacuum below $T_c$.
Confinement in $SU(7)$ then forces the ${\bf 7}$ loop
in $G(2)$ to vanish as well.

In both cases, even without center symmetry we are adding terms
to the nonperturbative potential which generate
an expectation value for the fundamental loop which is either small or
vanishing in the low temperature phase.
Confinement is thus driven not by the center symmetry, but through
the complete repulsion of eigenvalues.  

The crucial question is whether the Weyl invariance of $G(2)$ is
respected by such $SU(7)$ potentials.
The following simple argument is suggestive. 
The Weyl group of $SU(7)$ is the permutation group $S_7$. 
The Weyl group of $SO(7)$  is $O_h$, the group of 
rotations and reflections that leave the 
three dimensional cube invariant; it is of order $48$. 
The Weyl group of $G(2)$ is $D_6$, the dihedral group of order $12$. 
Now the latter two can be written as semi-direct products of 
$S_2$ with $S_4$, respectively $S_3$.  
This shows they are subgroups of $S_7$.  
So the answer to the question is yes, if one neglects 
the fact that $S_7$ acts in the six dimensional Cartan space of $SU(7)$, 
whereas $D_6$ acts in the two dimensional Cartan space of $G(2)$.
  
In Sec. (\ref{sec:rootso7g2}) we
discuss the three dimensional Cartan space of $SO(7)$, 
the embedded Cartan space of $G(2)$, and their roots.
Sec. (\ref{sec:su7type}) constructs confining potentials
for both $G(2)$ and $SO(7)$.
In Sec. (\ref{sec:weylsymsu7}) we give a simple proof that the
$SU(7)$ type potential respects the necessary Weyl symmetry.
Finally, in Sec. (\ref{sec:g2plots}) we give results for the possible
thermodynamic behavior of $G(2)$
on the basis of some representative models.

\subsection{The root systems of $SU(7), SO(7) ~\mbox{and}~ G(2)$}
\la{sec:rootso7g2}

We start with some  basic facts and notation. 
The Polyakov loop for $SU(7)$ is
$
{\bf L}=\exp(i2\pi{\bf q}_{SU7})
$
,
where
\beq
{\bf q}_{SU7}=\mbox{diag}(q_1, q_2, q_3, q_4, q_5, q_6, q_7) \; .
\eeq
As an element of $SU(7)$, this is a tracless matrix, 
\beq
q_1+\cdots+q_7=0 \; ,
\la{eq:su7constraint}
\eeq
and defines the six dimensional Cartan space of $SU(7)$.

$SU(7)$ has the subgroup $SO(7)$.  The Cartan space of $SO(7)$ is
three dimensional, where the $q$'s obey three extra constraints,
\beq
q_7=q_1+q_4=q_2+q_5=0.
\la{eq:so7constraint}
\eeq
Note that these constraints are identical to those 
which ensure that the trace of the fundamental
loop in $SU(7)$ is real, as illustrated in
Fig. (\ref{fig:su67realtrace}).

$G(2)$ is a subgroup of $SO(7)$.  Its Cartan subspace obeys one more 
constraint:
\beq
q_1+q_2+q_3=0,
\la{eq:g2constraint}
\eeq
We write the corresponding matrix as ${\bf q}_{G2}$,
and has two degrees of freedom, appropriate to the Cartan space of $G(2)$.

\begin{figure}
\includegraphics{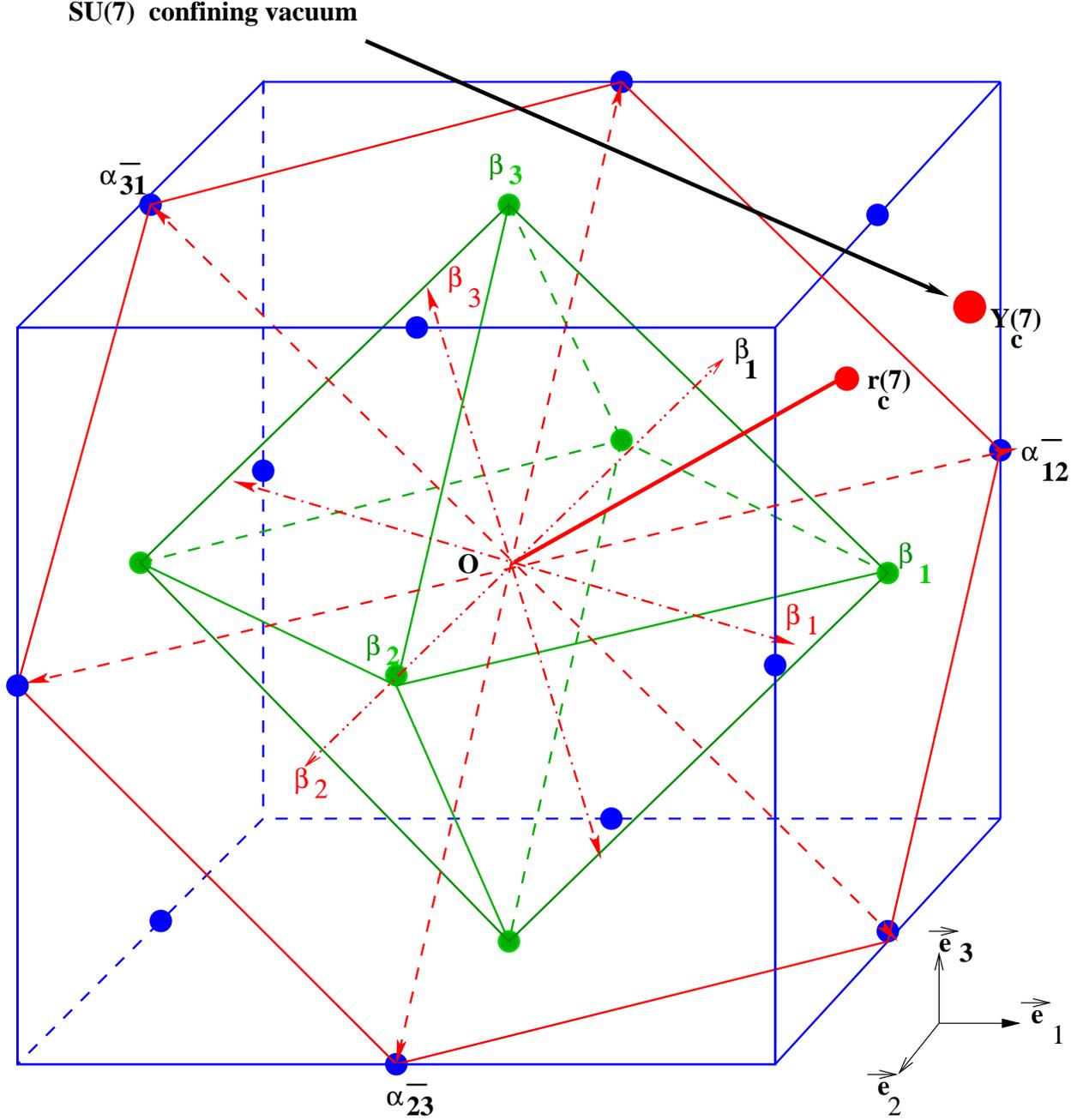}
\vspace*{-0.7cm}
\caption{The root system of $SO(7)$ and its $G(2)$ subgroup, see text.
The confining $SU(7)$ vacuum $\Y_c(7)$ 
is also shown. }\label{fig:so7g2project} 
\end{figure}

In Fig. (\ref{fig:so7g2project}) we show the three dimensional
root system of $SO(7)$,
and its $G(2)$ subgroup.  Here the
basis vector corresponding to $(q_1,q_2,q_3)$ are
$\vec e_{1,2,3}$.  This figure also illustrated the restriction to
the $G(2)$ plane, where $q_1 + q_2 + q_3 = 0$.
From this figure, and standard group theory \cite{georgi}, the roots
are given by
\beq
\vec\a^{\pm}_{ij}=\vec e_i\pm\vec e_j, \; \; \; ; \;\;\;
\vec\b_i =\vec e_i, \;\;\; ; \;\;\; i: 1 \ldots 3 \; ,
\la{eq:weylso2nrootprojq}
\eeq
together with the six roots with opposite sign.

Clearly the twelve 
roots $\vec\a^{\pm}$ lie on the edges of the cube, 
and the six roots $\vec\b$ on the vertices of the octahedron.
The Weyl group of this system is $O_h$, 
the group of rotations and reflections that leave the cube 
and the octahedron invariant.
Its order is $48$.  The group of rotations is identical 
to the permutation group $S_4$ of the four diagonal body axes of the cube.

The plane defined by  $q_1+q_2+q_3=0$ contains the six $SU(3)$ like  roots 
$\vec\a^{\, -}_{ij}$, to wit:
\beq
\vec\a^{\, -}_{12}=\vec e_1-\vec e_2 \;\;\; ; \;\;\;
\vec\a^{\, -}_{23}=\vec e_2-\vec e_3 \;\;\; ; \;\;\;
\vec\a^{\, -}_{31}=\vec e_3-\vec e_1 \; .
\la{eq:g2aroots}
\eeq
These are part of the $SO(7)$ root system.  
Then there are the six orthogonal projections of the 
short roots $\vec\b_i$ in $SO(7)$ onto the plane $q_1+q_2+q_3=0$. 
The resulting  projections  are denoted by $\b_i$ in 
Fig. (\ref{fig:so7g2project}).
Below we write them as $\vec{\hat{\b}}_i$
to avoid confusion with the corresponding $SO(7)$ roots $\vec\b_i=\vec e_i$:
\beq
\vec{\hat{\b}}_1={1\over{3\sqrt{2}}}(2\vec e_1-\vec e_2-\vec e_3) \; ; \;
\vec{\hat{\b}}_2={1\over{3\sqrt{2}}}(-\vec e_1 + 2\vec e_2-\vec e_3) \; ; \;
\vec{\hat{\b}}_2={1\over{3\sqrt{2}}}(-\vec e_1-\vec e_2+2\vec e_3) \; .
\la{eq:g2broots}
\eeq
The $\vec{\hat{\b}}$ roots are indeed $1/\sqrt{3}$ shorter 
then the $SU(3)$ like roots $\vec\a^-$, as behooves the 
root system of $G(2)$.  Note that the $SU(3)$ hypercharge matrices  
$\Y_1\sim\mbox{diag}(1,1,-2)$ and their
 permutations are generating the $\hat\b$ roots, 
in the same way as the $SU(2)$ matrices 
$\mbox{diag}(1,-1)$ generate the $\vec\a^-$ roots. 
 
As in Eq. (\ref{eq:sunargument}) we define
\beq
{\bf q}_{G2} =\vec{\q}_{G2} \cdot \vec{H} \; .
\la{eq:1psi}
\eeq
In the perturbative effective potential, Eq. (\ref{eq:potroot}),
the short and long roots of $G(2)$ appear
as  $\vec{\hat{\b}}_i\cdot \vec \q_{G2}$ and $\vec\a^{-}_{ij}\cdot \vec \q_{G2}$.
These arguments are easily found from 
Eqs. (\ref{eq:g2aroots}) and (\ref{eq:g2broots}). 
 They are, with the condition $q_1+q_2+q_3=0$:
\beq
\vec\a^-_{12}\cdot \vec{\q}_{G2}=q_1-q_2 \;\;\; ; \;\;\;
 \vec\a^-_{23}\cdot \vec{\q}_{G2}=q_2-q_3 \;\;\; ; \;\;\;
 \vec\a^-_{31}\cdot \vec{\q}_{G2}=q_3-q_1\; .
\eeq
\beq
\vec{\hat{\b}}_1\cdot \vec{\q}_{G2}=q_1 \;\;\; ; \; \;\;
\vec{\hat{\b}}_2\cdot \vec{\q}_{G2}=q_2 \;\;\; ; \;\;\;
\vec{\hat{\b}}_3\cdot \vec{\q}_{G2}=q_3 \; .
\la{eq:g2rootsprojq}
\eeq
 
The root lattice of $G(2)$, and its Weyl chamber, are illustrated in
Fig. (\ref{fig:g2lattice}).  The left panel is the root lattice,
with six long and six short roots.  
The Weyl group is generated by the two mirrors indicated
in the left panel, at an angle of $2\pi/12$. 
The product of the two reflections is a rotation over $2\pi/6$.  
The group is generated by this six fold rotation and by 
one of the reflections. 
This gives the dihedral group of order $12$, and so twelve Weyl chambers. 
These are shown in the right hand panel. 
We pick the Weyl chamber as
defined by the points ${\bf O}$, and 
$(1,-2,1)/3$ and $(1,-1,0)/2$. This is the upper half
of the Weyl chamber of $SU(3)$ in fig. (\ref{fig:su34}). 
We also indicate the confining vacuum for $SU(3)$, 
${\bf Y}_c(3) = (1,-1,0)/3$; the confining vacuum for $SU(7)$,
$\Y_c(7)=(2,-3,1)/7$, and by a black curve, the path where
trace of the fundamental loop in $G(2)$ vanishes.
The Weyl chamber for $G(2)$ is precisely half that of $SU(3)$, 
Fig. (\ref{fig:su34}).

\begin{figure}
\includegraphics[width=15cm]{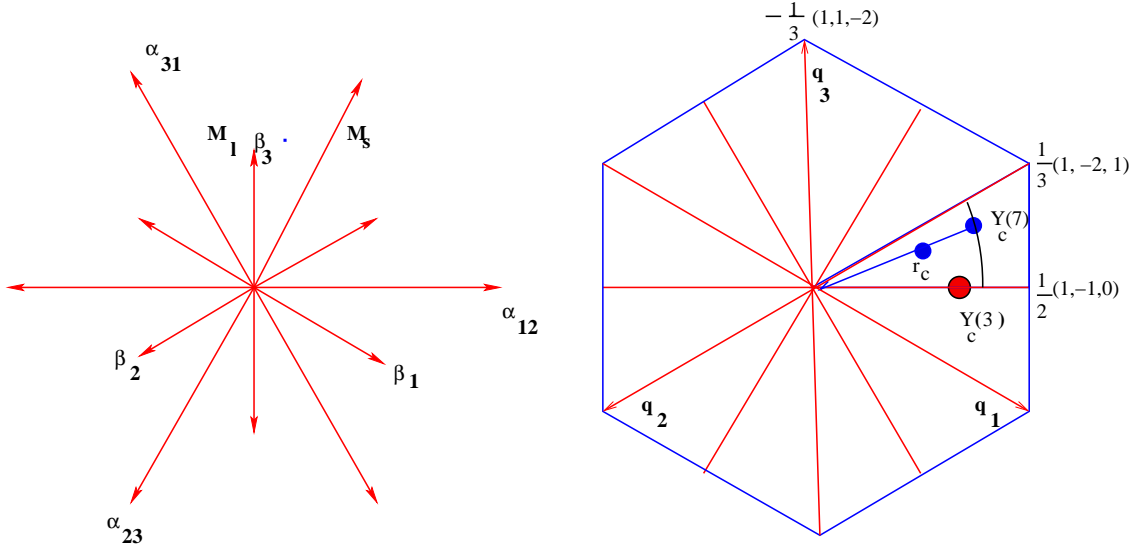}
\vspace*{-0.61cm}
\caption{ Left: the root lattice of $G(2)$.
Right: the twelve Weyl chambers of $G(2)$.
${\bf Y}_c(3)$ is the $SU(3)$ confined vacuum; ${\bf Y}_c(7)$, 
that for $SU(7)$.
}
\label{fig:g2lattice} 
\end{figure}

\subsection{Confining potentials for $G(2)$ and $SO(7)$}
\label{sec:su7type}

We drop an overall factor of $T^4$ from the potential, and
split it into perturbative and nonperturbative parts,
\beq
V_{tot}({\bf q}_{G2})=
V^{G(2)}_{pt}({\bf q}_{G2})
+V_{np}({\bf q}_{G2}) \; .
\la{eq:g2su7tot}
\eeq
As discussed in Sec. (\ref{sec:effectivepotential}) 
the perturbative $G(2)$ potential is
\beq
V^{G(2)}_{pt}({\bf q}_{G2})
= - \; {14\over{45}}\; \pi^2
+{4\pi^2\over{3t^2}}V_2^{G(2)}({\bf q}_{G2}) \; ,
\eeq
where $t = T/T_c$, and
\beq
2V_k^{G(2)}({\bf q}_{G2})=
\sum_{\a}B_{2k}(\vec \a\cdot \vec \q_{G2})+
\sum_{\b}B_{2k}(\vec\b\cdot \vec \q_{G2}),~k=1, 2 \; ,
\la{eq:g2vk}
\eeq
with the roots $\a$ and $\b$ those of $G(2)$.

We now consider the types of nonperturbative potentials which can
produce a confined phase at low temperature.
  
The first method is to take a sum as in 
Eq. (\ref{eq:generalformpot}), but sum over powers of the Wilson line in the
fundamental representation, the ${\bf 7}$.
The weights of the ${\bf 7}$ are precisely the six 
$\pm\vec{\hat{\b}}_i, ~i=1, 2, 3$ from
Eq. (\ref{eq:g2rootsprojq}).  This gives the potential
\beq
V^{{\bf 7}}_2({\bf q}_{G2})
= B_4(q_1)+B_4(q_2)+B_4(q_3) \; .
\la{eq:addingvector}
\eeq

The second method is to add potentials from $SU(7)$.
We {\it construct} the nonperturbative potential to
eliminate the contribution of the perturbative $G(2)$ term at the
transition temperature, $t = 1$.
\bea
V_{np}({\bf q}_{G2})=
&-&{4\pi^2\over {3t^2}}\left(
c_1^{G(2)} V_1^{G(2)}({\bf q}_{G2})
+ c_1^{SU(7)} V_1^{SU(7)}({\bf q}_{G2}) \right.
\nonumber\\
&+& \left. c_2^{G(2)} V_2^{G(2)}({\bf q}_{G2})
+c_2^{SU(7)} V_2^{SU(7)}({\bf q}_{G2})+c_3 \right).
\la{eq:g2su7nonpertA}
\eea
The potentials $V_k^{SU(7)}$ 
are defined by summing the corresponding 
Bernoulli polynomials $B_{2k}$ over 
the roots $\vec\a^{-}_{ij},~1\le i < j\le 7$  of $SU(7)$:
\beq
V_k^{SU(7)}({\bf q}_{G2})
=\sum_{1\le i < j\le 7}B_{2k}(\vec\a^{\, -}_{ij} \cdot \vec{\q}_{G2}).
\la{eq:vksu7g2}
\eeq

By appropriately adjusting the coefficients in the nonperturbative
potential, we show that we can ensure that at the critical
temperature, the system goes into the $SU(7)$ confining vacuum $Y_c(7)$.
We stress there is no elegance in our approach: we are manifestly
constructing a confined vacuum by hand.  Nevertheless,
our model gives testable predictions, as shown in
Sec. (\ref{sec:g2plots}).

Along these same lines we can also construct a
potential which generates confinement in $SO(7)$:
the perturbative potential is that for $SO(7)$, 
while in the $SU(7)$ like potential, Eq. (\ref{eq:vksu7g2}),
we change the argument to ${\bf q}_{SO7}$.  Doing this gives:
\beq
V_{tot}({\bf q}_{SO(7)})=V^{SO(7)}_{pt}({\bf q}_{SO(7)})
+V_{np}({\bf q}_{SO(7)}) \; ,
\la{eq:so7su7tot}
\eeq
where
\beq
V^{SO(7)}_{pt}({\bf q}_{SO(7)})
=- \; {21\over{45}}\, \pi^2
+{4\pi^2\over{3t^2}} \; V_2^{SO(7)}({\bf q}_{SO7}) \; ,
\eeq
and
\beq
2V_k^{SO(7)}({\bf q}_{SO(7)})=\sum_{\a}B_{2k}(\vec \a \cdot \vec \q_{SO7})
+\sum_{\b}B_{2k}(\vec\b \cdot \vec \q_{SO7}),~k=1, 2,
\la{eq:so7vk}
\eeq
with the roots $\a$ and $\b$ those of $SO(7)$.
The $SU(7)$ like potentials are
\beq
V_k^{SU(7)}({\bf q}_{SO(7)})
=\sum_{1\le i < j\le 7}B_{2k}(\vec\a^{\, -}_{ij} \cdot \vec{\q}_{SO7}).
\la{eq:vksu7so7}
\eeq

As discussed following
Eq. (\ref{eq:sympllongrootproj}), perturbatively the
${\bf q}$-potentials for $SO(7)$ and $Sp(6)$ are related by
duality.  If one generates maximal eigenvalue repulsion for $Sp(6)$, however,
one obtains confinement appropriate to $SU(6)$, while 
Eq. (\ref{eq:vksu7so7}) for $SO(7)$ gives confinement like
that of $SU(7)$.  If the law of maximal eigenvalue repulsion holds,
then, the nonperturbative potentials are not related by duality
\cite{KorthalsAltes2012}.

\subsection{Weyl symmetry of $SU(7)$ like potentials}
\label{sec:weylsymsu7}

In this subsection we show that the $SU(7)$
nonperturbative potentials introduced above
are invariant under the Weyl symmetries of $G(2)$ and $SO(7)$.
First observe the only terms which are not manifestly Weyl invariant 
are those in Eqs. (\ref{eq:vksu7g2}) and (\ref{eq:vksu7so7}).

In the case of $G(2)$ we use 
Eq. (\ref{eq:sunargument}) to write the potential as:
$$
V^{SU(7)}_k({\bf q}_{G2})=\frac{1}{2} \sum_{1\le i,j\le N}B_2(q_i-q_j)
= 2\left(B_{2k}(q_1-q_2)+B_{2k}(q_2-q_3)+B_{2k}(q_3-q_1)\right)
$$
\beq
+4\left(B_{2k}(q_1)+B_{2k}(q_2)+B_{2k}(q_3)\right)
+B_{2k}(2q_1)+B_{2k}(2q_2)+B_{2k}(2q_3) \; ,
\la{eq:g2su7vk}
\eeq
with $q_1+q_2+q_3=0$.

Now use Eq. (\ref{eq:g2rootsprojq}) 
to rewrite this in terms of the roots of $G(2)$ roots.  This gives:
\beq
2V^{SU(7)}_k({\bf q}_{G2})=4\sum_\a B_{2k}(\vec\a \cdot \vec \q_{G2})
+\sum_\b \left( 2 B_{2k}(\vec\b \cdot \vec \q_{G2})
+ B_{2k}(2\vec\b \cdot\vec \q_{G2}) \right)\; ,
\la{eq:g2weylinv}
\eeq
which is manifestly invariant under the Weyl symmetry of $G(2)$.

Applying the same method to $SO(7)$ gives
\beq
V_k^{SU(7)}({\bf q}_{SO(7)})=
B_{2k}(2q_1)+B_{2k}(2q_2)+B_{2k}(2q_3)
+ 2\left(B_{2k}(q_1)+B_{2k}(q_2)+B_{2k}(q_3)\right)
\la{eq:vkso7deformation}
\eeq
$$
+ 2\left(B_{2k}(q_1-q_2)+B_{2k}(q_2-q_3)+B_{2k}(q_3-q_1) 
+B_{2k}(q_1+q_2)+B_{2k}(q_2+q_3)+B_{2k}(q_3+q_1)\right) .
$$
Note that the constraint $q_1+q_2+q_3=0$ is absent for $SO(7)$. 
Imposing it gives the $G(2)$ result, Eq. (\ref{eq:g2su7vk}).
Using 
Eqs. (\ref{eq:so2nrootprojq}) and (\ref{eq:so2nplus1shortrootproj}) 
this can be rewritten as
\beq
2V_k^{SU(7)}({\bf q}_{SO(7)})=
2\sum_{\a}B_{2k}(\vec\a\cdot \vec \q_{SO7})
+\sum_{\b} \left ( 2B_{2k}(\vec\b\cdot \vec \q_{SO7})
+B_{2k}(2\vec\b\cdot \vec \q_{SO7}) \right)\; .
\eeq
This result is manifestly Weyl invariant as the sums are separately 
invariant: the Weyl transformations
are orthogonal, and so cannot transform $\a$ roots into $\b$ roots.

\subsection{Results for $G(2)$ }
\label{sec:g2plots}

In this section, we give results for the thermodynamics, assuming
parameters which give a confined low temperature phase. 
The total potential is
\beq
{\cal V}_{tot}^{\Gtwo}({\bf q}_{G2}) = T^4 \left(
\Vpt^{\Gtwo}({\bf q}_{G2}) \; + \; \Vnpt({\bf q}_{G2},t) \right)  \; ,
    \label{eq:Vpt_G2}
\eeq
where the perturbative potential is
\beq
\Vpt^{\Gtwo}({\bf q}_{G2}) = - \; \frac{14 \pi^2}{45} \; + \;
     \frac{4 \pi^2}{3} \; V_2^{\Gtwo}({\bf q}_{G2}) \; .
\eeq
For clarity, we write all functions in terms of
the two independent $q$'s for $G(2)$, $q_1$ and $q_2$.
The $G(2)$ potentials are
\bea
V_{n}^{\Gtwo}({\bf q}_{G2}) &=& B_{2 n}(q_1) + B_{2 n}(q_2) \nonumber \\
&+& B_{2 n}(q_1+q_2)
    + B_{2 n}(q_1-q_2) + B_{2 n}(2q_1+q_2) + B_{2 n}(q_1+2q_2)\; .
 \label{eq:V2G2}
\eea

We consider a nonperturbative potential
\bea
V_{np}({\bf q}_{G2})=
-{4\pi^2\over {3t^2}}&\left( \right. & 
c_1^{G(2)} \; V_1^{G(2)}({\bf q}_{G2})
+ c_1^{SU(7)} \; V_1^{SU(7)}({\bf q}_{G2}) 
\nonumber\\
&+& \left. c_2^{G(2)} \; V_2^{G(2)}({\bf q}_{G2})
+ c_2^{SU(7)} \; V_2^{SU(7)}({\bf q}_{G2})
+ d_2^{\, G(2)} \; V^{{\bf 7}}_2({\bf q}_{G2}) +c_3 \right).
\la{eq:g2su7nonpertB}
\eea
The $SU(7)$ potentials are
\bea
V^{SU(7)}_{n}(q_1,q_2) &=&
B_{2n}(2 q_1) + B_{2n}(2 q_2) + B_{2 n}(2 q_1 + 2 q_2) \nonumber\\
&+& 2 \left( B_{2 n}(q_1 - q_2) + \, B_{2 n}(2 q_1 + q_2) 
+ \, B_{2 n}(q_1 + 2 q_2) \right) \nonumber\\
&+& 4 \left( B_{2 n}(q_1) + \, B_{2 n}(q_2) +  \, B_{2 n}(q_1 + q_2) 
\right)
\; ,
\label{SU7nonpert_2}
\eea
while the potential from summing over Eq. (\ref{eq:generalformpot})
using the fundamental representation, the ${\bf 7}$, is
\beq
V^{{\bf 7}}_2({\bf q}_{G2})
= B_4(q_1)+B_4(q_2)+B_4(q_1+q_2) \; .
\la{eq:addingvector2}
\eeq

Besides thermodynamic quantities, such as the pressure and the
interaction measure, it is also possible to measure loops.
In $G(2)$, the fundamental representation is related to the $SU(3)$
embedding as
${\bf 7} = {\bf 1} + {\bf 3} + \overline{{\bf 3}}$
\cite{Holland:2002vk}, so that
the fundamental loop is given by
\cite{Wellegehausen:2009rq, *Wellegehausen:2010ai, *Wellegehausen:2011sc}
\beq
{\ell}_{\bf{7}} = \frac{1}{7}
\left( 1 + 2 \, \cos(2 \pi q_1) + 2 \, \cos(2 \pi q_1) 
+ 2 \, \cos(2 \pi (q_1+q_2) ) \right) \; .
\eeq
The adjoint representation is related to the $SU(3)$ embedding as
${\bf 14} = {\bf 3} + \overline{{\bf 3}} + {\bf 8}$, so that 
\cite{Wellegehausen:2009rq, *Wellegehausen:2010ai, *Wellegehausen:2011sc}
\bea
{\ell}_{\bf{14}} = \frac{1}{7}
\left( 1 \right. &+& 
\cos(2 \pi q_1) + \cos(2 \pi q_1) + \cos(2 \pi (q_1+q_2) )\nonumber\\
&+& \left.\cos(2 \pi (q_1 - q_2)) + \cos(2 \pi (2 q_1 +  q_2) ) 
+ \cos(2 \pi (q_1+ 2 q_2) \right) \; .
\eea

In Eq. (\ref{eq:g2su7nonpertB}) we have a model with six parameters.
Even imposing two conditions --- that the transition occur at $T_c$,
and that the pressure vanishes there --- we are left with four free
parameters.  

Instead of investigating the entire four dimensional space, we consider
some representative models.  

The first is a minimal $G(2)$ model,
\beq
c_1^{G(2)} \, , \, c_3 \neq 0 \;\;\; ; \;\;\;
c_2^{G(2)} = c_1^{SU(7)} = c_2^{SU(7)} = d_2^{\, G(2)} =  0 \; .
\label{minimal_G2_model}
\eeq
This is the $G(2)$ analogy 
of the analogue of 
zero parameter $SU(N)$ model of Ref. \cite{Meisinger:2001cq}.
We introduce
terms $\sim c_1^{G(2)} B_2({\bf q}_{G2})$ 
to drive the theory to a Higgs phase.   
We could also introduce a term $\sim c_2^{G(2)} B_4({\bf q}_{G2})$,
which would be like our one parameter $SU(N)$ model
\cite{Dumitru:2010mj}.  We have done
so, and find that the results are similar 
to the minimal model of Eq. (\ref{minimal_G2_model}).

The next is a model with a single fundamental loop,
\beq
d_2^{\, G(2)} \, , \, c_1^{G(2)} \, , \,  c_2^{G(2)} \, , \, 
c_3 \neq 0 \;\;\; ; \;\;\;
c_1^{SU(7)} = c_2^{SU(7)} =  0 \; .
\label{loop7_model}
\eeq
The specific parameters chosen were
$d_2^{\, G(2)} =-0.210$ and $c_2^{G(2)}= 0.3$;
the values of $c_1^{G(2)}$ and $c_3$ follow as for $SU(N)$,
Sec. (\ref{sec:thermo_model}), and are
$c_1^{G(2)} = 0.278$ and $c_3 = -0.364$.

Finally, we consider a $SU(7)$ type model, 
\beq
c_2^{G(2)} = 1 \;\;\; ; \;\;\;
c_1^{SU(7)} \; , \; c_2^{SU(7)} \; , \; 
c_3 \neq 0 \;\;\; ; \;\;\;
c_1^{G(2)} = d_2^{\, G(2)} =  0 \; .
\label{SU7_model}
\eeq
Notice that we have fixed the parameter $c_2^{G(2)}=1$; with this
value, the $G(2)$ part of the nonperturbative potential
cancels, {\it identically}, the perturbative $G(2)$ potential at $T_c$.
This ensures that the confining effects of the $SU(7)$ potential
are maximized at $T_c$.  We then considered two 
representative values of $c_2^{SU(7)}$:
$c_2^{SU(7)} = 2.0$, for which 
$c_1^{SU(7)}= 0.623$, and $c_3 = -1.093$;
and $c_2^{SU(7)} =  4.0$, for which
$c_1^{SU(7)}= 1.246$ and $c_3= -1.952$.

In Fig. (\ref{fig:G2Loopsfund}) 
show the expectation value of the Polyakov loops in the fundamental
representations.  At present, there is
only data on histograms for the expectation
value for the bare, fundamental loop
\cite{Holland:2002vk, Pepe:2006er, Cossu:2007dk, HoyosBadajoz:2007ds, Wellegehausen:2009rq, *Wellegehausen:2010ai, *Wellegehausen:2011sc, Caselle2012}.
This shows the expectation value of the bare loop is very small in the
low temperature phase.  Including renormalization should not alter this
conclusion.  Presumably, more careful lattice studies will show that
the expectation value is small, but nonzero, as true for the adjoint
loop in $SU(3)$ in the confined phase \cite{Gupta:2007ax}.  

Nevertheless, the minimal $G(2)$ model appears to be excluded. It is
negative at $T_c^-$, with a value $\sim - 0.2$.  
In Fig. (\ref{fig:weylchamberg2}) we illustrate the Weyl chamber for
$G(2)$, indicating both the vacua at $T_c^+$ and $T_c^-$.  The
low temperature phase at $T_c^-$ is close, but not coincident with
${\bf Y}_c(7)$, the confining vacuum for $SU(7)$.  

The other models give fundamental loops which are small in the low
temperature phase.  In the fundamental
loop model, this expectation value is $\sim + 0.05$ at $T_c^-$,
which may be compatible with the lattice results.  Lastly,
the $SU(7)$ models automatically give zero fundamental loop below $T_c$.

In Fig. (\ref{fig:G2Loopsadj}) we
give the expectation value of the adjoint loop.
This also provides 
a way of distinguishing between different models.  On the lattice,
the bare adjoint loop is strongly suppressed, but with effort can
be measured. In the minimal $G(2)$ model, it is positive at $T_c^-$,
but then becomes negative.  It is negative in the low temperature
phase for the fundamental loop model, and essentially zero in the 
$SU(7)$ model.

\begin{figure}[htbp]
\includegraphics[width=0.8\textwidth]{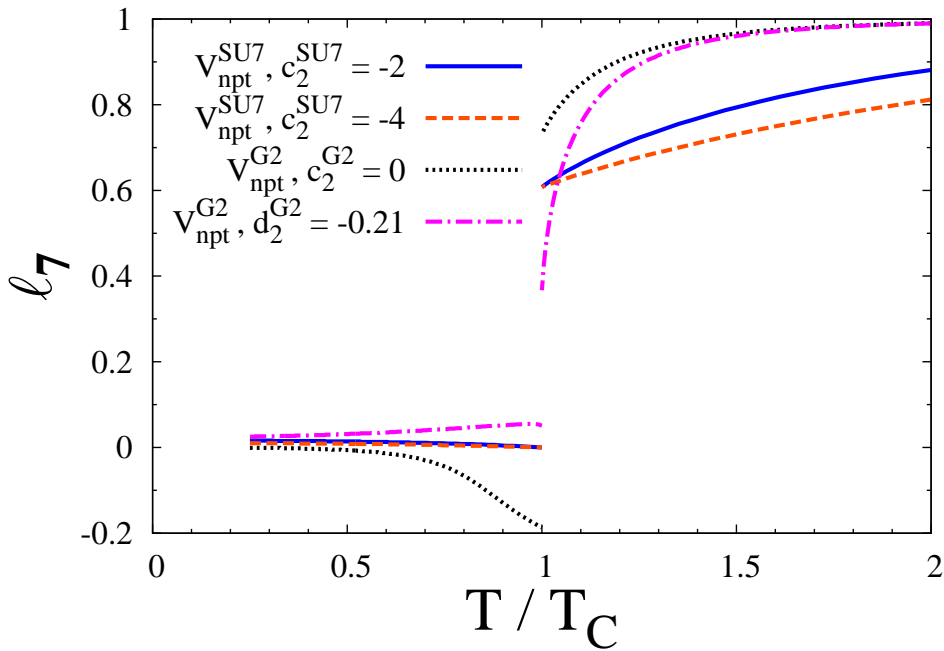}
\caption{\label{fig:G2Loopsfund}Expectation value of the Polyakov loop in
  the fundamental representation of $G(2)$ for
  the minimal $G(2)$ model, Eq. (\ref{minimal_G2_model});
  the fundamental loop model, Eq. (\ref{loop7_model});
  and two $SU(7)$ models, Eq. (\ref{SU7_model}).
}
\end{figure}

\begin{figure}
\includegraphics[width=0.8\textwidth]{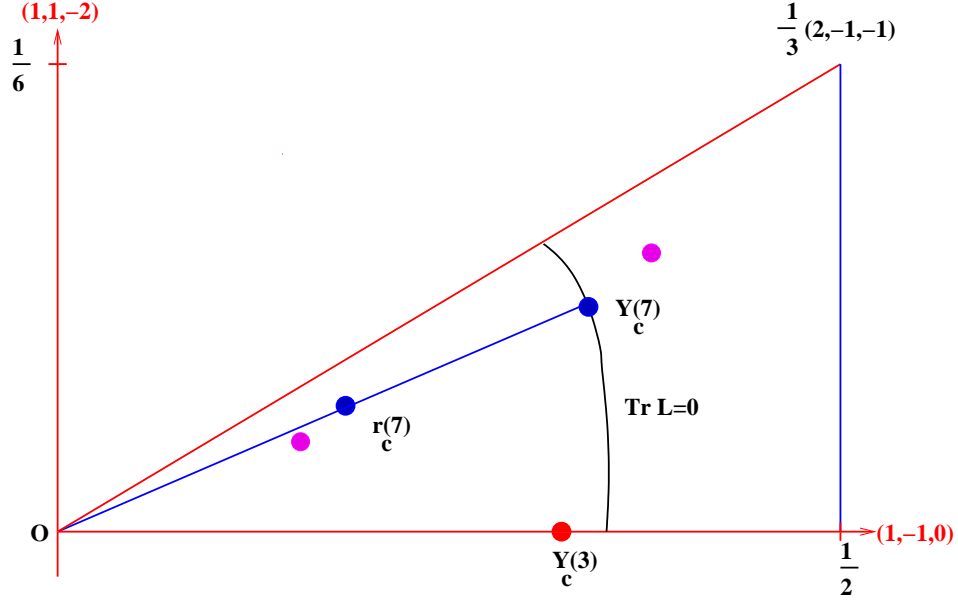}
\vspace*{-0.61cm}
\caption{Trajectories in the Weyl chamber of $G(2)$,
Fig.(\ref{fig:g2lattice}).
The crosses denote the vacua at $T_c^\pm$ for the minimal
$G(2)$ model; ${\bf Y}_c(7)$ is the confining $SU(7)$ vacuum.
}
\label{fig:weylchamberg2} 
\end{figure}

\begin{figure}[htbp]
\includegraphics[width=0.8\textwidth]{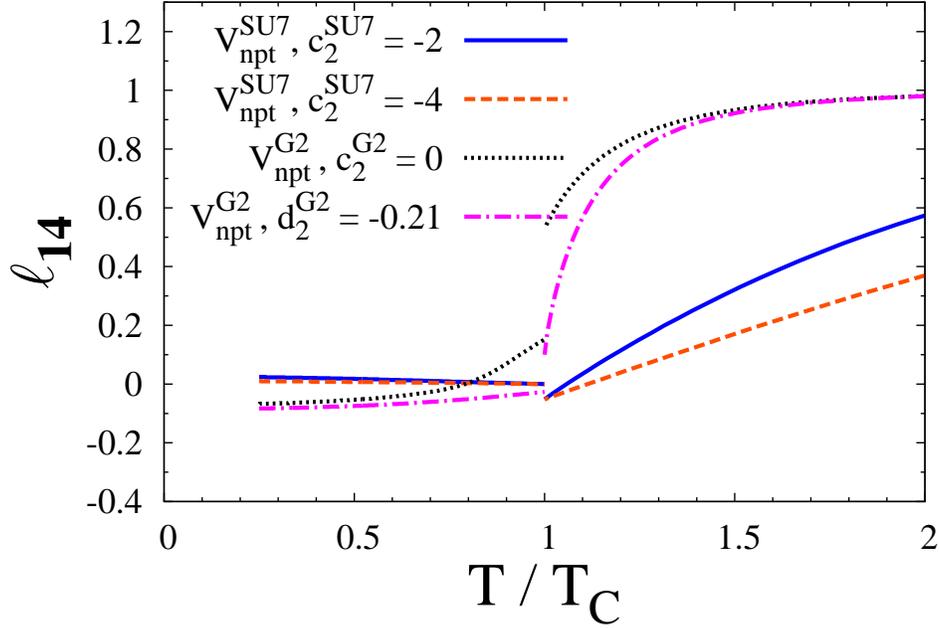}
\caption{\label{fig:G2Loopsadj}Expectation value of the Polyakov loop in
  the adjoint representation of $G(2)$ for
  the minimal $G(2)$ model, Eq. (\ref{minimal_G2_model});
  the fundamental loop model, Eq. (\ref{loop7_model});
  and two $SU(7)$ models, Eq. (\ref{SU7_model}).
}
\end{figure}

Fig.~(\ref{fig:G2q1q2})
shows the evolution of the eigenvalues of
the fundamental Polyakov loop with temperature for
the minimal $G(2)$ model, Eq. (\ref{minimal_G2_model}), 
and for the $SU(7)$ model, Eq. (\ref{SU7_model}), with
$c_2^{SU(7)} = -2$.  
We find that they remain close to although not
exactly on the \SUseven\ path $|q_1/q_2|=2$.

\begin{figure}[htbp]
\includegraphics{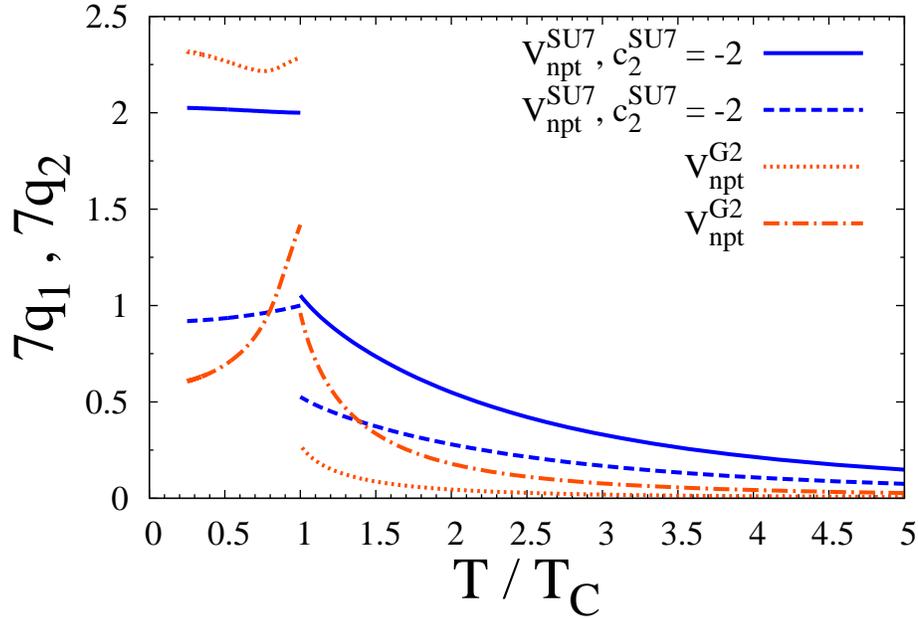}
\vspace*{-1cm}
\caption{\label{fig:G2q1q2}Eigenvalues of the Polyakov loop for the 
minimal $G(2)$ model (dotted and
  dashed-dotted lines) 
and for one $SU(7)$ model (solid and dashed lines).}
\end{figure}

In Fig.~(\ref{fig:G2pressure}) we show the pressure obtained from the
four $G(2)$ models, Eqs. (\ref{minimal_G2_model}), (\ref{loop7_model}),
and (\ref{SU7_model}).  This shows that the pressure itself is not
very useful for differentiating between models.  The pressure of
the minimal $G(2)$ model is negative below $T_c$, but this is a limitation
of our assumption that the pressure vanishes at $T_c$.

\begin{figure}[htbp]
\includegraphics{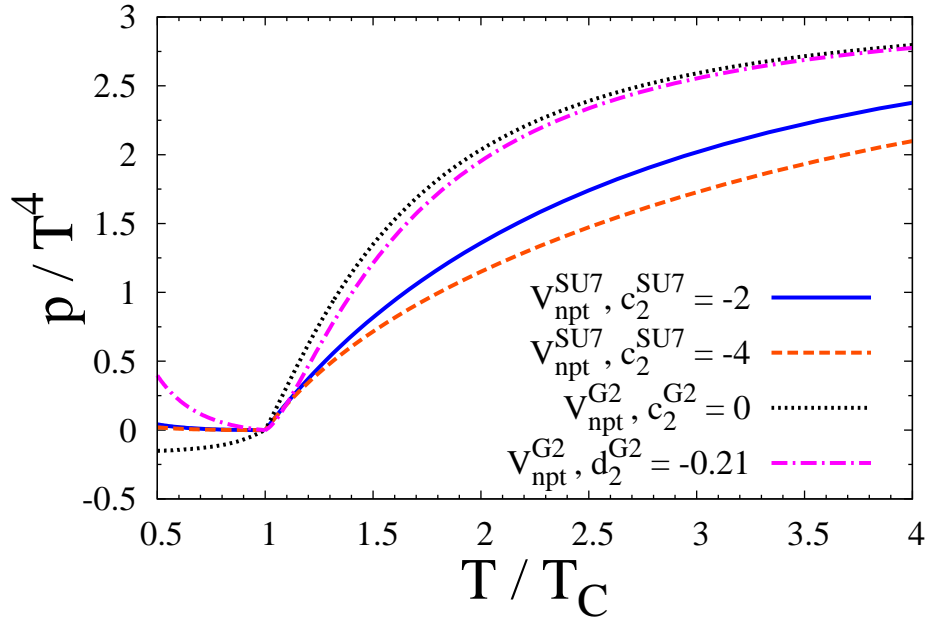}
\vspace*{-1cm}
\caption{\label{fig:G2pressure}
Pressure for our four $G(2)$ models.}
\end{figure}

In Fig.~(\ref{fig:G2e-3pA}) we show the interaction measure $(e-3p)/T^4$
obtained from the $G(2)$ models.  All of the transitions are of first
order.  
The minimal $G(2)$ model looks most like that of $SU(N)$, with a large
latent heat and a sharp peak in $(e-3p)/T^4$ near $T_c$.
The fundamental loop model also has a sharp
peak in $(e-3p)/T^4$, but its latent heat is small.  
Notice also that for the fundamental loop model,
the expectation of the fundamental and adjoint loops,
Figs. (\ref{fig:G2Loopsfund}) and (\ref{fig:G2Loopsadj}), 
is much smaller than the other models.  
This suggests that the fundamental loop model 
may be near a critical endpoint.

For the two $SU(7)$ models, the interaction measure
$(e-3p)/T^4$ does not exhibit a peak near $T_c$, but instead
drops rather slowly as the temperature increases.

\begin{figure}[htbp]
\includegraphics[width=0.8\textwidth]{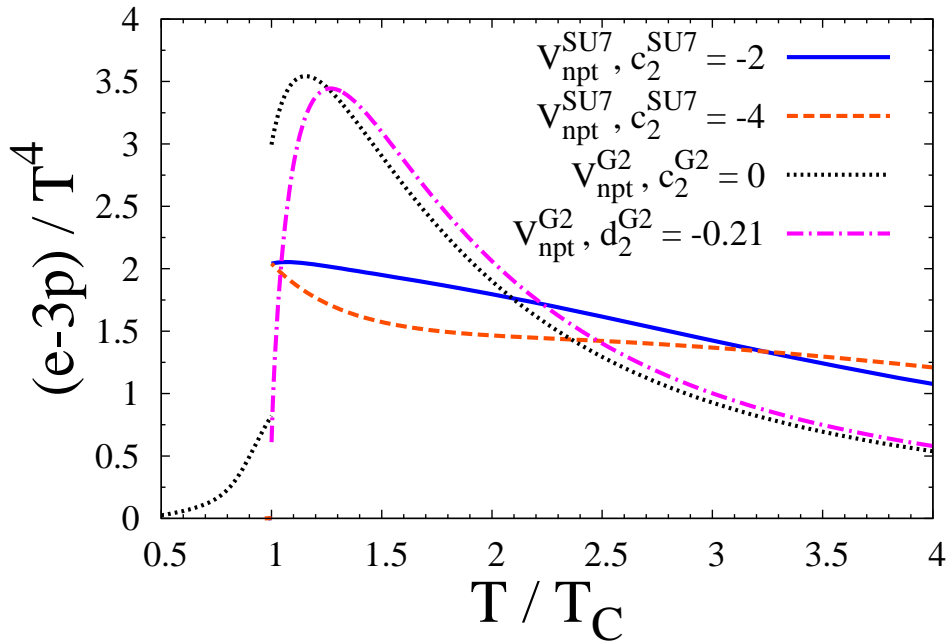}
\vspace*{-.5cm}
\caption{\label{fig:G2e-3pA} Interaction measure, $(e-3p)/T^4$, 
for our four $G(2)$ models.}
\end{figure}

Lastly, in fig. (\ref{fig:G2e-3pB}) we show the rescaled interaction
measure, $(e-3p)/T^2T_c^2$, for the four different models.
That for the minimal $G(2)$ and the fundamental loop models are
flat, while that for the $SU(7)$ model is not.  

\begin{figure}[htbp]
\includegraphics[width=0.8\textwidth]{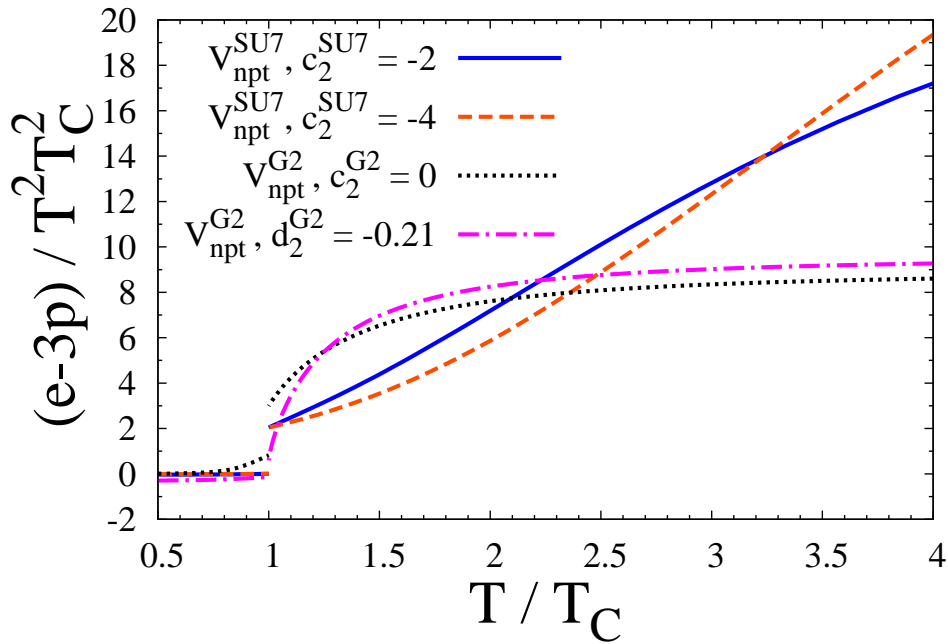}
\vspace*{-.5cm}
\caption{\label{fig:G2e-3pB} 
Rescaled interaction measure, $(e-3p)/T^2T_c^2$, 
for our four $G(2)$ models.}
\end{figure}

In conclusion, more detailed numerical simulations in a lattice
$G(2)$ gauge theory \cite{Caselle2012} will enable us to fix the parameters of
our effective model, Eq. (\ref{eq:g2su7nonpertB}).

\section{Conclusions}
\la{sec:conclusion}

In this paper we generalized the model of Ref. 
\cite{Meisinger:2001cq} and \cite{Dumitru:2010mj} to a theory with two
parameters.  The two parameters were chosen by comparing to the
interaction measure.  We then obtain results for the 't Hooft loop, or
the order-order interface tension, and for the 
order-disorder interface tension at $T_c$.  While
for most quantities the agreement is good, within $10 \%$ or so.

There is one glaring problem with the model: the results for the
Polyakov loop, Figs. (\ref{fig:nc3_compare}) and (\ref{fig:SUNLoop}),
look nothing like the lattice results, Refs. 
\cite{Gupta:2007ax} and \cite{Mykkanen:2012ri}.  The discrepancy is not
minor: on the lattice, the renormalized loop indicates a broad semi-QGP,
from $T_c$ to at least $3 \, T_c$.  In our model, the width is extremely
narrow, to only $1.2 \, T_c$.  Further, latest lattice measurements
compute the renormalized loop by explicit subtraction of the zero point
energy at zero temperature, which is theoretically an unambiguous proceedure.

Can corrections to our model
change this?  These are of two types.  One is corrections in
the coupling constant.  For example, there are corrections to
$\sim g^2$ to the perturbative potential.   It would also be possible
to compute corrections $\sim g^2$ 
to our potential, which includes both perturbative
and non-perturbative terms.  (We ignore obvious questions
of principle, namely, to what extent does the non-perturbative potential
include perturbative corrections?)
At next to 
leading order, it is known that the ${\bf q}$'s shift by an amount
$\sim g^2$, \cite{Bhattacharya:1992qb}, and the same happens in
our effective model \cite{Dumitru2013}.  Such a shift is $\sim g^2$,
and should be relatively small.  More to the point, however, in our
model the width of the transition region in the $\bf q$'s is tied
intimately to the width of the interaction measure.  No matter how
the $\bf q$'s shift, it would seem unavoidable that the width will
broaden greatly, and still give the same sharp peak in the interaction
measure.

Other corrections are those $\sim 1/N$.  From lattice measurements, though,
the behavior found for three colors \cite{Gupta:2007ax} is very similar
for four and five colors \cite{Mykkanen:2012ri}.
While the value of the Polyakov loop at the critical temperature changes
modestly with $N$, in all cases the transition region, as seen from the
renormalized Polyakov loop, remains broad.

This discrepancy must be considered the outstanding problem in the model.
We can only suggest that the effects of the non-perturbative potential
broaden what is measured as the (renormalized) Polyakov loop, although the
width of the transition remains narrow.  For example, for three
\cite{Gupta:2007ax,Megias:2007pq}, as well as for four and five colors
\cite{Mykkanen:2012ri}, there are terms $\sim 1/T^2$ in the
logarithm of the Polyakov loop.  Perhaps there is some 
non-perturbative term which broadens the measured value of the Polyakov loop,
but not the width of the transition region for the ${\bf q}$'s?

We conclude by pointing out
that there are two others ways in which
the width of the transition region can be measured,
although indirectly.  One was discussed previously
\cite{Dumitru:2010mj}.  
When the $\bf q$'s develop an expectation value, the theory
is an adjoint Higgs phase.  In our model, in principle this happens
for all $T$, but in practice it is only numerically large below $1.2 \, T_c$.
While there is no order parameter to distinguish such an
adjoint Higgs phase, it generates a characteristic splitting of masses.  
One may have to work very close to $T_c$ to see this, but it
is a necessary consequence of our model.  

The 't Hooft loop provides another way of testing a narrow transition region.
In Sec. (\ref{sec:interface}) we gave results
for the 't Hooft loop in the semi-QGP.  Better measurements of
these quantities should provide a stringent test of our model.

For four or more colors, there are further tests.
For four colors, besides the simplest 't Hooft loop, 
between a Polyakov loop with phase $1$ and $i$, there is also
a loop between $1$ and $-1$.  To $\sim g^3$, the ratio
of these 't Hooft loops satisfy
Casimir scaling \cite{Giovannangeli:2002uv, *Giovannangeli:2004sg}.
This Casimir scaling is observed to a very good precision over a
wide range in temperature \cite{deForcrand:2005rg}.
We have not computed the ratio of these 't Hooft loops in our model,
but suggest that it will produce small, but measureable deviations from
Casimir scaling in the ratio of such 't Hooft loops, certainly
in the region close to $T_c$, for $T < 1.2 \, T_c$.

In conclusion, we suggest that detailed measurements 
of the pure glue theory, especially near $T_c$,
can help illuminate how the law of maximal 
eigenvalue repulsion acts to generate confinement.

\begin{acknowledgments}
The research of A.D.\ was supported by the U.S. Department of
Energy under contract \#DE-FG02-09ER41620, and
by PSC-CUNY research grant 64132-00~42;
of Y.H. by a Grant-in-Aid for Scientific Research (No.23340067) from
the Ministry of Education, Culture, Sports, Science and Technology (MEXT) of
Japan; of R.D.P.,
by the U.S. Department of Energy under contract \#DE-AC02-98CH10886.
C.P.K.A. thanks the Nuclear Theory Group at 
BNL for their warm hospitality. 
We thank T. Umeda and the WHOT collaboration for sharing their data with us
\cite{Umeda:2008bd}, which enabled us to plot Fig. (\ref{figwhot}).
R.D.P. thanks J. Pawlowski for numerous discussions on the model
of Ref. \cite{Braun:2007bx}, which also exhibits a narrow transition region;
V. Begun, for bringing \cite{Begun:2010eh} to his attention;
and A. Bazavov, F. Karsch, M. Panero, and P. Petreczky for discussions.
C.P.K.A. thanks Oleg Ogievetsky
and Loic Poulain d'Andecy for their explanations about classical groups.

\end{acknowledgments}

\bibliography{eff}

\end{document}